\documentclass[11pt]{article}

\usepackage[usenames,dvipsnames,svgnames,table]{xcolor} 
\usepackage[obeyspaces,hyphens,spaces]{url}
\usepackage{jcapmod}
\usepackage[english]{babel}
\usepackage{amsmath, amssymb, amsbsy, amstext, amsthm,mathtools}
\usepackage{graphicx}
\usepackage{amsfonts}
\usepackage{fouridx}
\usepackage{cases}
\usepackage{bm}
\usepackage{bbm}
\usepackage{cleveref}
\usepackage{slashed}
\usepackage{simpler-wick}

\usepackage{setspace}

\usepackage{braket}

\usepackage{subcaption}

\definecolor{Blue}{rgb}{0.25, 0.41, 0.88}
\definecolor{Red}{rgb}{0.92,0.,0.}
\definecolor{darkorange}{rgb}{1.0,0.549,0.}
\definecolor{cobalt}{RGB}{44, 98, 120}
\definecolor{Mathematica1}{rgb}{0.368417, 0.506779, 0.709798}
\definecolor{Mathematica2}{rgb}{0.880722, 0.611041, 0.142051}
\definecolor{Mathematica3}{rgb}{0.560181, 0.691569, 0.194885}
\definecolor{Mathematica4}{rgb}{0.922526, 0.385626, 0.209179}
\definecolor{Mathematica5}{rgb}{0.528488, 0.470624, 0.701351}
\definecolor{Mathematica6}{rgb}{0.772079, 0.431554, 0.102387}
\definecolor{Mathematica7}{rgb}{0.363898, 0.618501, 0.782349}
\definecolor{Mathematica8}{rgb}{1, 0.75, 0}
\definecolor{Mathematica9}{rgb}{0.647624, 0.37816, 0.614037}
\definecolor{plotBlue}{RGB}{94, 130, 181}
\definecolor{plotRed}{RGB}{233, 85, 54}
\definecolor{plotGreen}{RGB}{142, 176, 50}
\definecolor{plotPurple}{RGB}{135, 120, 178}

\DeclareMathOperator{\re}{Re}

\newcolumntype{C}[1]{>{\centering\let\newline\\\arraybackslash\hspace{0pt}}m{#1}}

\makeatletter
\newlength{\apb@width}
\newcommand{\autoparbox}[2][c]{\settowidth{\apb@width}{#2}\parbox[#1]{\apb@width}{#2}}

\makeatother

\makeatletter
\newsavebox\myboxA
\newsavebox\myboxB
\newlength\mylenA

\newcommand*\xoverline[2][0.75]{
    \sbox{\myboxA}{$\m@th#2$}%
    \setbox\myboxB\null
    \ht\myboxB=\ht\myboxA%
    \dp\myboxB=\dp\myboxA%
    \wd\myboxB=#1\wd\myboxA
    \sbox\myboxB{$\m@th\overline{\copy\myboxB}$}
    \setlength\mylenA{\the\wd\myboxA}
    \addtolength\mylenA{-\the\wd\myboxB}%
    \ifdim\wd\myboxB<\wd\myboxA%
       \rlap{\hskip 0.5\mylenA\usebox\myboxB}{\usebox\myboxA}%
    \else
        \hskip -0.5\mylenA\rlap{\usebox\myboxA}{\hskip 0.5\mylenA\usebox\myboxB}%
    \fi}
\makeatother

\numberwithin{equation}{section}

\def\beq{\begin{equation}}
\def\eeq{\end{equation}}

\def\bea{\begin{eqnarray}}
\def\eea{\end{eqnarray}}

\newcommand{\ud}{\mathrm{d}}
\newcommand{\lab}[1]{{\mathrm{#1}}}
\newcommand{\slab}[1]{{\textsc{#1}}}
\newcommand{\mb}[1]{{\mathbf{#1}}}
\newcommand{\minus}{{\scalebox {0.75}[1.0]{$-$}}}
\newcommand{\sminus}{{\scalebox {0.6}[0.85]{$-$}}}

\newcommand{\dc}{{\hat{n}}}

\newcommand{\nord}[1]{{:\mathrel{#1}:}}

\theoremstyle{definition}

\DeclareRobustCommand{\SkipTocEntry}[4]{}

\newcommand{\es}{\hspace{0.5pt}}

\definecolor{blue2}{cmyk}{1, 0.1, 0.1, 0.1}

\definecolor{pyBlue}{RGB}{31, 119, 180}
\definecolor{pyRed}{RGB}{214, 39, 40}
\definecolor{pyGreen}{RGB}{44, 160, 44}
\definecolor{pyBlue2}{RGB}{0, 111, 237}
\definecolor{pyRed2}{RGB}{224, 52, 36}
\definecolor{Mathematica1}{rgb}{0.368417, 0.506779, 0.709798}
\definecolor{Mathematica2}{rgb}{0.880722, 0.611041, 0.142051}
\definecolor{Mathematica3}{rgb}{0.560181, 0.691569, 0.194885}
\definecolor{Mathematica4}{rgb}{0.922526, 0.385626, 0.209179}
\definecolor{Mathematica5}{rgb}{0.528488, 0.470624, 0.701351}

\newcommand{\subp}{{\scriptscriptstyle +}}
\newcommand{\subm}{{\scriptscriptstyle -}}
\newcommand{\vac}{\Psi}

\DeclareMathOperator{\arctanh}{arctanh}

\DeclareMathOperator{\sign}{sgn}

\begin{document}

\pagenumbering{roman}
\begin{titlepage}
\baselineskip=15.5pt \thispagestyle{empty}

\bigskip\

\vspace{1cm}
\begin{center}
{\LARGE \selectfont  {\bfseries  Infinite Distance Limits and  Information Theory}}
 
\end{center}
\vspace{0.1cm}
\begin{center}
{\fontsize{12}{18}\selectfont John Stout} 
\end{center}

\begin{center}
\vskip8pt
\textit{Department of Physics, Harvard University, Cambridge, MA 02138, USA}

\end{center}

\vspace{1.2cm}
\hrule \vspace{0.3cm}
\noindent {\bf Abstract}\\[0.1cm]
	The classical information metric provides a unique notion of distance on the space of probability distributions with a well-defined operational interpretation: two distributions are far apart if they are readily distinguishable from one another. The quantum information metric generalizes this to the space of quantum states, and thus defines a notion of distance on an arbitrary continuous family of quantum field theories via their vacua that is proportional to the metric on moduli space when restricted appropriately. In this paper, we study this metric and its operational interpretation in a variety of examples. We specifically focus on why and how infinite distance singularities appear. We argue that two theories are infinitely far apart if they are hyper-distinguishable: that is, if they can be distinguished from one another, with certainty, using only a few measurements. We explain why such singularities appear for the simple harmonic oscillator yet are absent for quantum field theories near a typical quantum critical point, and show how an infinite distance point can emerge when a tower of fields degenerates in mass. Finally, we use this perspective to provide a potential bottom-up motivation for the Swampland Distance Conjecture and indicate how we might extend it beyond current lampposts.
	
\vskip10pt
\hrule
\vskip10pt

\emailAdd{johnstout@g.harvard.edu}

\end{titlepage}

\thispagestyle{empty}
\setcounter{page}{2}
\begin{spacing}{1.03}
\tableofcontents
\end{spacing}

\clearpage
\pagenumbering{arabic}
\setcounter{page}{1}

\newpage

\section{Introduction}

	One of the central aims of theoretical physics is to characterize the space of theories consistent with a given set of fundamental principles and to understand how they are related to one another~\cite{Douglas:2010ic}. For instance, we might be interested in the general space of quantum field theories, i.e. the set of theories that are simultaneously consistent with both quantum mechanics and special relativity. Or we could restrict our attention to a more constrained class, such as those theories that are conformal or enjoy some amount of supersymmetry. Provided such a space, how do we study it? Intuitively, we might try to encode its behavior and the relationship between different theories geometrically. Is there a natural notion of distance on this space? What does this distance encode, and what does it mean when two theories are infinitely far apart? The goal of this paper is to borrow techniques from information theory to study these questions, with a specific focus on infinite distance limits in families of quantum field theories.

	These questions are well-studied in specific classes of theories. Great progress has been made in understanding and classifying the moduli spaces of vacua that naturally arise in supersymmetric theories. These families are parameterized by the vacuum expectation values of scalar operators and are naturally equipped with a metric compatible with the underlying supersymmetry, i.e. the metric derived from the pre- or K\"{a}hler potential used to define the theory \cite{Seiberg:1994bz,Seiberg:1994rs,Intriligator:1995au,Dijkgraaf:1997ip,Klemm:1997gg,DHoker:1999yni,Strassler:2003qg,Freedman:2012zz}.  Similarly, continuous families of conformal field theories (CFTs)---related to one another by exactly marginal deformations---also come equipped with a natural metric, the so-called Zamolodchikov metric~\cite{Zamolodchikov:1986gt,Seiberg:1988pf,Kutasov:1988xb,dijkgraaf1989geometrical,Ranganathan:1993vj,Cecotti:1991me,Leigh:1995ep,Seiberg:1999xz,deBoer:2008ss,Papadodimas:2009eu,Green:2010da,Gerchkovitz:2014gta,Behan:2017mwi}. These two metrics can sometimes be related \cite{Seiberg:1988pf,Cvetic:1989ii,Candelas:1989qn,Candelas:1989ug,Greene:1996cy,Polchinski:1998rq,Polchinski:1998rr}.

	Unfortunately, these notions of distance on theory space are so intrinsically tied to the underlying symmetries of these families that it is not clear what general lessons can be learned from them. This problem becomes sharper in light of the intense current interest in the Swampland Distance Conjecture \cite{Ooguri:2006in,Heidenreich:2018kpg,Grimm:2018ohb,Brennan:2017rbf,Palti:2019pca,vanBeest:2021lhn} and its CFT analog \cite{Baume:2020dqd,Perlmutter:2020buo}. These conjectures propose quantum gravitational constraints on the geometry of these theory spaces, and are specifically focused on theories at infinite distance from all others. There is a general expectation that these theories are somehow ``different'' from others in the moduli space. This is borne out by examples where, for instance, these points correspond to the restoration of a global symmetry and the appearance of a tower of exponentially light fields. Are these infinite distance points a lamppost effect or are they a general phenomenon? What is their interpretation? Can we quantify ``different?''

	In this paper, we study the quantum information metric \cite{provost1980,Wootters:1981ki,Berry:1984jv,Wilczek:Phases,Zanardi_2006,GU_2010,Venuti:2007qcs,Carollo:2019ygj} associated to the vacua of arbitrary continuous families of quantum field theories. This information metric is proportional to both the metric on moduli space and the Zamolodchikov metric when appropriately restricted. However, this metric generalizes these distances since it can be defined for any continuous family. Furthermore, the quantum information metric is naturally related to the unique metric on the space of classical probability distributions, called the classical information metric. This provides a precise operational interpretation for these distances based on distinguishability. We will specifically focus on cases where this so-called statistical distance diverges, and we will argue that two theories are infinitely far apart if they are \emph{hyper-distinguishable}: they can be distinguished with \emph{certainty} using only a few measurements. We then use this perspective to provide a potential bottom-up motivation for these quantum gravitational distance conjectures.

	The quantum information metric has appeared in the condensed matter literature as a way of identifying and characterizing continuous quantum phase transitions. Since we expect that the different phases of a theory exhibit qualitatively distinct behavior, they should be readily distinguishable from one another. Typically, these qualitative differences are detected by studying the analytic structure of some carefully chosen order parameter, as a function of the parameters or Wilson coefficients of the theory. Crossing a phase boundary causes this order parameter to change discontinuously and so we can identify different phases with its different regions of analyticity. The information metric, instead, quantifies the difference between two nearby theories by how well we can distinguish them using \emph{any} measurement---there is no need to identify a discontinuous order parameter, nor does there need to be one. 

	Since different phases are qualitatively distinct, the information metric places them far apart from one another: phase boundaries are tied to metric singularities. For instance, a typical quantum critical point is characterized by a vanishing energy gap, wherein some finite number of fields become massless, the correlation length diverges, and the theory becomes scale-invariant. Such critical points generally obey the scaling hypothesis, wherein different observables (like the energy gap or correlation length) exhibit definite and interrelated scaling behavior, characterized by a set of critical exponents. As we might expect, the information metric can diverge in some power of the correlation length. However, we will see that this divergence is never strong enough to place the critical theory at infinite distance from its neighbors, assuming that it obeys typical scaling laws. That is, an infinite distance point in the information metric requires a violation of the ubiquitous scaling behavior found at quantum critical points, and so these infinite distance points represent a class of critical theories that are qualitatively distinct from the ones typically encountered. For example, will show that one way these infinite distance points can emerge is if an infinite tower of non-interacting fields degenerates in mass.

	Why, then, should infinite distance points---like those associated to the restoration of a continuous global symmetry and a corresponding exactly conserved charge---appear in quantum gravitational theories? This interpretation in terms of the information metric suggests a simple answer: because they may be distinguished from the others in the family \emph{with certainty} by an observation of charge non-conservation. While we do not prove that these infinite distance points are always generated by towers of fields (and in fact, we provide counter-examples to show that they are not in quantum field theory), we show that this is a sufficient mechanism to generate the infinite distance point, and so the existence of the tower may be intrinsically tied to insuring that these points remain hyper-distinguishable. 

	The aim of this work is to understand the nature of such infinite distance points without reference to specific families of string compactifications or conformal field theories, and thus to push beyond these well-studied lampposts. As such, it represents a first step towards a more rigorous understanding of these infinite distance limits in general, information-theoretic terms and aims to provide intuition about them through a variety of examples in diverse dimensions.

	\newpage

	\subsubsection*{Outline}

		We recognize that many of the concepts we use in this paper are not immediately familiar to our intended audience of high energy theorists, especially in the way that we use them. For this reason, in Section~\ref{sec:cim}, we provide a quick introduction to relevant concepts in classical information theory. Here, we introduce the notions of statistical length and distinguishability, focusing on classical probability distributions with few degrees of freedom where we can use familiar toy examples to illustrate and study infinite statistical distance limits. We extend these notions to the quantum realm in Section~\ref{sec:qim}, where we discuss the definition, interpretation, computation, and regulation of the quantum information metric in field theories. Its behavior around quantum critical points is analyzed in \S\ref{sec:metricScaling}, where we argue that such points are always at finite distance as long as obey standard scaling relations.  We compute the quantum information metric for a variety of examples in Section~\ref{sec:examples}. After studying the metric for the simple harmonic oscillator (\S\ref{sec:sho}), we compute the metric associated to mass deformations of free scalar~(\S\ref{sec:scalar}) and fermionic~(\S\ref{sec:dirac}) fields in arbitrary spatial dimensions. We then discuss the equivalence of the information metric to the Zamolodchikov metric (\S\ref{sec:cft}) and the metric on field space for a general bosonic nonlinear sigma model (\S\ref{sec:nlsm}). Our free field results can be easily extended to a tower of non-interacting fields and, in Section~\ref{sec:towers}, we describe how an infinite distance point can emerge from their collective behavior. We describe potential implications of this picture for the Swampland program in Section~\ref{sec:swamp} and present our conclusions in Section~\ref{sec:conclusions}.

		\subsubsection*{Conventions}

			We work in $(d+1)$-dimensional spacetime, where $d$ is the spatial dimension. We will generally work in Euclidean signature and use $\tau$ to denote Euclidean time, while bold-faced Roman letters like $\mb{x}$ denote spatial points. We use Greek indices at the end of the alphabet, $\mu, \nu, \ldots = 0, \ldots\!\es\es\es, d$, as spacetime indices, while those at the beginning $\alpha, \beta, \ldots = 1, \ldots \!\es\es\es, 2^{\lfloor \frac{d+1}{2} \rfloor}$ denote Dirac spinor indices. We use $N$ to denote the number of samples taken from a distribution, $\mathcal{N}$ to denote the number of degrees of freedom, $n$ to denote the dimension of parameter or moduli space, and $\dc$ to denote lattice dimension. We will generally use $\varphi^a$, where Roman letters at the beginning of the alphabet ${a = 1, \ldots\!\es\es\es, n}$ to denote coordinates on our $n$-dimensional parameter or moduli space, or equivalently coordinates on the statistical manifold. We will suppress the index of these coordinates when they appear in the argument of a function, i.e.~$m^2(\varphi^a) = m^2(\varphi)$. We will use $\mathcal{D}[\,\cdot\,|\,\cdot\,]$ to denote statistical divergences, which are asymmetric in their arguments, and $\mathcal{D}(\,\cdot\,,\,\cdot\,)$ to denote statistical distances, which are symmetric. There are a variety of such measures, and we will distinguish them with subscripts.

\newpage
	\section{The Classical Information Metric} \label{sec:cim}

			It will be helpful to first introduce relevant concepts in classical information theory before we move onto quantum information theory. This will help us motivate the quantum information metric and understand its general behavior. Those interested in more detail expositions should consult \cite{cover2006elements, Caticha:2008eso,amari2016information,nielsen2020elementary} for introductions slanted towards information theorists.

			The main goal of this section is to introduce the notion of statistical length and the associated classical information metric. A family of probability distributions $p(x\es |\es \varphi)$ for a random variable $x$ forms a so-called statistical manifold, parameterized by the continuous coordinates $\varphi^a$. The classical information metric endows this manifold with a distance based on distinguishability---two distributions are far apart in this metric if they can be distinguished from one another with certainty with only a few measurements of $x$. Infinite distance points correspond to those distributions that are qualitatively different from others in the family and are thus hyper-distinguishable.

			We will illustrate these concepts in the familiar exponential family of probability distributions, which we review in \S\ref{sec:exp}. This family forms a statistical manifold, and in~\S\ref{sec:divergences} we describe how to define notion of distance, a divergence, on this manifold based on distinguishability. In \S\ref{sec:statLength}, we discuss how the classical information metric emerges from this divergence for infinitesimally separated points, and we describe its properties and the interpretation of its associated statistical length. This provides a precise interpretation of what is happening in the family of theories when this length diverges, which we provide in \S\ref{sec:infStatLength}. Finally, we study a toy model, evocative of those found in quantum and statistical field theory, of these infinite distance singularities in \S\ref{sec:toy}.

			\subsection{The Exponential Family} \label{sec:exp}
			
			To aide intuition, we will first introduce relevant concepts in classical information theory using the \emph{exponential family} of distributions, since they are the most familiar to physicists. However, the conclusions of this discussion will apply more generally and are not restricted to just this~family.

			The exponential family is comprised of probability distributions with the form
			\begin{equation}
				p(x\es |\es \lambda) = \frac{1}{\mathcal{Z}(\lambda)} \lab{e}^{-k(x) + \lambda_a f^a(x)}\,, \label{eq:expFam}
			\end{equation}
			where we take $x \in \mathbb{R}^N$ and use the function $k(x)$ to define an intrinsic measure on the space. The family depends on the provided functions $f^a(x)$ and the corresponding constants, called the Lagrange multipliers, $\lambda_a \in \mathbb{R}$, with $a = 1, \dots, n$. Dividing by $\mathcal{Z}(\lambda) = \int\!\ud^N \!x\, p(x\es|\es\lambda)$, the so-called partition function, ensures that the distribution is properly normalized. 

			These distributions can be parameterized in two different ways. The first is clear from our notation---we may use the Lagrange multipliers $\lambda_a$ to unambiguously specify a probability distribution for $x$. However, we can arrive at another parameterization by recognizing a fact fundamental \cite{Wightman:1979} to both statistical mechanics and thermodynamics: the logarithm of the partition function
			\begin{equation}
				\mathcal{F}(\lambda) = \log \mathcal{Z}( \lambda) \label{eq:freeEn}
			\end{equation} 
			is convex. This can be shown by checking that its Hessian is positive definite for all $\lambda_a$. Convex (and concave) functions are very special since there is a one-to-one correspondence between the $\lambda_a$ and the derivatives
			\begin{equation}
				\varphi^a \equiv \langle  f^a(x) \rangle = \frac{\partial \log \mathcal{Z}(\lambda)}{\partial \lambda_a}\,. \label{eq:fLambda}
			\end{equation}
			As written here, these expectation values are functions of the Lagrange multipliers, $\varphi^a = \varphi^a(\lambda)$. Convexity implies that there is a unique inverse $\lambda_a = \lambda_a(\varphi)$, so instead of using the Lagrange multipliers $\lambda_a$ to parameterize this family, we can alternatively use the expectation values $\varphi^a$.\footnote{That the exponential family (\ref{eq:expFam}) admits these two equivalent parameterizations can be seen most naturally by noting that they are the unique distributions that maximize the relative entropy or Kullback-Leibler divergence (\ref{eq:klDiv}), subject only to the constraints $\langle f^a(x) \rangle = \varphi^a$. Since this objective is (by design) concave, its maximum is unique and so these distributions can be parameterized by the expectation values $\varphi^a$. Furthermore, the $\lambda_a$ show up as Lagrange multipliers in this maximization procedure (justifying their name). It then naturally follows from this constrained optimization procedure that they too uniquely parameterize this family of distributions.}  Specifying either $n$-tuple $\lambda_a$ or $\varphi^a$ uniquely fixes a particular probability distribution for $x$, and the relationship between the two parameterizations is determined by~(\ref{eq:fLambda}). A natural object to then define is the Legendre transform of $\mathcal{F}(\lambda)$,
			\begin{equation}
				\tilde{\mathcal{F}}(\varphi) = \lambda_a \varphi^a -  \mathcal{F}(\lambda)\,,
			\end{equation}
			whose negative is called the entropy $S(\varphi) = -\tilde{\mathcal{F}}(\varphi)$, where we implicitly take $\lambda_a = \lambda_a(\varphi)$.

			Familiar examples of this family are the Gaussian and Boltzmann distributions. The Gaussian of mean $\mu$ and variance $\sigma^2$,
			\begin{equation}
				p(x \es |\es \mu, \sigma) = \frac{1}{\sqrt{2 \pi \sigma^2}} \lab{e}^{-(x - \mu)^2/(2 \sigma^2)}\,, \label{eq:gaussian}
			\end{equation}
			has a two-dimensional parameter space $(\varphi^1, \varphi^2) = (\mu, \sigma^2 + \mu^2)$ corresponding to expectation values of $\big(f^1(x), f^2(x)\big) = (x, x^2)$. As written above, the Lagrange multipliers are $(\lambda_1, \lambda_2) = \big( \mu/\sigma^2, \minus 1/(2 \sigma^2)\big)$ and the intrinsic measure is zero, $k(x) = 0$. A simple calculation yields the Legendre pair
			\begin{equation}
				\mathcal{F}(\lambda_1, \lambda_2) = -\frac{\lambda_1^2}{4 \lambda_2} + \frac{1}{2} \log \left(\frac{\pi}{\minus \lambda_2}\right) \quad \text{and}\quad \tilde{\mathcal{F}}(\varphi^1, \varphi^2) = -\frac{1}{2} \left(1 + \log 2\pi\left[\varphi^2 - (\varphi^1)^2\right]\right)\,,
			\end{equation}
			which are both convex as long as $\lambda_2 \leq 0$ and $\varphi^2 \geq (\varphi^1)^2$, or equivalently $\sigma^2 > 0$. 

			This is an unusual presentation for the simple Gaussian distribution, though this language is more familiar for the Boltzmann distribution. For instance, if $x \in \mathbb{R}^{6 N}$ represents the positions and momenta of $N$ particles in three spatial dimensions, i.e. the position of the system in phase space, the Boltzmann distribution
			\begin{equation}
				p(x \,|\,\beta) = \frac{1}{\mathcal{Z}(\beta)} \lab{e}^{-\beta \mathcal{H}(x)}
			\end{equation}
			represents a family of models parameterized by either the Lagrange multiplier $\lambda_1 = -\beta$ (the coolness or inverse temperature) or the associated expectation value $\varphi^1 = \langle \mathcal{H}(x) \rangle = E$ (the average energy). The function $-\beta \mathcal{F}(\beta) = A(\beta)$ is then the (Helmholtz) free energy and its Legendre dual $-\tilde{\mathcal{F}}(E) = S(E)$ is the entropy of the system. The extra signs and factors of $\beta$ in these definitions follow from an unfortunate historical accident. That these functions are convex (or concave) follows from the requirement that the system is thermodynamically stable, which is necessary for us to be able to describe the system using thermodynamics \cite{Wightman:1979}.

			\subsection{Divergences and Distinguishability} \label{sec:divergences}

\begin{figure}
				\centering
				\includegraphics{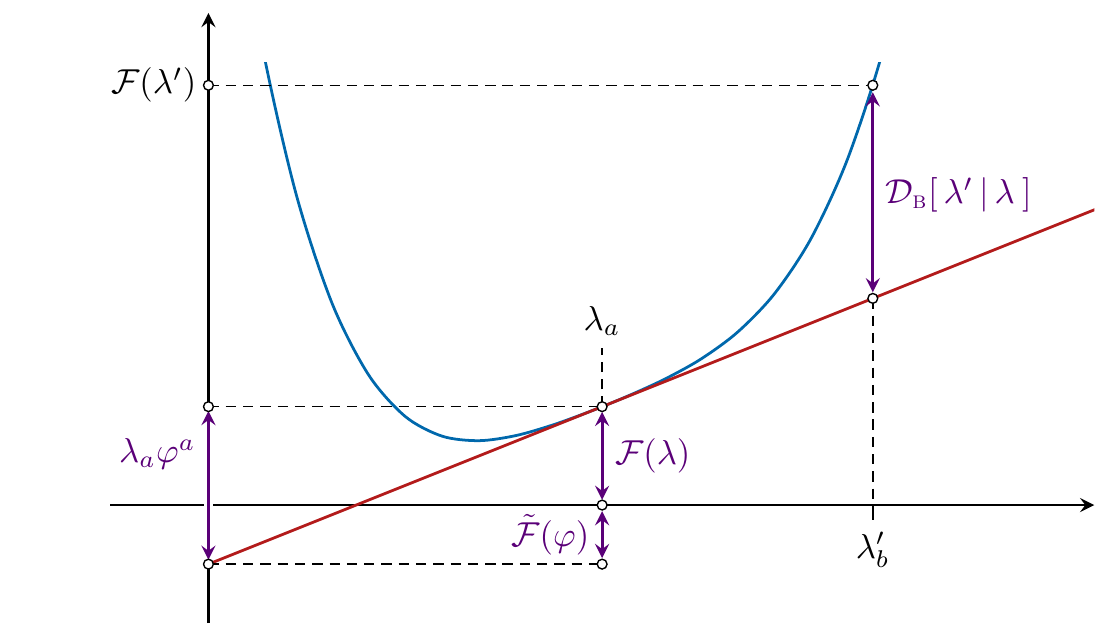}
				\caption{The Bregman divergence $\mathcal{D}_\slab{b}[\lambda', \lambda]$ measures the separation between a reference point $\lambda_a$ and another point $\lambda_b'$ with respect to a convex function $\mathcal{F}(\lambda)$. It is defined geometrically as the vertical distance (shown here as the right-most purple line) between $\mathcal{F}(\lambda')$ and the tangent plane (shown here in red) that uniquely intersects $\mathcal{F}(\lambda)$. \label{fig:bregman}}
			\end{figure}

			We can think of this family of distributions as comprising a \emph{statistical manifold}, the points of which are probability distributions that are either specified by the expectation values $\varphi^a$ or the Lagrange multipliers $\lambda_a$. We would now like to define a notion of distance on this space that places both parameterizations on an equal footing. Given the geometric relationship between these two sets of coordinates, it is natural to define this distance geometrically.
			Given any convex function $\mathcal{F}(\lambda)$, we can define the \emph{Bregman divergence} between the two points $\lambda_a$ and $\lambda'_{a}$ by 
			\begin{equation}
				\mathcal{D}_{\textsc{b}}[\, \lambda' \,|\, \lambda \,] = \mathcal{F}(\lambda') - \mathcal{F}(\lambda) -   \frac{\partial \mathcal{F}(\lambda)}{\partial \lambda_a} \left(\lambda'_a - \lambda_a\right)\,. \label{eq:bregman}
			\end{equation}
			As illustrated in Figure~\ref{fig:bregman}, this can be interpreted geometrically as the vertical distance between the point $\big(\lambda', \mathcal{F}(\lambda')\big)$ and the tangent plane that intersects $\big(\lambda, \mathcal{F}(\lambda)\big)$. The Legendre transform $\tilde{\mathcal{F}}(\varphi)$ has a similar interpretation. Since $\mathcal{F}(\lambda) + \tilde{\mathcal{F}}(\varphi) = \varphi^a \lambda_a$, $\tilde{\mathcal{F}}(\varphi)$ is just the vertical distance between this tangent plane and the plane $\mathcal{F} = 0$. The divergence then takes a similar form when expressed in terms of its Legendre transform,
			\begin{equation}
				\mathcal{D}_{\textsc{b}}[\,\varphi' \, |\, \varphi\,] = \tilde{\mathcal{F}}(\varphi) - \tilde{\mathcal{F}}(\varphi') - \frac{\partial \mathcal{F}(\varphi')}{\partial \varphi'^a} \left(\varphi^a - \varphi'^a\right)\,,
			\end{equation}
			though the position of its arguments interchange.

			This Bregman divergence successfully measures separation between two points on the statistical manifold and treats the dual variables $\lambda$ and $\varphi$ on equal footing. Unfortunately, it fails to define an actual distance on this manifold since it is asymmetric in its arguments---this is why we called it a divergence instead of a distance. However, it still satisfies several properties that we expect of a measure of separation---it is non-negative, $\mathcal{D}[\,\lambda'\,|\, \lambda\,] \geq 0$, and only vanishes when the two points are equal, $\lambda' = \lambda$. Both of these properties are readily apparent from Figure~\ref{fig:bregman}, and they will be key to defining an actual notion of statistical distance between the two points $\lambda$ and $\lambda'$, or equivalently their duals $\varphi$ and $\varphi'$.

			The exponential family's Bregman divergence defined via the convex function (\ref{eq:freeEn}) is also called the Kullback-Leibler (KL) divergence or relative entropy,
			\begin{equation}
				\mathcal{D}_{\textsc{kl}}[\, p(x)\,|\, q(x)\,] = \int\!\ud^N x\, p(x) \log \frac{p(x)}{q(x)}\,. \label{eq:klDiv}
			\end{equation}
			This quantity plays a preferred role\footnote{Though there are many information theoretic divergences like the so-called $f$-divergences, which replace the $\log$ in~(\ref{eq:klDiv}) with a generic function $f\big(p(x)/q(x)\big)$, the KL divergence is the unique divergence (up to an overall scale) that satisfies some relatively simple axioms, see e.g.~\cite{Hobson:1969ant,Caticha:2008eso}. Fortunately, the Hessian of these divergences are all equivalent (again up to an overall scale) and so the precise choice of this divergence does not matter in defining the statistical distance.} in information theory and this definition applies for arbitrary distributions $p(x)$ and $q(x)$, not necessarily in the exponential family. Furthermore, it also applies to distributions over discrete variables, in which case the integral is replaced by a sum. It is a special feature of the exponential family that the Bregman and KL divergences are equivalent,
			\begin{equation}
				\mathcal{D}_{\textsc{kl}} \big[ \, p(x \es|\es \varphi') \,|\, p(x \es |\es \varphi)\,\big] = \mathcal{D}_{\textsc{b}} [\, \varphi' \, |\, \varphi\,]\,,
			\end{equation}
			and so, at least in the exponential family, the KL divergence also has the simple geometric interpretation shown in Figure~\ref{fig:bregman}.

			Importantly, rephrasing things in terms of the relative entropy provides us with a useful operational interpretation of the divergence between two distributions. We can illustrate this with an elementary example, also in the exponential family. Let us consider a biased coin, which either lands heads or tails with probability $p$ and $1-p$, respectively. We may then ask: assuming that we flip the coin $N$ times, what is the probability that we are \emph{fooled} into thinking that the coin instead follows the probability distribution $(f, 1-f)$? The probability that the coin lands heads $m = N f$ times is
			\begin{equation}
				P_N( f \, |\, p) = \binom{N}{m}\, p^m (1-p)^{N - m}\,.
			\end{equation}
			In the limit of a large number of samples, we find that this distribution is extremely strongly peaked about $f = p$\,,
			\begin{equation}
				P_N(f \, |\,p) \sim \frac{\exp\big(\minus N \mathcal{D}_{\textsc{kl}}[\es f\, |\,p\es]\big)}{\sqrt{2 \pi N f (1 - f)}}\,, \mathrlap{\qquad N \to \infty\,,}
			\end{equation}
			where the KL divergence between these distributions is
			\begin{equation}
				\mathcal{D}_\slab{kl}[\es f\, |\, p\es] = f \log \tfrac{f}{p} + (1 - f) \log \tfrac{1 - f}{1- p}\,.
			\end{equation}
			One way to interpret this is that the KL divergence measures the rate at which we are able to distinguish between two probability distributions, in the limit that we  have access to a large number of samples. That is, it measures the rate at which the predictions made by the two distributions \emph{diverge}. This is the content of Sanov's theorem \cite{cover2006elements}, which is derived for the general multinomial family of probability distributions and applies to general continuous distributions via quantization.\footnote{By this, we mean separating the continuous variable $x$ into different bins which we consider as discrete variables with finite probability, and not anything having to do with $\hbar$.} We will, by convention, say that the distribution $p(x\, |\, \varphi)$ can be \emph{distinguished} from $p(x\, |\, \varphi')$ in $N$ samples if their KL divergence satisfies 
			\begin{equation}
				N\mathcal{D}_\slab{kl}[\, \varphi \,| \, \varphi' \, ] \geq 1\,, \label{eq:distinguishability}
			\end{equation}
			where the lower bound $1$ is a matter of convention.

			The KL divergence successfully ties the ``distance'' between the distributions at $\varphi$ and $\varphi'$ to how distinguishable they are. It describes the asymptotic rate at which the probability of empirically observing the distribution $p(x \es |\es \varphi)$ decays if the true distribution is $p(x\es|\es\varphi')$, as we draw more and more samples from $p(x\es|\es\varphi')$. Unfortunately, it is still asymmetric in its arguments, so our next task will be to define a Riemannian metric on this manifold that maintains this connection to distinguishability, the so-called classical information metric.

		\subsection{Statistical Length and Distinguishability} \label{sec:statLength}

			To introduce a metric, it will be helpful to consider the distinguishability criterion (\ref{eq:distinguishability}) for nearby distributions.  We say that two nearby probability distributions, $p(x\es|\es\varphi+ \ud \varphi)$ and $p(x\es|\es\varphi)$, can be distinguished after $N$ samples if 
			\begin{equation}
				\frac{1}{2} \, g_{ab}(\varphi)\, \ud \varphi^a \es \ud \varphi^b \gtrsim \frac{1}{N} \,, \label{eq:ellipsoid}
			\end{equation}
			where we have introduced the Hessian
			\begin{equation}
				g_{ab}(\varphi) \equiv \minus \!\left.\frac{\partial^2 \mathcal{D}_{\textsc{kl}}[\,\varphi\,|\,\varphi'\,]}{\partial \varphi^a \partial \varphi'^b}\right|_{\varphi' = \varphi} \!\!\!= \frac{\partial^2 \tilde{\mathcal{F}}(\varphi)}{\partial \varphi^a \partial \varphi^b} = \int\!\ud^N x\, p(x\es|\es\varphi) \, \frac{\partial \log p(x\es |\es\varphi)}{\partial \varphi^a} \frac{\partial \log p(x \es |\es \varphi)}{\partial \varphi^b}\,. \label{eq:fim}
			\end{equation}
			As is clear from Figure~\ref{fig:bregman}, the KL divergence vanishes at both leading and first order in $\ud \varphi^a$ when expanded about the reference point $\varphi^a$. Furthermore, since the divergence is convex, its Hessian is necessarily positive definite, so we may interpret it as a metric induced on the statistical manifold by the family of probability distributions $p(x\es | \es \varphi)$. It is known as the \emph{Fisher information metric} or \emph{classical information metric}.

			\begin{figure}
				\centering
				\includegraphics{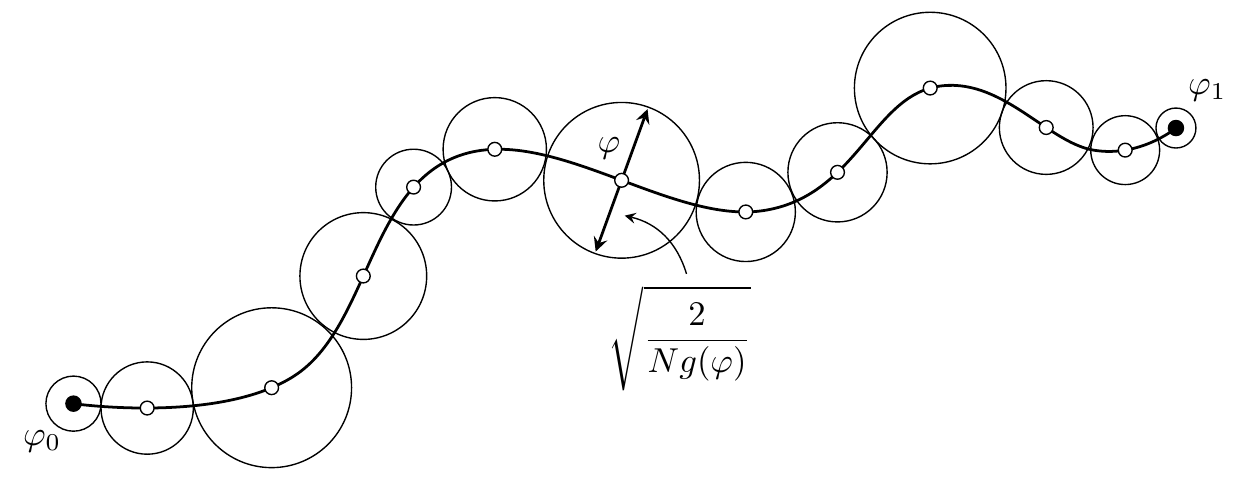}
				\caption{At finite $N$, we may draw an ``ellipsoid of distinguishability'' around the point $\varphi$, defined by $N \mathcal{D}_\slab{kl}[\varphi + \ud \varphi \es | \es \varphi] \approx \frac{1}{2}N g_{ab}(\varphi)\es \ud \varphi^a \ud \varphi^b = 1$. Here, we have assumed that the metric is isotropic $g_{ab}(\varphi) = g(\varphi)\es \delta_{ab}$, so that these ellipsoids are spheres.  The statistical length of the curve between two endpoints $\varphi_0$ and $\varphi_1$ is then the number of such ellipsoids we can fit along the curve, or equivalently the number of distinguishable distributions, divided by a conventional~$\sqrt{N/2}$. \label{fig:statLength}}
			\end{figure}

			Lengths measured with this metric have a precise interpretation inherited from the KL divergence. Let us consider a curve in the statistical manifold, parameterized by $\varphi^a(t)$ with $t \in [0, 1]$, $\varphi^a(0) = \varphi^a_0$, and $\varphi^a(1) = \varphi_1^a$. As illustrated in Figure~\ref{fig:statLength}, we can divide this curve into distinguishable segments using ellipsoids (\ref{eq:ellipsoid}) of constant ``radii'' $\sqrt{2/N}$. This allows us count the number of distinguishable distributions that fit along the curve in $N$ trials, which we label $N_\lab{dist}(N)$.  The length of this curve is simply the number of these distinguishable distributions \cite{Wootters:1981ki},
			\begin{equation}
				d(\varphi_1, \varphi_0) = \int_0^1 \!\ud t\, \sqrt{g_{ab}(\varphi(t))\,  \dot{\varphi}^a(t)\, \dot{\varphi}^b(t)}\, = \lim_{N \to \infty} \frac{N_\lab{dist}(N)}{\sqrt{N/2}}, \label{eq:statLength}
			\end{equation}
			scaled by a conventional $\sqrt{N/2}$ (inherited from our definition of ``distinguishable'') to render the limit finite. The minimal statistical length between the two points $\varphi_0$ and $\varphi_1$, i.e. the statistical length along the geodesic connecting them, then measures how different the two distributions at those points are from one another by counting the number of distinguishable distributions that can be fit on a line.

			It will be helpful to study a few properties of this classical information metric before we introduce its quantum mechanical analogue. We first discuss an alternative derivation of the metric which will help make contact with the quantum information metric. We then discuss how the information metric associated to the Lagrange multipliers $\lambda_a$ is related to the one associated to the expectation values $\varphi^a$, and how we should interpret the information metric's lack of units.

			\subsubsection*{Probability Distributions as Spheres}
				As we mentioned previously, though there are many divergences that measure the separation between two probability distributions, almost all of them reduce to the information metric when considering two nearby points in the statistical manifold. That is, up to an overall scaling, the information metric is unique\footnote{In particular, up to a scale it is the unique metric metric tensor that preserves inner products under statistical mappings called Markov maps, which are mappings of the random variable $x$ that preserve information. The simplest example of such maps are those that are one-to-one. This is a desirable property to have---our measure of how distinguishable two probability distributions are should not depend on the parameterization we choose.} \cite{cencov2000statistical,Cambpell:1986aec,Ay_2014,Bauer_2016}. We introduced it using the KL divergence since it provided a clear operational definition for statistical lengths. However, it will also be useful to motivate the metric a different way so that we can more easily make contact with the quantum information metric in Section~\ref{sec:qim}.

				Let us consider a probability distribution with $N+1$ outcomes, characterized by the probabilities $p^i$, with $i = 1, \dots, N+1$. These distributions may be labeled with the probabilities themselves, which must be normalized
				\begin{equation}
					\sum_{i = 1}^{N+1} p^i = 1\,.
				\end{equation}
				The set of discrete probability distributions thus forms a $N$-dimensional simplex in $\mathbb{R}^{N+1}$. If we further change coordinates to $\xi^i = \sqrt{p^i}$, then the normalization condition defines a $N$-dimensional sphere $\lab{S}^N$ embedded in $\mathbb{R}^{N+1}$. That the probabilities are all positive restricts us into the positive hyperoctant of this hypersphere.

				If we take the standard metric on this sphere, then it is natural to define the distance between two distributions $\xi^i$ and $\zeta^i$ as the arc length between them, measured along the great circle that connects them. This defines the \emph{Bhattacharyya distance} $\mathcal{D}_\lab{Bhatt}(\xi, \zeta)$ between the two points,
				\begin{equation}
					\cos \mathcal{D}_\lab{Bhatt}(\xi, \zeta) = \xi \cdot \zeta = \sum_{i = 1}^{N+1} \xi^i \zeta^i\,.
				\end{equation}
				When expressed in terms of the probabilities $p^i = \left(\xi^i\right)^2$ and $q^i = \left(\zeta^i\right)^2$, this reduces to
				\begin{equation}
					\cos \mathcal{D}_\lab{Bhatt}(p, q) = \sum_{i = 1}^{N+1} \sqrt{p^i q^i}\,, \label{eq:bhatt}
				\end{equation}
				where the right-hand side is sometimes called the \emph{classical fidelity}. Furthermore, if we take the two distributions to be separated infinitesimally, $q^i = p^i + \ud p^i$, we find that this distance reduces to
				\begin{equation}
					\left[\mathcal{D}_\lab{Bhatt}(p, p+\ud p)\right]^2 = \ud s^2 = \frac{1}{4} \frac{\delta_{ij}}{p^i} \, \ud p^i \,\ud p^j\,.
				\end{equation} 
				That is, the natural metric on the $N$-sphere induces a metric
				\begin{equation}
					g_{ij} = \frac{1}{4 p^i} \delta_{ij}
				\end{equation}
				on the probability simplex, or equivalently on the space of probability distributions. If we then pull this metric back to a hypersurface $p^i(\varphi)$ parameterized by the coordinates $\varphi^a$, this reduces to a discrete version of classical information metric (\ref{eq:fim})
				\begin{equation}
					g_{ab}(\varphi) = \frac{1}{4} \sum_{i=1}^{N+1} p^i(\varphi) \, \frac{\partial \log p^i(\varphi)}{\partial \varphi^a}\, \frac{\partial \log p^i(\varphi)}{\partial \varphi^b}\,,
				\end{equation}
				up to a factor of $1/4$. This is a general theme: many measures of separation between two probability distributions are proportional to the statistical length when restricted to nearby points. 

				It is clear from this derivation that there is a major caveat we should be aware of when we interpret statistical length in terms of distinguishability. We argued that the statistical length could be determined by counting the number distributions $N_\lab{dist}(N)$ that could be distinguished in $N$ measurements along the curve in the statistical manifold. There is, however, the possibility\footnote{We thank Matt Reece for raising this point.} that the submanifold parameterized by the coordinates $\varphi^a$ is complicated enough that two distributions could be close to one another on the hypersphere, in the sense that the Bhattacharya distance is small and their predictions are similar, yet separated by a large distance when we are constrained to move only along the submanifold traced out by $p^i(\varphi)$. It is still the case that we are able to discern many distributions along the curve, but we can not guarantee that the curve does not nearly loop back on itself, as it could if the submanifold is like a tightly-coiled helix. This complication is avoided if we focus on infinitesimally separated points. Indeed, we will discuss later how the metric provides a (local) measure of the fundamental uncertainty involved in parameter estimation---this is another way of interpreting distinguishability among different probability distributions in a statistical family.

			\subsubsection*{The Inverse Metric}

				So far, we have focused on the metric defined with respect to the expectation values $\varphi^a$. However, the Lagrange multipliers $\lambda_a$ also parameterize the exponential family's statistical manifold and we could just as easily use these to define a metric,
				\begin{equation}
					g^{ab}(\lambda) = \int\!\ud x\, p(x \es |\es \lambda) \, \frac{\partial \log p(x\es|\es \lambda)}{\partial \lambda_a} \frac{\partial \log p(x\es|\es \lambda)}{\partial \lambda_b}\,. \label{eq:invMetric}
				\end{equation}
			 	Fortunately, as our index structure would suggest, this turns out to be the inverse of the metric associated with the dual coordinates $\varphi^a$, ${g_{ab} g^{bc} = \delta_a^c}$. This is not such a surprising connection, since Legendre duality implies that $g^{ab} = \partial^2 \mathcal{F}/(\partial \lambda_a \partial \lambda_b) = \partial \varphi^a/\partial \lambda_b$ and $g_{ab} = \partial^2 \tilde{\mathcal{F}}/(\partial \varphi^a \partial \varphi^b) = \partial \lambda_a/\partial \varphi^b$. Interestingly, metric singularities become metric zeros, and vice versa, when using dual coordinates. We will use this fact later to gain a better understanding of how and why metric singularities appear. 

			\subsubsection*{Units}

				Statistical lengths are dimensionless, as they arise from counting the number of distinguishable distributions along a particular curve. What, then, are we to make of the information metric's units? We can clarify this~\cite{Caticha:2015fqr} first by considering the metric for the Gaussian distribution (\ref{eq:gaussian}). Interestingly, this is the metric on Euclidean $\lab{AdS}_2$,
				\begin{equation}
					\ud s^2 = \frac{1}{\sigma^2}\left(\ud \mu^2 + 2 \es \ud \sigma^2\right)\,. \label{eq:gaussianMetric}
				\end{equation}
				The classical information metric measures distances in units of the uncertainty implied by the distribution. 

				The same is true in the more general exponential family. The inverse metric (\ref{eq:invMetric}) is simply the covariance of the distribution,
				\begin{equation}
					g^{ab} = \frac{\partial^2 \log \mathcal{Z}(\lambda)}{\partial \lambda_a \, \partial \lambda_b} = \langle \big(f^a(x) - \varphi^a\big)\big(f^b(x) - \varphi^b \big)\rangle\,.
				\end{equation}
				The metric itself will diverge when this covariance vanishes. Intuitively, this makes sense---it is much easier to distinguish between two distributions if their inherent uncertainty is small. This relation is a special case of the infamous Cram\'{e}r-Rao bound~\cite{amari2016information}, which states that the variance of an unbiased estimator $f^a(x)$, with $\langle f^a(x) \rangle = \varphi^a$, is bounded from below by the inverse of the classical information metric,
				\begin{equation}
						\langle \big(f^a(x) - \varphi^a\big)\big(f^b(x) - \varphi^b \big)\rangle \geq g^{ab}(\varphi)
				\end{equation}
				for any family of probability distribution. It is a special property of the exponential family that this bound is saturated.

			\subsection{Infinite Distance Points and Hyper-Distinguishability} \label{sec:infStatLength}

				The primary goal of this work is to study and understand the appearance of infinite distance metric singularities in quantum field theories. While we still need to discuss how to define the information metric in such theories, it will be helpful to first understand the meaning of infinite distance in much simpler, classical statistical models.

				We may think of probability distributions that are infinitely far away from one another in this metric, as measured along a geodesic, are hyper-distinguishable. That is, they can be distinguished from one another \emph{with certainty} via a finite number of discerning measurements.\footnote{As we discussed previously, this interpretation assumes that the statistical manifold is a well-behaved submanifold in the space of probability distributions. It is possible that one can realize an infinite distance curve between two nearby points on the probability simplex by choosing an appropriate submanifold---a simple example is if the submanifold is space-filling, similar to a Hilbert or Peano curve. However, such curves are necessarily non-differentiable~\cite{MichalMorayne1987} and we would find it extremely surprising if such bizarre submanifolds arise naturally in families of physical theories.}  We see already that  (\ref{eq:gaussianMetric}) exhibits infinite distance points at both $\sigma = 0$ and $\sigma = \infty$. These points represent a qualitative change in behavior of the distribution. At $\sigma =0$, the support of the distribution collapses to a single point $x = \mu$, while at $\sigma = \infty$ the distribution has become flat. Both points are easy to distinguish from distributions with finite and non-zero variance. For instance, a single observation of $x \neq  \mu$ immediately rules out the distribution at $\sigma =0$.   

				These two cases represent fairly generic examples of infinite distance points---changes in either the support or tails of a distribution are easily detected, and so they correspond to large distances in the information metric. This sensitivity to a general, qualitative difference in the behavior between probability distributions is at the core of what makes the classical and quantum information metrics useful diagnostics of phase transitions and the appearance of gapless modes in the spectrum. This will be the main focus of Sections~\ref{sec:qim} and~\ref{sec:examples}.

			\subsection{A Familiar Toy Model} \label{sec:toy}

\begin{figure}
				\centering
				\includegraphics{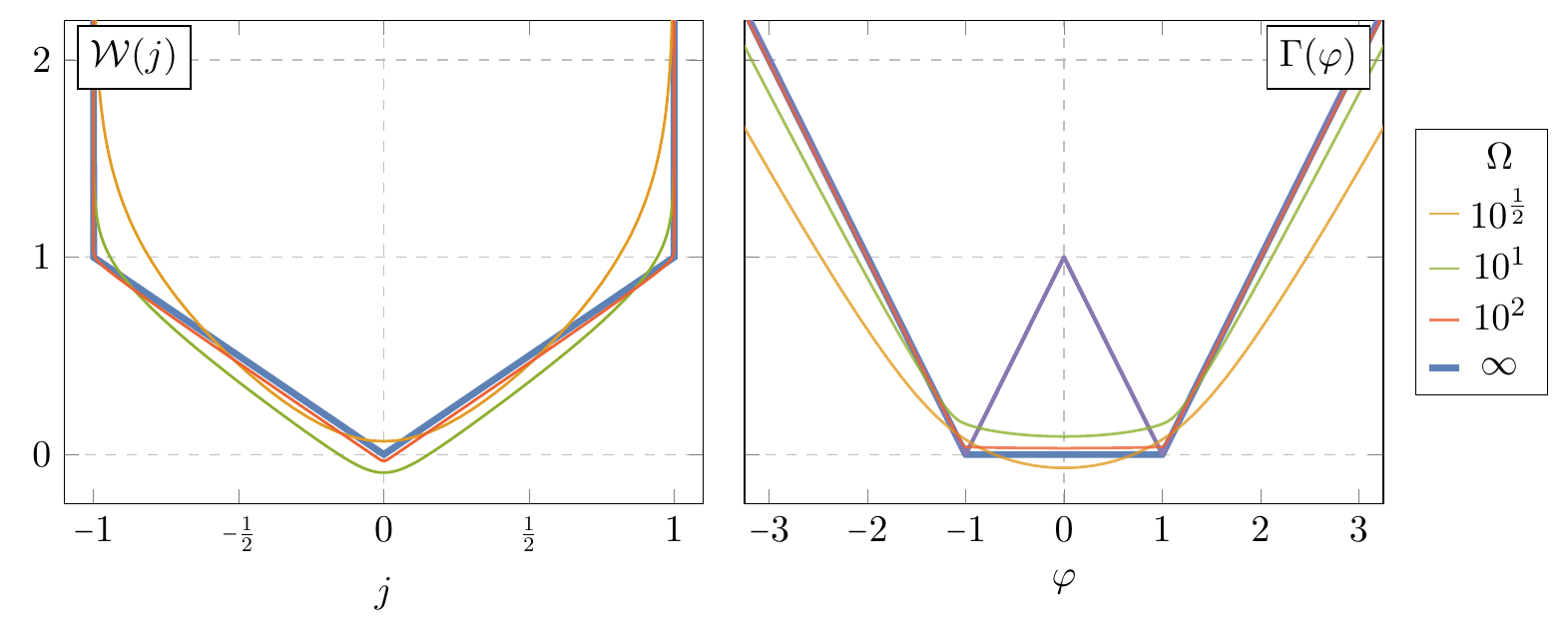}
				\caption{The generating function $\mathcal{W}(j)$ (\emph{left}) and effective potential $\Gamma(\varphi)$ (\emph{right}) for the toy potential~(\ref{eq:toyPotential}). In the $\Omega \to \infty$ limit, the effective potential {\color{Mathematica1}[blue]} is simply the convex hull of the potential $V(\phi)$ {\color{Mathematica5}[purple]}. The toy potential is not monotonic for $|\phi| \leq 1$, and so its convexification $\Gamma(\varphi)$ must have a flat face there, which in turn forces its Legendre transform $\mathcal{W}(j)$ to have a cusp, or discontinuous first derivative, at $j = 0$. \label{fig:wgPlot}}
			\end{figure}

			We can illustrate these concepts with a simple, calculable toy model reminiscent of those encountered in quantum field theory and statistical mechanics. We focus on the family of probability distributions with the form
			\begin{equation}
				p(\phi \es|\es j ) = \frac{1}{\mathcal{Z}(j)}\exp\!\Big(\Omega\big[j \phi - V(\phi)\big]\Big), \label{eq:toyPT}
			\end{equation}
			with the potential \cite{ORaifeartaigh:1986axd}
			\begin{equation}
				V(\phi) = \begin{cases}
							|\phi - 1 | & \phi \geq 0 \\
							|\phi + 1 | & \phi < 0
						 \end{cases}\,, \label{eq:toyPotential}
			\end{equation}
			which we plot in Figure~\ref{fig:wgPlot}.
			For familiarity's sake, we have replaced the random variable $x$ in our previous discussion with $\phi$ and used $\Omega$ to denote an overall ``spacetime volume'' which we consider to be a fixed albeit large constant. This forms an exponential family, which we can parameterize by either using the ``source'' $j$, with $|j| < 1$, or the expectation value $\varphi = \langle \phi \rangle$.  Usefully, we can compute the partition function exactly,
			\begin{equation}
				\log \mathcal{Z}(j) = \log\!\left[2 \cosh \Omega j - \lab{e}^{-\Omega}\right] + \log \left[\frac{2}{\Omega ( 1 - j^2)}\right]\,.
			\end{equation}
			We will find that the information metric for this family of models can diverge in the $\Omega \to \infty$ limit, and it will be useful to first understand the cause of these divergences in this simple model to gain intuition for the more involved models presented in the following sections.

			\begin{figure}
				\centering
				\includegraphics{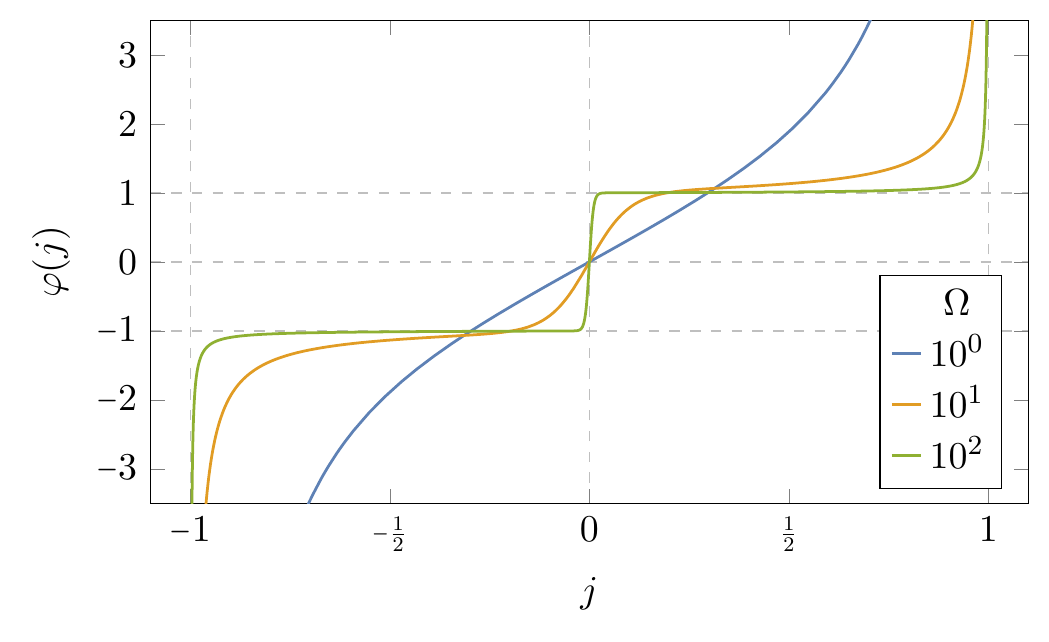}
				\caption{The expectation value $\varphi$ as a function of the source $j$. It becomes discontinuous at $j =0$ as $\Omega \to \infty$, a manifestation of spontaneous symmetry breaking in the potential (\ref{eq:toyPotential}). For moderate values of $|j|$, $\varphi(j)$ hovers $\varphi = \pm 1$.  When the source becomes strong enough, $|j| \sim 1 - \mathcal{O}(\Omega^{-1})$, it can overpower the potential and the system can explore larger values of $\phi$. \label{fig:phij}}
			\end{figure}

			This model is simple enough that we compute the expectation value of $\phi$ exactly,
			\begin{equation}
				\varphi(j)  = \langle \phi \rangle = \frac{1}{\Omega} \frac{\partial \log \mathcal{Z}(j)}{\partial j} = \frac{2j}{\Omega(1- j^2)} + \left(1 + \frac{\lab{e}^{-\Omega}}{2 \cosh \Omega j - \lab{e}^{-\Omega}}\right) \tanh \Omega j \,,
			\end{equation}
			which, in the large $\Omega$ limit, reduces to
			\begin{equation}
				\varphi(j) \sim 
					\begin{dcases}
						\tanh \Omega j & |j| \lesssim 1 \\
						\frac{2 j}{\Omega (1 - j^2)} + \sign j & |j| \sim 1 - \mathcal{O}(\Omega^{\sminus 1})
					\end{dcases}\,, \mathrlap{\qquad \Omega \to \infty\,,} \label{eq:toyVarphi}
			\end{equation}
			where $\sign j \equiv j/|j|$ the sign of the source $j$. We plot $\varphi(j)$ for a few decades of $\Omega$ in Figure~\ref{fig:phij}. As~$\Omega \to \infty$, we find that $\phi$ becomes ``trapped'' into either well of the potential, $\phi = \pm 1$. Which well the distribution selects depends on how the source $j$ approaches $0$. A similar situation occurs in more realistic models of spontaneous symmetry breaking, where the symmetry breaking relies on the addition of a small, non-zero source to specify which vacuum, or super-selection sector, the theory is in. Said differently, a slight biasing of the theory is necessary for ergodicity breaking~\cite{Goldenfeld:2018lop}. Since $j=0$ represents a point of  hyper-distinguishability---that is, we can determine whether $j$ is infinitesimally positive or negative via a single measurement of $\phi$ in the $\Omega \to \infty$ limit---we expect that the information metric diverges here. Indeed, we can compute the metric exactly,
			\begin{equation}
				g^{jj} = \Omega \frac{\partial \varphi(j)}{\partial j} = \frac{1}{(1+j)^2} + \frac{1}{(1-j)^2} + \frac{2 \Omega^2(2 - \lab{e}^{- \Omega} \cosh \Omega j)}{(2 \cosh \Omega j - \lab{e}^{-\Omega})^2}
			\end{equation}
			and confirm that it scales as $g^{jj} \sim \Omega^2$ at $j =0$, as $\Omega \to \infty$. In Section~\ref{sec:qim}, we will discuss similar ``super-extensive'' behavior in higher-dimensional systems, where metric divergences scale with particular powers of the IR and UV cutoffs.

\begin{figure}
				\centering
				\includegraphics{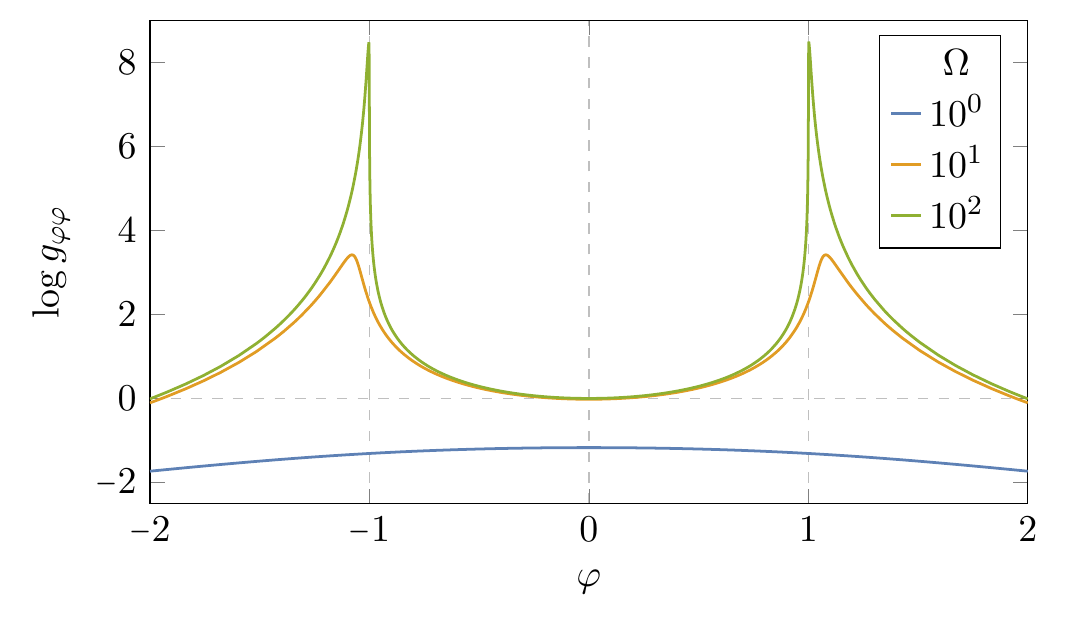}
				\caption{The logarithm of the metric $g_{\varphi \varphi}(\varphi)$ is sharply peaked about the minima $\varphi =\pm 1$, and relatively small elsewhere. Models near $\varphi = \pm 1$ are thus highly distinguishable from one another. \label{fig:toyMetric}}
			\end{figure}

			We can also study the metric with respect to the expectation value $\varphi$. To calculate this, we must invert (\ref{eq:toyVarphi}) to find
			\begin{equation}
				 j(\varphi) \sim 
							\begin{dcases}
								\quad\qquad \Omega^{\sminus 1} \arctanh \varphi & |\varphi|  \lesssim 1 \\ 
								\frac{\sqrt{(\varphi- \sign \varphi )^2 + \Omega^{\sminus 2}} - \Omega^{\sminus 1}}{\varphi - \sign \varphi} & |\varphi| \gtrsim 1
							\end{dcases}\,, \mathrlap{\qquad \Omega \to \infty\,.} \label{eq:toyJ}
			\end{equation}
			We can compute the metric by taking a derivative
			\begin{equation}
				g_{\varphi \varphi} = \Omega \frac{\partial j(\varphi)}{\partial \varphi} \sim \begin{dcases}
					\left(1 - \varphi^2\right)^{\sminus 1} & \varphi \lesssim 1 \\
					\left[(\varphi - \sign \varphi)^2 + \Omega^{\sminus 2} \left(1 + \sqrt{\Omega^2 (\varphi - \sign \varphi)^2 +1}\right)\right]^{\sminus 1} & \varphi \gtrsim 1
				\end{dcases}\,,
			\end{equation}
			and we plot its logarithm in Figure~\ref{fig:toyMetric}.  We see that it is sharply peaked around the wells $\varphi = \pm 1$ as $\Omega \to \infty$, and diverges there as $\Omega^2/2$. In this limit, we make it extremely improbable for the system to occupy anything other than $\phi = \pm 1$, and so the local variance when $\varphi = \langle \phi \rangle = \pm 1$ becomes infinitesimal. Recalling that the metric is measured in units of the local uncertainty (\ref{eq:gaussianMetric}), this localization of probability at $\phi = \pm 1$ is responsible for these metric singularities.

			It is also apparent from Figure~\ref{fig:toyMetric} that values of $\varphi$ away from the minima $\varphi = \pm 1$ are relatively indistinguishable. For $|\varphi| > 1$, the source has overpowered the potential (\ref{eq:toyPotential}), $j \phi - V(\phi) \sim 1/\Omega$. This effectively nullifies the localizing power of the $\Omega \to \infty$ limit and treats models with different $\varphi$ on equal footing, so they are harder to distinguish. For $|\varphi| < 1$, the statistical model has essentially reduces to a mixture distribution with values pulled equally from $\phi = +1$ and $\phi = -1$. It costs the same ``energy'' (or equivalently has the same probability) to draw equal numbers of $+1$'s and $-1$'s (to measure an expectation value $\langle \phi \rangle = 0$) as it does to draw all $+1$'s, and so we expect all distributions with $|\varphi| \leq 1$ to also be relatively indistinguishable from one another.

			Another way of understanding how these singularities arise is to instead consider the convex function
			\begin{equation}
				\mathcal{W}(j) = \frac{1}{\Omega} \log \mathcal{Z}(j)\,,
			\end{equation}
			the analog of the generating functional for connected diagrams in quantum field theory.
			From our discussion about Bregman divergences, we know that we can also introduce a notion of distance based on this function, parameterized by $j$. The derivative of this function yields the expectation value
			\begin{equation}
				\varphi = \langle \phi \rangle = \frac{\partial \mathcal{W}(j)}{\partial j}\,,
			\end{equation}
			and we can introduce the dual convex function via the Legendre transform,
			\begin{equation}
				\Gamma(\varphi) = \sup_{j} \left(j \varphi - \mathcal{W}(j)\right)\,,
			\end{equation}
			usually called the effective potential. Taking two derivatives of this defining relation and keeping track of factors of $\Omega$, we see that the metric computed with $\Gamma(\varphi)$ is related to the one computed above by an additional power of $\Omega$, $\tilde{g}_{\varphi \varphi} = \partial^2 \Gamma(\varphi)/\partial \varphi^2 = \Omega^{\sminus 1} g_{\varphi \varphi}$. 

			\newpage
			The benefit of working with these functions, instead of $\log \mathcal{Z}(j)$ and its Legendre transform, is that they take relatively simple forms in the $\Omega \to \infty$ limit. The partition function is readily approximated using the method of steepest descent, in which case the generating function $W(j)$, plotted on the left of Figure~\ref{fig:wgPlot}, becomes the Legendre transform of $V(\phi)$, 
			\begin{equation}
				W(j) \sim |j|\,,\mathrlap{\qquad \Omega \to \infty\,,}
			\end{equation}
			and the effective potential is its convexification,
			\begin{equation}
				\Gamma(\varphi) = \begin{cases} |\varphi| - 1 & |\varphi| \geq 1 \\ 0  & |\varphi| < 1 \end{cases}\,.
			\end{equation}
			This picture geometrizes the behavior of the metric in this simple class of models. The singular behavior of $g^{jj} \propto \partial^2\mathcal{W}/\partial j^2$ at $j = 0$ is a simple consequence of the fact that the potential (\ref{eq:toyPotential}) is not monotonic for $|\phi| \leq 1$, so its convexification must have a flat face for $|\varphi| \leq 1$. This flat face both induces cusp in $\mathcal{W}(j)$ and forces models with $|\varphi| < 1$ to be relatively indistinguishable.

			While the singularity in the source's information metric $g^{jj}$ is a generic consequence of $V(\phi)$'s non-monotonicity, we can appreciate that the localized singularities in $\tilde{g}_{\varphi \varphi}$ at $\varphi = \pm 1$ are only present because $V(\phi)$ is not smooth. Indeed, if we consider a more well-behaved potential like the familiar double-well $V(\phi) = (\phi^2 - 1)^2$, the effective potential is again its convex hull
			\begin{equation}
				\Gamma(\varphi) = \begin{cases} (\varphi^2 - 1)^2 & |\varphi| \geq 1 \\ 0 & |\varphi| < 1 \end{cases}
			\end{equation}
			and the associated metric is
			\begin{equation}
				\tilde{g}_{\varphi \varphi}(\varphi) = \frac{\partial^2 \Gamma}{\partial \varphi^2} = \begin{cases} 12 \varphi^2 -4 & |\varphi| \geq 1 \\ 0 & |\varphi| < 1 \end{cases}\,.
			\end{equation}
			This is merely discontinuous, not singular, at $\varphi = \pm 1$. It is that $\tilde{g}_{\varphi\varphi}$ vanishes for $|\varphi| < 1$ that is a generic consequence of $V(\phi)$'s non-monotonicity. Since the two metrics are inverses of one another, it is not so surprising that the singularities of one are related to the zeros of the other. When one parameterization fails to distinguish between different models in the family, the dual parameterization is extremely successful. 

			Finally, we note that the behavior of the metric $g_{\varphi \varphi}$ with this smooth potential is no longer super-extensive: it scales extensively $\propto \Omega$ at $\varphi = \pm 1$ instead of $\Omega^{2}$. This is the same behavior that would be present when the symmetry $\phi \to \minus \phi$ is restored, i.e. when $V(\phi) = (\phi^2 + 1)^2$, then $\Gamma(\varphi) = (\varphi^2 + 1)^2$, $\tilde{g}_{\varphi \varphi}(\varphi) = 12 \varphi^2 + 4$, and so $g_{\varphi \varphi} \propto \Omega$ in both phases. Instead, it is only the metric with respect to the source $j$, $g^{jj}(j)$, that is super-extensive $g^{jj}(0) \sim \Omega^2$ in the symmetry-broken phase and extensive $g^{jj}(0) \sim \Omega$ in the symmetry-restored phase. This is evocative of behavior we will encounter in the next section, where the existence of metric singularities---or equivalently metric super-extensivity---imply the existence of gapless modes, but not vice versa. We now turn to understanding the analog of this metric in quantum field theories.

		\section{The Quantum Information Metric} \label{sec:qim}

			In the previous section, we introduced the classical information metric and reviewed its properties. This metric provides a notion of distance between two probability distributions that quantifies how readily they may be distinguished from one another, in a way made precise by its definition (\ref{eq:statLength}). In this section, we discuss its quantum mechanical counterpart, the \emph{quantum information metric}, and describe its interpretation, the many ways of computing it, and how it behaves when we approach criticality.

			In \S\ref{sec:statLength}, we argued that the set of  classical probability distributions with $N+1$ outcomes had the geometry of a hypersphere $\lab{S}^N$, and that the statistical length could be understood as the distance induced by the standard round metric on this sphere. In \S\ref{sec:metricDef}, we will similarly motivate the quantum information metric using the standard metric on $\mathbb{CP}^N$, the $(N+1)$-dimensional space of quantum states, called the Fubini-Study metric. We then describe how this quantum information metric is related to the classical one in \S\ref{sec:metricInterp} and discuss various ways to compute it in \S\ref{sec:metricComp}.  This metric is divergent in field theories and must be regulated with both an IR and UV cutoff, whose meaning we discuss in \S\ref{sec:metricReg}. We will find that the metric typically scales extensively in the number of degrees of freedom. However, in \S\ref{sec:metricScaling}, we will describe how it may scale super-extensively when the correlation length of the theory diverges. There, we use  critical scaling arguments to preclude the existence of infinite distance points in $(d > 0)$-dimensional quantum field theories, and we take this as an indication that such points must be qualitatively different from the types of quantum critical points typically considered. We will provide supporting examples in Sections~\ref{sec:examples} and~\ref{sec:towers}.

			\subsection{Defining the Metric} \label{sec:metricDef}

			In the previous section, we argued that the set of probability distributions over a discrete set of $N+1$ outcomes has the geometry of an $N$-dimensional hypersphere $\lab{S}^N$. We then used the standard geodesic distance on this sphere to define a distance between two distributions, called the Bhattacharyya distance. This distance reduced to the classical information metric in the limit that the distributions were infinitesimally separated. We can use a similar strategy to define the quantum information metric. We will follow the phenomenal \cite{bengtsson2017geometry}.

			The quantum mechanical analogs of these discrete probability distributions are states in a finite-dimensional Hilbert space. We can represent an arbitrary state $|\psi \rangle$ in an $(N+1)$-dimensional Hilbert space as a complex vector $\psi^i \in \mathbb{C}^{N+1}$, with $i= 1, \dots, N+1$. Since the overall phase and magnitude of these vectors are unphysical, we must quotient this $\mathbb{C}^{N+1}$ out by an overall complex scaling to properly identify the space of states. The space of quantum states in an $(N+1)$-dimensional Hilbert space can thus be identified with $N$-dimensional complex projective space~$\mathbb{CP}^N$. 

			Complex projective space can be equipped with a distance that is naturally related to the Bhattacharyya distance. Since $\mathbb{CP}^N = \lab{S}^{2N+1}/\lab{S}^1$, we can ``horizontally lift'' a geodesic in complex projective space to the odd-dimensional sphere $\lab{S}^{2N+1}$ and use the standard distance defined there to define a notion of distance the original space. This distance is  called the \emph{Fubini-Study distance}. Given two states, say $|\xi\rangle$ and $|\zeta \rangle$, the Fubini-Study distance $\mathcal{D}_{\textsc{fs}}(\xi, \zeta)$ is defined as
			\begin{equation}
				\cos \mathcal{D}_{\textsc{fs}}(\xi, \zeta) = |\langle \xi | \zeta \rangle|\,,
			\end{equation}
			which is clearly evocative of the Bhattacharyya distance (\ref{eq:bhatt}). The right-hand side of this equation is called the \emph{quantum fidelity}, and is another popular measure of the distance between quantum states.

			As in the classical case, this notion of distance endows a metric on $\mathbb{CP}^N$ called the \emph{Fubini-Study metric}. It will often be useful to work with unnormalized vectors and express everything in combinations that are invariant under complex rescalings. In terms of the coordinate $|\zeta\rangle = \zeta^i \in \mathbb{C}^{N+1}$, the Fubini-Study line element on $\mathbb{CP}^N$ can be written as
			\begin{equation}
				\ud s^2 = \frac{\zeta^j \bar{\zeta}^{\bar{\jmath}}}{\big(\zeta \cdot \bar{\zeta}\big)^2} \left[ \delta_{i \bar{\imath}} \delta_{j \bar{\jmath}} - \delta_{i \bar{\jmath}} \delta_{j \bar{\imath}} \right] \ud \zeta^i \ud \bar{\zeta}^{\es \bar{\imath}}\,,
			\end{equation}
			where repeated indices are summed over and $\zeta \cdot \bar{\zeta} = \delta_{i \bar{\imath}} \,\zeta^i \bar{\zeta}^{\es \bar{\imath}}$.
			Since $\mathbb{CP}^n$ is a K\"{a}hler manifold, we can also define the (real) K\"{a}hler form
			\begin{equation}
				J = \frac{i \, \zeta^j \bar{\zeta}^{\bar{\jmath}}}{\big(\zeta \cdot \bar{\zeta}\big)^2} \left[\delta_{i \bar{\imath}} \delta_{j \bar{\jmath}} - \delta_{i \bar{\jmath}} \delta_{j \bar{\imath}}\right] \ud \zeta^i \wedge \ud \bar{\zeta}^{\es \bar{\imath}}.
			\end{equation}
			Both can be derived from the K\"{a}hler potential
			\begin{equation}
				\mathcal{K}(\zeta, \bar{\zeta}) = \log\,  \langle \zeta | \zeta \rangle =  \log\! \big(\es\bar{\zeta} \cdot \zeta\es\big)
			\end{equation}
			with
			\begin{equation}
				g_{i \bar{\jmath}} = \frac{1}{2} \partial_i \partial_{\bar{\jmath}}\, \mathcal{K}(\zeta, \bar{\zeta}) \qquad\text{and}\qquad J = \frac{i}{2} \partial \bar{\partial}\es \mathcal{K}(\zeta, \bar{\zeta})\,. \label{eq:fubiniDeriv}
			\end{equation}
			Comparing this expression with (\ref{eq:fim}), we see that the (negative) K\"{a}hler potential plays the same role as the Bregman or Kullback-Leibler divergence. We will identify the Fubini-Study metric as the \emph{quantum information metric}.

			So far, this discussion is too general to be of much use---so what if we can define a distance on the generic space of quantum states? As in the classical case, its utility arises when we consider a submanifold in this space, which we will continuously parameterize using the coordinates $\varphi^a$, $a = 1, \dots, n$, and write as $|\vac(\varphi)\rangle$. Throughout, we will think of $|\vac(\varphi)\rangle$ as the vacuum state associated to the Hamiltonian $\mathcal{H}(\varphi)$ that varies continuously with the parameters $\varphi^a$, and defines the so-called \emph{vacuum bundle}~\cite{Hori:2003ic}. We will assume that this vacuum state is unique for generic $\varphi^a$, though we will allow for it to degenerate along surfaces of codimension 1 or higher. 

			The Fubini-Study metric in the ambient complex projective space then defines a metric on this submanifold. Generally, the parameters $\varphi^a$ are real-valued---for instance, they represent Wilson coefficients in an action---so this generally a real submanifold. We can pull both the metric and K\"{a}hler form back onto this submanifold to define
			\begin{equation}
				T_{ab}(\varphi) = \langle \partial_a \vac | \big[\mathbbm{1} - |\vac \rangle \langle \vac|\big] |\partial_b \vac \rangle = g_{ab}(\varphi) + \frac{i}{2} F_{ab}(\varphi)\,, \label{eq:qim}
			\end{equation}
			where we denote $|\partial_a \vac\rangle = \frac{\partial}{\partial \varphi^a} |\vac(\varphi)\rangle$. Here we have implicitly assumed that these states are normalized. This object $T_{ab}(\varphi)$ is known as the quantum geometric tensor \cite{provost1980}, whose real part $g_{ab}(\varphi)$ is symmetric and called the quantum information metric and whose imaginary part $F_{ab}(\varphi)$ is antisymmetric and called the Berry curvature \cite{Berry:1984jv,Wilczek:Phases}.

			It can be convenient to complexify the parameters $\varphi^a$ and consider $|\vac(\varphi^a)\rangle$ as a complex submanifold in the Hilbert space, onto which we can naturally pull-back the ambient Fubini-Study metric and K\"{a}hler form. We can then restrict to the real submanifold defined by $\varphi^a = \bar{\varphi}^{\bar{a}}$ to arrive at the metric and Berry curvature defined on the ``physical'' set of states. This is a choice, whose main benefit is that we can derive (\ref{eq:qim}) from the K\"{a}hler potential
			\begin{equation}
				\mathcal{K}(\varphi, \bar{\varphi}) = \log \,\langle \vac(\varphi) | \vac(\varphi)\rangle\,, \label{eq:kahlerPotential}
			\end{equation}
			where the complex conjugate coordinates $\bar{\varphi}$ arise in the definition of the ket $\langle \vac(\varphi)|$. We can make this choice even when the theory does not provide a natural complexification of the parameters $\varphi^a$, as it does in supersymmetric theories. 

			However, in our examples we will not need to rely on this complexification.  We will instead find it more convenient to derive the information metric from the log fidelity between the vacuum states at $\varphi$ and $\varphi'$,
			\begin{equation}
				\minus \mathcal{D}(\varphi, \varphi') =  \log \, \langle \vac(\varphi) | \vac(\varphi') \rangle\,. \label{eq:logFidelity}
			\end{equation} 
			In the absence of Berry curvature, we can always take ${|\vac(\varphi)\rangle \to \lab{e}^{\sminus i \alpha(\varphi)} |\vac(\varphi)\rangle}$ and rephase the family of vacuum states so that the argument of the logarithm is real-valued and less than one for all $\varphi$ and $\varphi'$. Non-trivial Berry curvature characterizes a topological obstruction\footnote{In fact, the Berry curvature provides another way of characterizing a quantum phase transition~\cite{Carollo:2005gpc,Pachos:2006gpc,Kolodrubetz_2017,Carollo:2019ygj}.}  to such a rephasing, but will not appear in the models we consider. The quantum information metric may then be derived from
			\begin{equation}
				g_{ab}(\varphi) = -\!\left.\frac{\partial^2 \mathcal{D}(\varphi, \varphi')}{\partial \varphi^a \partial \varphi'^b}\right|_{\varphi' = \varphi}\,\,, \label{eq:qimLogFidelity}
			\end{equation}
			which is familiar from the definition of the classical information metric (\ref{eq:fim}).
			This type of geometry is in fact the real-valued analog of K\"{a}hler geometry, known as Hessian geometry~\cite{shima2007geometry,zhang2014divergence,nielsen2020elementary}, and shows up naturally in the study of statistical manifolds.

			\subsection{Interpreting the Metric} \label{sec:metricInterp}

			How is the quantum information metric related to the classical one? Perhaps the simplest way to understand their relation is to project the family of vacuum states onto a measurement basis~$|x\rangle$,
			\begin{equation}
				\langle x | \vac(\varphi) \rangle = \sqrt{p(x\es\es |\es\es \varphi)}\, \lab{e}^{i \alpha(x \es |\es \varphi)}\,, \label{eq:vacProb}
			\end{equation}
			so that we can write these vacua in terms of an induced probability distribution $p(x\es|\es\varphi)$ and phase factor $\alpha(x\es |\es \varphi)$. We then find \cite{Facchi_2010} that line element can be written as
			\begin{equation}
				\ud s^2 = g_{ab}(\varphi) \,\ud \varphi^a \ud \varphi^b = \frac{1}{4}  \mathbb{E}_p\!\left[(\ud \log p)^2\right] + \mathbb{E}_p\!\left[(\ud \alpha)^2\right] - \big(\mathbb{E}_p[\ud \alpha]\big)^2
			\end{equation}
			and 
			\begin{equation}
				F = \frac{1}{2} F_{ab}(\varphi)\,  \ud\varphi^a \wedge \ud \varphi^b = \mathbb{E}_p\!\left[\ud \log p \wedge \ud \alpha\right]
			\end{equation}
			where we use $\mathbb{E}_p[\,\cdot\,]$ to denote expectation values with respect to the probability distribution~$p(x\es|\es\varphi)$. As we discussed before, unless there is non-zero Berry curvature we can rephase these states so that $\alpha = 0$ for all $\varphi^a$, so that the quantum information metric coincides with the classical one for the vacuum probability distribution\footnote{The authors of \cite{Balasubramanian:2014bfa} studied the classical information metric associated with the probability distributions for Euclidean field configurations, i.e. $p(\phi) \propto \lab{e}^{-S_\slab{E}[\phi]}$, where $S_\slab{E}[\phi]$ is the Euclidean action. While this is also an interesting object to study in its own right, it is more difficult to relate the properties of this metric to the structure of the Hilbert space, and it is not obvious how well this metric measures the phase structure of a family of theories.} up to a familiar factor of $1/4$.

			In the previous section, we described how statistical lengths have an operational interpretation in terms of the number of distinguishable distributions along a curve. There, we had a clear notion of what constituted a measurement---it was a realization of $x$ from the probability distribution $p(x|\varphi)$ or some function of it. However, quantum states provide a whole host of distributions, depending on which operator is being measured. If we say that the quantum information metric measures the distinguishability of two different vacuum states, it is natural to ask: distinguishability with respect to which measurements? 

			We can answer this question by considering two states in a finite dimensional Hilbert space, projected onto a measurement basis $\mathcal{O}_i|i \rangle = o_i |i \rangle$,
			\begin{equation}
				|\zeta \rangle = \sum_{i = 1}^{N+1} |i \rangle \langle i | \zeta \rangle \qquad \text{and}\qquad |\xi \rangle = \sum_{i = 1}^{N+1} |i \rangle \langle i | \xi \rangle\,.
			\end{equation} 
			These define two probability distributions for the outcomes $o_i$ and we can define the Bhattacharyya distance 		with respect to the operator $\mathcal{O}$ between them,
			\begin{equation}
				\cos \mathcal{D}_\lab{Bhatt}(\xi, \zeta; \mathcal{O}) = \sum_{i = 1}^{n+1} |\langle \xi |i \rangle| |\langle i | \zeta \rangle|\,. \label{eq:bhattOp}
			\end{equation}
			If we vary this operator,  we will find a variety of classical statistical distances. We would like to choose the operator which makes the distance as large as possible, and is thus the most discerning. This occurs \cite{Wootters:1981ki} when the left-hand side is as small as possible, and an appropriate choice is to select an operator with either $|\zeta \rangle$ or $|\xi \rangle$ as an eigenstate. This collapses (\ref{eq:bhattOp}) to the Fubini-Study distance,
			\begin{equation}
				\min_{\mathcal{O}} \cos \mathcal{D}_\lab{Bhatt}(\zeta, \xi; \mathcal{O}) = \cos \mathcal{D}_{\textsc{fs}}(\zeta, \xi) = |\langle \zeta | \xi \rangle |\,.
			\end{equation}
			Intuitively, this makes sense---if we want to distinguish $|\xi\rangle$ from $|\zeta\rangle$, we can project one onto the other. If they are different states, this will always be less than 1. The Fubini-Study distance, and equivalently the Fubini-Study metric, thus represent an optimal statistical distance. That is, theoretically and with no limits on the type of experiment we can do, this distance describes how well can we distinguish the two quantum-mechanical states. Equivalently, it describes how well we can distinguish between two different theories, assuming that both are in their ground state. This is how we will use it.

			\subsection{Computing the Metric} \label{sec:metricComp}

				Now that we have established that this metric encodes something interesting about a family of theories and is a natural object to define on the vacuum submanifold, we will now discuss how to compute it.

				The defining relation (\ref{eq:qim}) provides a particularly simple way of computing the metric, and can be put into a more palatable form as follows. We will assume that $|\vac(\varphi) \rangle$ is defined as the vacuum state of the Hamiltonian $\mathcal{H}(\varphi)$, such that $\mathcal{H}(\varphi) |\vac(\varphi)\rangle = E_\lab{vac}(\varphi)|\vac(\varphi)\rangle$. At each point $\varphi^a$, the Hamiltonian also has a complete set of eigenstates $\mathcal{H}(\varphi)|\nu(\varphi)\rangle = E_\nu(\varphi) |\nu(\varphi)\rangle$, which we label with an abstract index $\nu$ that represents possibly multiple discrete and continuous labels.  By inserting this complete set of states and using $\langle \nu | \partial_a \!\es\vac \rangle = (E_\lab{vac} - E_\nu)^{\sminus 1} \langle \nu | \partial_a \mathcal{H} | \vac \rangle$, the metric may written in the spectral representation
				\begin{equation}
					g_{ab}(\varphi) = \sum_{\nu \neq \vac} \re\! \left[ \frac{\langle \vac | \partial_a \mathcal{H} | \nu \rangle \langle \nu |\partial_b \mathcal{H} | \vac\rangle}{(E_\lab{vac}(\varphi) - E_\nu(\varphi))^2}\right]\,. \label{eq:qimPert}
				\end{equation}
				From this expression, it is clear why singularities in the metric can be tied to presence of a phase transition or a drastic, qualitative change in behavior of the theory---the metric may diverge when excited states become degenerate with the vacuum. Whether or not the matrix elements $\langle \nu | \partial_a \mathcal{H} | \vac \rangle$ vanish quickly enough as $E_\nu(\varphi) \to E_\lab{vac}(\varphi)$ to remove this singularity depends on the details of the system.

				The metric takes a very similar form to a classic measure of a phase transition: the vacuum energy's Hessian,
				\begin{equation}
					\frac{1}{2}\frac{\partial^2 E_\lab{vac}(\varphi)}{\partial \varphi^a \partial \varphi^b } =  \sum_{\nu \neq \vac} \re\! \left[\frac{\langle \vac | \partial_a \mathcal{H} |\nu \rangle \langle \nu | \partial_b \mathcal{H} |\vac\rangle}{E_\lab{vac}(\varphi) - E_\nu(\varphi)}\right]\,. \label{eq:hessianPert}
				\end{equation}
				This Hessian is also sensitive to the vacuum structure of the theory and may diverge when there is a vacuum degeneracy, though it is not sensitive to all types of phase transitions. For example, in a continuous (or second order) phase transition, the numerator approaches zero fast enough as we approach the degeneracy that (\ref{eq:hessianPert}) remains finite. The metric has an additional power of the energy difference in the denominator, so we might expect that it can detect a wider range of transitions.\footnote{This observation motivated the introduction of a set of \emph{generalized adiabatic susceptibilities} \cite{De_Grandi_2010}, which differ from (\ref{eq:qimPert}) and (\ref{eq:hessianPert}) by an increased power of the energy difference. Presumably, these are even more sensitive to the phase structure of the theory than (\ref{eq:qimPert}), though we will not consider them here.}

				Though (\ref{eq:qimPert}) is very useful for relating the metric's behavior to the spectrum of the theory, it is rarely useful for actually computing the metric in non-trivial quantum field theories. To make progress in that direction, it is helpful to first relax the constraint that the vacuum state is normalized in the the metric's definition (\ref{eq:qim}) and write the line element as
				\begin{equation}
					\ud s^2 = g_{ab}(\varphi) \, \ud \varphi^a \es \ud \varphi^b = \frac{\langle \ud \vac | \ud \vac \rangle}{\langle \vac | \vac \rangle} - \frac{\langle \vac | \ud \vac \rangle}{\langle \vac | \vac \rangle} \frac{\langle \ud \vac | \vac \rangle}{\langle \vac | \vac \rangle}\,, \label{eq:qimUnnormalized}
				\end{equation}
				where $|\ud \Psi \rangle \equiv \big(\frac{\partial}{\partial \varphi^a} |\Psi \rangle \big)\ud \varphi^a$. We will assume that the unnormalized ground state can be prepared using the Euclidean path integral over ``fundamental fields,'' the collection of which we denote $\eta$,
				\begin{equation}
					\langle \tilde{\eta}| \Psi(\varphi) \rangle = \int^{\tilde{\eta}}\!\mathcal{D} \eta \, \lab{e}^{\sminus S_\slab{e}^{\es \scalebox{0.45}[0.5]{$-$}}[\varphi, \eta]} \quad \text{and} \quad \langle \Psi(\varphi) |\tilde{\eta} \rangle = \int_{\tilde{\eta}} \!\mathcal{D} \eta\, \lab{e}^{\sminus S_\slab{e}^{\es \scalebox{0.5}[0.5]{$+$}}[\varphi, \eta]}\,.
				\end{equation}
				Here, we have introduced the quantities
				\begin{equation}
					S_\slab{e}^{\scriptscriptstyle \pm}[\varphi, \eta] = \mp \int_{0}^{\pm}\!\ud \tau \int\!\ud^d x\, \mathcal{L}[\varphi, \eta]\,,
				\end{equation}
				the Euclidean action integrated over all positive (negative) Euclidean time. We have also assumed that the initial state of the path integral is washed out by the semi-infinite Euclidean evolution, so that $\langle \tilde{\eta} |\Psi(\varphi) \rangle$ and its conjugate only depend on the Euclidean action and the boundary value of the integral at $\tau = 0$, $\tilde{\eta}(\mb{x})$.

				Assuming that this boundary value does not vary with the parameters $\varphi^a$, we can write
				\begin{equation}
					\langle \tilde{\eta} |\es \ud \Psi \rangle = \minus \ud \varphi^a \int^{\tilde{\eta}}\!\mathcal{D} \eta\left[\frac{\partial S_\slab{e}^{\subm}}{\partial \varphi^a}\right] \lab{e}^{\sminus S_\slab{e}^{\es \scalebox{0.45}[0.5]{$-$}}[\varphi, \eta]},
				\end{equation}
				while a similar expression involving  $\big(\partial S_\slab{e}^{\subp}/\partial \varphi^b\big) \ud \varphi^b$ holds for $\langle \ud \Psi| \tilde{\eta} \rangle$. This expression allows us to rewrite the line element~(\ref{eq:qimUnnormalized}) in terms of expectation values in the theory at the point $\varphi^a$,
				\begin{equation}
					g_{ab}(\varphi)\es  \ud \varphi^a \, \ud \varphi^b = \left[\left\langle\frac{\partial S_\slab{e}^\subp}{\partial \varphi^a} \frac{\partial S_\slab{e}^{\subm}}{\partial \varphi^b}\right\rangle - \left\langle \frac{\partial S_\slab{e}^\subp}{\partial \varphi^a}\right\rangle \left\langle \frac{\partial S_\slab{e}^{\subm}}{\partial \varphi^b}\right \rangle \right] \ud \varphi^a \, \ud \varphi^b\,. 
				\end{equation}
				In particular, if the Euclidean action $S_\slab{e}[\varphi, \eta]$ shifts by a local operator under $\varphi^a \to \varphi^a + \ud \varphi^a$,
				\begin{equation}
					S_\slab{e}[\varphi+ \ud \varphi, \eta] = S_\slab{e}[\varphi, \eta] + \ud \varphi^a  \int\!\ud^{d+1} x\, \mathcal{O}_a(x)\,, \label{eq:defLag}
				\end{equation}
				then the information metric can be extracted from the two-point function $\langle \mathcal{O}_a(x) \, \mathcal{O}_b(x) \rangle$,
				\begin{equation}
					g_{ab} = \int_0^{\infty}\!\ud \tau_1 \int_{\sminus \infty}^0\!\ud \tau_2 \int\!\ud^d x_1\, \ud^d x_2 \left[\langle \mathcal{O}_a(\tau_1, \mb{x}_1) \mathcal{O}_b(\tau_2, \mb{x}_2) \rangle - \langle \mathcal{O}_a(\tau_1, \mb{x}_1) \rangle \langle \mathcal{O}_b(\tau_2, \mb{x}_2) \rangle \right]\,. \label{eq:qimTwoPoint}
				\end{equation}
				If the correlators of the $\mathcal{O}_a$ are not symmetric under time-reversal, this expression will contain a piece that is antisymmetric in $a$ and $b$, proportional to the Berry curvature. In this case, we must explicitly symmetrize. The same expression was derived in \cite{MIyaji:2015mia,Alvarez-Jimenez:2017gus,Trivella:2016brw,Bak:2015jxd,Bak:2017rpp} by considering the information metric as the fidelity susceptibility 
				\begin{equation}
					|\langle \vac(\varphi + \ud \varphi) | \vac(\varphi) \rangle | = 1 - \frac{1}{2}\es g_{ab}(\varphi)\, \ud \varphi^a \, \ud \varphi^b + \cdots\,, \label{eq:fidelitySusceptibility}
				\end{equation}
				though the derivation presented here applies to more generic $\varphi$-dependence.

		Finally, we can derive the information metric by computing the K\"{a}hler potential or log fidelity directly from the Euclidean path integral. We promote the parameter $\varphi^a$ to a spacetime-dependent background field $\varphi^a(\tau, \mb{x})$, and compute the Euclidean path integral
		\begin{equation}
			\lab{e}^{-\Gamma[\varphi(\tau)]} = \int\!\mathcal{D} \eta\, \exp\!\big(\minus S_\slab{e}[\varphi(\tau, \mb{x}), \eta]\big)\,.
		\end{equation} 
		We assume that the background field approaches a constant spacetime-independent value as $\tau \to \pm \infty$ so that the path integral starts and ends in a vacuum state. Though there are interesting questions that may arise if we consider arbitrary space-time dependent background fields \cite{Belin:2018fxe,Belin:2018bpg,Cordova:2019jnf,Cordova:2019uob}, in which case the K\"{a}hler potential and associated metric can be defined for a much more general class of states, we restrict ourselves to study the information metric associated to the vacua of a family and so assume that the background field is of the form
		\begin{equation}
			\varphi^a(\tau, \mb{x}) = \begin{cases} \varphi^a_\subp & \tau > 0 \\ \varphi^a_\subm & \tau < 0 \end{cases}\,.
		\end{equation}
		We can then identify the logarithm of this Euclidean path integral with the \emph{log fidelity}, or simply the \emph{divergence},
		\begin{equation}
			\minus \mathcal{D}(\varphi_\subp, \varphi_\subm) =  \Gamma[\varphi(\tau)]\,,
		\end{equation}
		and compute the information metric by taking the appropriate derivatives. 

		\subsection{Regulating the Metric} \label{sec:metricReg}

		When applied to a family of quantum field theories, each of the above prescriptions produces a divergent result which must be both UV and IR regulated. This is not so surprising, since such field theories contain an infinite number of degrees of freedom $\mathcal{N}$ and the metric is typically extensive in this number. A simple way to see this can happen is to consider the metric's definition as the fidelity susceptibility (\ref{eq:fidelitySusceptibility}), and assume that the vacuum states at both $\varphi^a$ and $\varphi^a + \ud \varphi^a$ are product states. The fidelity susceptibility is then an infinite product of these individual states, and any infinitesimal collective difference between the states will drive the overlap to zero, and thus force the metric to diverge. This is the physics of Anderson's orthogonality catastrophe \cite{anderson1967infrared,Zanardi_2006,GU_2010} and is a generic feature of theories with an infinite number of degrees of freedom. We will instead be interested in the rate at which this orthogonalization occurs in the ``thermodynamic limit''~$\mathcal{N} \to \infty$.

		We will use a variety of schemes to regulate the information metric. Each will be characterized by an IR and a UV energy scale, $\Lambda_\slab{ir} = L^{\sminus 1}$ and  $\Lambda_\slab{uv} = \epsilon^{\sminus 1}$, respectively. Generally, we will imagine that we have placed the theory in a box of side-length $L$ and on a lattice with spacing $\epsilon$, though in other regulation schemes they will not have such a precise interpretation. Of course, the metric we derive will depend on the regulation scheme we choose. However, we will only be interested on the metric's functional dependence on $\varphi^a$ and its parametric dependence on the two cutoffs, in the limits that $\Lambda_\slab{ir} \to 0$ and $\Lambda_\slab{uv} \to \infty$. We will confirm that a different choice of regulator only changes the overall constant scaling of the metric, and does not affect these more universal dependencies, at least for the examples we consider. 

		This is similar to the scheme-dependence of the Zamolodchikov metric on families of conformal field theories \cite{Kutasov:1988xb,Friedan:2012hi}, where different renormalization schemes are related to one another by diffeomorphisms of the $\varphi^a$. Different renormalization schemes produce different coordinate systems on the space of conformal field theories.  In general, though, the scheme dependence of the metric cannot be absorbed into a change of coordinates---we are no longer considering such a highly restricted class of theories and so a change in scheme generally changes the class under consideration. This dependence should not be surprising, since this metric characterizes families of probability distributions and predictions that are made by a theory with an infinite number of degrees of freedom depends on the precise scheme chosen. As long as one has a family of theories wherein unambiguous predictions can be made, this metric is well-defined.

		What, then, are we to make of these divergences? For a quantum field theory, the probability distribution defined by the vacuum state (\ref{eq:vacProb}) is a distribution on \emph{field configurations}. That is, a single draw from that distribution yields a field configuration on $d$-dimensional space---with our UV and IR regulators, this amounts to roughly $\propto (L \Lambda_\slab{uv})^d$ individual pieces of data. Thus, in the generic case that the metric scales\footnote{Here, we have assumed that all components of the metric scale as $(L \Lambda_\slab{uv})^d$. While this is not necessarily always the case---for instance, at a critical point the metric can scale super-extensively---this occurs generically when we take the $\varphi^a$ to be dimensionless Wilson coefficients, and in all of the examples we consider in Section~\ref{sec:examples}.}   with ``number of lattice sites'' $(L \Lambda_\slab{uv})^d$ as it does away from a critical point, we can define the \emph{intensive metric} $\tilde{g}_{ab}(\varphi)$, 
		\begin{equation}
			g_{ab}(\varphi) = (L \Lambda_\slab{uv})^d \es \tilde{g}_{ab}(\varphi)\,, \label{eq:intMet}
		\end{equation}
		This object characterizes how well, on average, we can distinguish between theories at $\varphi^a$ and $\varphi + \ud \varphi^a$ using a single ``local'' degree of freedom \cite{Balasubramanian:2014bfa}. We place ``local'' in quotes because, in principle, these measurements could be local in real space, momentum space, or some other parameterization of the theory's spatial dependence. That said, if the information metric is associated to the vacua of a family of Poincar\'{e}-invariant theories, we expect real and momentum space to be preferred. Unfortunately, the information metric does not indicate which measurements are discerning. It instead quantifies the \emph{average distinguishability per degree of freedom} which is not fine-grained enough to discern between different types of measurements. This is a crucial point that will be key to how we interpret divergences in the statistical length calculated with the intensive metric $\tilde{g}_{ab}(\varphi)$.

	\subsection{The Metric At and Away From Criticality} \label{sec:metricScaling}

		It will be useful to consider the scaling of the metric with respect to the UV and IR cutoffs generally. We will imagine deforming the action by a set of local operators $\mathcal{O}_a$ with scaling or mass dimension $\Delta_a$,
		\begin{equation}
			S \to S +  \ud \varphi^a \left[\Lambda_\slab{uv}^{d+1 - \Delta_a}\! \int\!\ud^{d+1} x\, \mathcal{O}_a(x)\right]\,, \label{eq:defLag2}
		\end{equation}
		where here we have scaled $\ud \varphi^a$, when compared to (\ref{eq:defLag}), by appropriate powers of the UV cutoff $\Lambda_\slab{uv}$ to make it dimensionless. This is natural if we consider the coordinates $\varphi^a$ as a set of Wilson coefficients, where $\Lambda_\slab{uv}$ is some fixed---albeit high---UV scale used to define the effective theory.

		Let us first consider how the metric behaves when the theory is at a critical point. Assuming that our local operators $\mathcal{O}_a$ have definite scaling dimension and vanishing one-point functions $\langle \mathcal{O}_a(x) \rangle = 0$, their normalized Euclidean correlators
		\begin{equation}
			\langle \mathcal{O}_a(\tau_1, \mb{x}_1) \mathcal{O}_b(\tau_2, \mb{x}_2)\rangle = \frac{\delta_{\Delta_a, \Delta_b}}{\left[(\tau_1 - \tau_2)^2 + (\mb{x}_1 - \mb{x}_2)^2\right]^{\mathrlap{\Delta_a}}}\, \label{eq:eucScaleInvarCorr}
		\end{equation}
		can be used to compute the metric (\ref{eq:qimTwoPoint}) with respect to $\varphi^a$ \cite{MIyaji:2015mia,Trivella:2016brw,Bak:2015jxd,Bak:2017rpp}
		\begin{equation}
			g_{ab}(\varphi) = \Lambda_\slab{uv}^{2 + 2 d - 2\Delta_a} \int_{\epsilon}^{L}\!\!\ud \tau_1 \int_{\sminus L}^{\sminus \epsilon}\!\ud \tau_2\int_{L^d}\!\ud^d x_1 \,\ud^d x_2\, \frac{\delta_{\Delta_a, \Delta_b}}{\left[(\tau_1 - \tau_2)^2 + (\mb{x}_1 - \mb{x}_2)^2\right]^{\mathrlap{\Delta_a}}}\quad\,.
		\end{equation}
		These integrals may be either UV or IR divergent, depending on the scaling dimension $\Delta_a$, and we have regulated them by the appropriate UV and IR cutoffs, $\epsilon = \Lambda_\slab{uv}^{\sminus 1}$ and $L = \Lambda_{\slab{ir}}^{\sminus 1}$ respectively. In the $L \to \infty$ and $\epsilon \to 0$ limits, the metric scales as \cite{Venuti:2007qcs} 
		\begin{equation}
			g_{ab}(\varphi) \propto 
				\begin{dcases}
					(L \Lambda_\slab{uv})^d  & 2 \Delta_a > d+2 \\
					(L \Lambda_\slab{uv})^d \log (L \Lambda_\slab{uv}) & 2 \Delta_a = d +2 \\
					(L \Lambda_\slab{uv})^{d + (2 + d - 2 \Delta_a)}  & 2 \Delta_a < d+2
				\end{dcases}\,. \label{eq:metAtCrit}
		\end{equation}
		We see that IR divergences can cause the metric to scale \emph{super-extensively} in the number of degrees of freedom $N  = (L \Lambda_\slab{uv})^d$ when $2 \Delta_a \leq d + 2$. This is akin to the behavior we found in the toy model in Section~\ref{sec:cim}, whose metric diverged super-extensively with the ``spacetime volume'' $\Omega^2$ at metric singularities.

		Away from the critical point, the Euclidean correlator (\ref{eq:eucScaleInvarCorr}) is exponentially suppressed at large distances,
		\begin{equation}
			\langle \mathcal{O}_a(x_1) \mathcal{O}_b(x_2) \rangle \sim \frac{\lab{e}^{-|x_1 - x_2|/\xi}}{|x_1 - x_2|^{2\Delta_a}}\,,\mathrlap{\qquad |x_1 - x_2| \to \infty\,,} \label{eq:eucCorrAsymp}
		\end{equation}
		where we have assumed the operators have the same dimension $\Delta_a$ and introduced the correlation length of the theory $\xi$. The metric then scales as \cite{Venuti:2007qcs,Carollo:2019ygj}
		\begin{equation}
			g_{ab}(\varphi) \propto 
				\begin{dcases}
					(L \Lambda_\slab{uv})^d & 2\Delta_a > d + 2 \\
					(L \Lambda_\slab{uv})^d \log (\xi \Lambda_\slab{uv}) & 2\Delta_a = d+2 \\
					(L \Lambda_\slab{uv})^d (\xi \Lambda_\slab{uv})^{d+2 - 2 \Delta_a} & 2\Delta_a < d+2 
				\end{dcases}\,. \label{eq:metAwayFromCrit}
		\end{equation}
		As we approach the critical point, the correlation length diverges $\xi \to \infty$ and the metric (\ref{eq:metAwayFromCrit}) seemingly diverges for $2 \Delta_a \leq d+2$. At the critical point, the correlation length is effectively replaced with the IR cutoff and we recover the super-extensive scaling seen in (\ref{eq:metAtCrit}).

		Assuming that the vacuum state is translationally invariant and that the theory is perturbed only by local operators (\ref{eq:defLag}) under the shift $\varphi^a \to \varphi^a + \ud \varphi^a$, one can prove \cite{Venuti:2007qcs}  that a metric singularity, or equivalently super-extensive scaling, implies that theory is gapless and has divergent correlation length. This is what is so useful about the quantum information metric as tool for detecting (quantum) phase transitions---it is a general measure of the qualitative difference between two nearby theories. We should able to easily distinguish between two different phases as long as we pay attention to the right observables, and the information metric measures how distinguishable two nearby theories are using the most discerning observable.  We should note, however, the metric is not a perfect measure and that the converse is not necessarily true: the correlation length can diverge without an analogous metric divergence.

		Let us understand what this scaling analysis says about infinite distance points in the intensive metric $\tilde{g}_{ab}(\varphi)$. Let us restrict to a one-dimensional parameter space $\varphi$, in which the critical theory is perturbed by an operator of dimension $\Delta$. Assuming that the metric is super-extensive at $\varphi_0$, the correlation length of the theory must diverge as
		\begin{equation}
			\xi(\varphi) \propto |\varphi - \varphi_0|^{\sminus\nu}\,,
		\end{equation}
		where $\nu$ is the correlation length critical exponent. As long as the metric diverges, $2 \Delta < d+2$, this implies that \cite{Venuti:2007qcs}
		\begin{equation}
			g_{\varphi \varphi} \propto |\varphi - \varphi_0|^{\sminus \nu(d+2 - 2 \Delta)}\,,
		\end{equation}
		i.e. the metric scaling is determined by its relationship to the correlation length $\xi$ and the scaling of that length as we approach the critical point $\varphi_0$.
		This point is then at infinite distance if
		\begin{equation}
			-\nu(d+2 - 2 \Delta)  \leq -2\,.
		\end{equation}
		Furthermore, this critical exponent can be related to the scaling dimension of the perturbing operator. Under the usual assumptions about scaling at a critical point, $\nu^{\sminus 1} = d+ 1 -\Delta$, the above inequality can be written as
		\begin{equation}
			\frac{(d+2 - 2 \Delta)}{d + (d + 2 - 2 \Delta)} \geq 1\,.
		\end{equation}
		The terms in parenthesis are always positive if we demand that the metric behaves super-extensively, $2 \Delta < d +2$, and so this inequality can never be satisfied in field theories with non-zero dimension~$d \neq 0$. 
		In quantum mechanical theories, $d = 0$, this constraint is trivially satisfied---we will find that the simple harmonic oscillator, for example, exhibits such an infinite distance point. We conclude that, in order to realize an infinite distance point in this intensive metric, there must be some violation of the naive scaling arguments that are usually valid around a quantum critical point. In the following sections, we will find that such infinite distance points arise not when a single field becomes gapless but instead from the collective behavior of an infinite tower of fields.

\section{Some Illustrative Examples} \label{sec:examples}
		
		The goal of this section is to study the quantum information metric in a variety of examples with finitely many fields. We will be mainly concerned with the behavior of the metric associated with the masses of free scalar and fermionic fields, in the limit the field becomes massless. As a warmup to this, we study the metric for the simple harmonic oscillator as a function of its frequency $\omega$ in~\S\ref{sec:sho}, and compute it in a variety of ways. Like the Gaussian distribution (\ref{eq:gaussianMetric}), this simple model has infinite distance singularities in the zero- and infinite-frequency limits. In~\S\ref{sec:scalar}, we show that free scalar field, the harmonic oscillator's higher-dimensional analog, does not. However, it does provide a useful testing ground for computing the information metric in quantum field theories, and we apply the techniques we learn there to the Dirac fermion in \S\ref{sec:dirac}. 

		We conclude this section by studying the information metric in two interesting cases, already equipped with a metric. In \S\ref{sec:cft}, we study the information metric associated to marginal deformations of a conformal field theory, which take us along its moduli space. The CFT moduli space is already equipped with a metric, the Zamolodchikov metric, and we find that this is proportional to the intensive metric (\ref{eq:intMet}) we defined in the previous section. We then study the information metric associated to the moduli of a $(d+1)$-dimensional bosonic nonlinear sigma model in~\S\ref{sec:nlsm}. These theories are similarly equipped with a metric on field space, and again we find that this is proportional to the intensive information metric.

	\subsection{The Harmonic Oscillator} \label{sec:sho}

		Since free fields are nothing more than collections of (bosonic or fermionic) springs, it will be useful to first compute the information metric for the simple harmonic oscillator. We will compute it in several different ways, all of which will be used as warm-ups to the field theory computations that we will perform later.

	\subsubsection*{From the Ground State}
		We argued in \S\ref{sec:metricInterp} that the quantum information metric was equivalent to the classical information metric associated to the probability distribution defined by the ground state. Since we know this ground state wavefunction explicitly, we can use it to easily compute the metric.

		The simple harmonic oscillator Hamiltonian
		\begin{equation}
			\mathcal{H} = \frac{p^2}{2m} + \frac{1}{2} m \omega^2 q^2 \label{eq:shoHam}
		\end{equation}
		has the normalized ground state wavefunction $\langle q |\vac(\omega) \rangle = (m \omega/\pi)^{1/4} \exp\big(\minus m \omega q^2/2\big)$ and, using the definition of the information metric (\ref{eq:qim}) and taking derivatives of this state with respect to $\omega$, we quickly find that
		\begin{equation}
			g_{\omega \omega} =  \langle \partial_\omega \vac| \big(\mathbbm{1} - | \vac \rangle \langle \vac| \big) | \partial_\omega \vac \rangle = \frac{1}{8 \omega^2}\,. \label{eq:shoQim}
		\end{equation}
		Since the associated probability distribution is a Gaussian, it is no wonder we recover a metric similar to (\ref{eq:gaussianMetric}). However, we should note that the variance of this Gaussian is $\sigma^{2} = 1/(2 m \omega)$, so we only recover the same form as (\ref{eq:gaussianMetric}) since $\ud s^2 = (\ud \log \sigma)^2$.

		We can similarly compute the overlap of the unnormalized ground state,
		\begin{equation}
			\langle \vac(\omega_\subp) | \vac(\omega_\subm) \rangle = \int_{\sminus \infty}^{\infty}\!\ud q\, \lab{e}^{- \tfrac{1}{2} m (\omega_\subp + \omega_\subm) q^2} = \sqrt{\frac{2 \pi/m}{\omega_\subp + \omega_\subm}}
		\end{equation}
		to form the divergence
		\begin{equation}
			\minus \mathcal{D}(\omega_\subp, \omega_\subm) = \log \, \langle \vac(\omega_\subp) | \vac(\omega_\subm) \rangle = \frac{1}{2} \log\!\left(\frac{2 \pi/m}{\omega_\subp + \omega_\subm}\right) \label{eq:shoDivergence}
		\end{equation}
		from which we can also derive the metric
		\begin{equation}
			g_{\omega \omega} = -\!\!\es \left.\frac{\partial^2 \mathcal{D}(\omega_\subp, \omega_\subm)}{\partial \omega_\subp \,\partial \omega_\subm}\right|_{\omega} = \frac{1}{8 \omega^2}\,,
		\end{equation}
		where we use $|_{\omega}$ as short-hand setting $\omega_\subp = \omega_\subm = \omega$.
		As the mass term clearly only serves to set dimensions and does not effect the form of the metric, we will take $m = 1$ in what follows.

	\subsubsection*{From the Spectrum}

		It will also be useful to consider a more perturbative approach in computing the metric. Since we will later be interested in providing a ``second-quantized'' picture of this metric for the scalar field, let us confirm that (\ref{eq:qimPert}) reproduces the metric found above. By changing the frequency $\omega$, the Hamiltonian (\ref{eq:shoHam}) is deformed by $\partial_\omega \mathcal{H} = \omega q^2$.
		We can then use (\ref{eq:qimPert}) to write the metric in terms of the standard energy eigenstates,  $\mathcal{H}|\nu \rangle = \omega\big(\nu + \frac{1}{2}\big)|\nu \rangle$,
		\begin{equation}
			g_{\omega \omega} = \sum_{\nu =1}^{\infty} \frac{\omega^2 \langle 0 | q^2 | \nu \rangle \langle \nu| q^2 | 0 \rangle}{\omega^2 \nu^2}\,.
		\end{equation}
		The only non-zero matrix element is $\langle 2 | q^2 |0 \rangle = 1/(\sqrt{2} \omega)$, so the sum collapses to a single term and we recover (\ref{eq:shoQim})\,.

	\subsubsection*{From the Correlation Function}

		Since quantum field theories are often best understood using their correlation functions, with which it is much easier to incorporate interactions, it will be helpful to also compute the metric using the integrated two-point function (\ref{eq:qimTwoPoint}). For the harmonic oscillator,  this reduces to
		\begin{equation}
			g_{\omega \omega} = \omega^2 \int_{0}^{\infty}\!\ud \tau_1 \, \int_{\minus \infty}^0\!\ud \tau_2 \left[\langle q^2(\tau_1) q^2(\tau_2)\rangle - \langle q^2(\tau_1) \rangle \langle q^2(\tau_2)\rangle\right]\,,
		\end{equation}
		where $\tau_{1,2}$ are Euclidean times.
		
		With an eye towards field theory, we will compute the correlation functions of these ``composite operators'' using point-splitting, i.e. by writing
		\begin{equation}
			g_{\omega \omega} = \omega^2  \int_{0}^{\infty}\!\ud \tau_1 \, \int_{\minus \infty}^0\!\ud \tau_2 \, \Big[\langle q(\tau_1 + \tilde{\epsilon}) q(\tau_1) q(\tau_2) q(\tau_2 - \tilde{\epsilon})\rangle - \langle q(\tau_1 + \tilde{\epsilon}) q(\tau_1) \rangle \langle q(\tau_2) q(\tau_2 - \tilde{\epsilon})\rangle\Big]\,,
		\end{equation}
		and taking the $\tilde{\epsilon} \to 0$ limit at the end. Since this theory is free, Wick's theorem allows us to rewrite the integrand as
		\begin{equation}
			\langle q(\tau_1 + \tilde{\epsilon}) q(\tau_2) \rangle \langle q(\tau_1)  q(\tau_2 - \tilde{\epsilon})\rangle +  \langle q(\tau_1 + \tilde{\epsilon}) q(\tau_2 - \tilde{\epsilon}) \rangle \langle q(\tau_1)  q(\tau_2)\rangle \underset{\tilde{\epsilon} \to 0}{=} 2 \langle q(\tau_1) q(\tau_2)\rangle^2\,.
		\end{equation}
		The possible divergence associated with the two operators coinciding in Euclidean time is cancelled, a fact that will persist in our field theory computations. By using the Euclidean correlator
		\begin{equation}
			\langle q(\tau_1) q(\tau_2) \rangle = \frac{1}{2 \omega} \lab{e}^{-\omega (\tau_1 - \tau_2)}\,,
		\end{equation}
		we find that
		\begin{equation}
			g_{\omega \omega} = \omega^2 \int_{0}^{\infty}\!\ud \tau_1 \int_{\sminus \infty}^0 \!\ud \tau_2 \left[\frac{1}{2 \omega^2} \lab{e}^{-2\omega(\tau_1 - \tau_2)}\right] = \frac{1}{8 \omega^2}\,,
		\end{equation}
		which is exactly the same expression we derived using the explicit form of the ground state.

	\subsubsection*{From the Euclidean Path Integral}

		Finally, it will be useful to extract the metric from a Euclidean path integral performed in the presence of a background field. That is, we define
		\begin{equation}
			\lab{e}^{\sminus \Gamma[\omega^2(\tau)]} = \int\!\mathcal{D} q(\tau) \, \lab{e}^{-S_\slab{e}[q(\tau), \omega(\tau)]}\,,
		\end{equation}
		where the Euclidean action is
		\begin{equation}
			S_\slab{e}[q(\tau), \omega(\tau)] = \frac{1}{2}\int\!\ud \tau \left[\dot{q}^2 + \omega^2(\tau) q^2\right]\,,
		\end{equation}
		$\omega(\tau)$ is an arbitrary, time-dependent frequency, and $\dot{q} \equiv \partial_\tau q$. 

		This path integral is Gaussian, and so can be immediately evaluated to yield the effective action
		\begin{equation}
			\Gamma[\omega(\tau)] = \frac{1}{2} \log \!\left[\frac{\det(-\partial_\tau^2 + \omega^2(\tau))}{\det(-\partial_\tau^2)}\right], 
		\end{equation}
		where we have regularized the determinant by that of the ``free'' operator $-\partial_\tau^2$. 
		The divergence $\mathcal{D}(\omega_\subp, \omega_\subm)$ can then be extracted from $\Gamma[\omega(\tau)]$ by taking $\omega(\tau)$ to discontinuously jump from $\omega_\subm$ to $\omega_\subp$ at $\tau = 0$,
		\begin{equation}
			\omega(\tau) = \begin{cases} \omega_\subp & \tau > 0 \\ \omega_\subm & \tau < 0 \end{cases}\,,
		\end{equation}
		thus forming the Euclidean path integral representation of the overlap $\langle \vac(\omega_\subp) | \vac(\omega_\subm) \rangle$. In field theories, we will need to regularize this expression with a UV cutoff $\Lambda_\slab{uv}$, or alternatively smoothly interpolate between the analogs of $\omega_\subm$ and $\omega_\subp$ on a timescale of order $\epsilon = \Lambda_\slab{uv}^{\sminus 1}$. However, this regularization is unneeded for the simple harmonic oscillator. 

		Perhaps the most straightforward way of computing this determinant is via the Gel'fand-Yaglom method \cite{Coleman:1985rnk,Dunne:2007rt}. We first place the two operators $-\partial_\tau^2 + \omega^2(\tau)$ and $-\partial_\tau^2$ on the finite interval $\tau \in [-T, T]$. We then compute the ratio of the determinants by first solving the initial value problems
		\begin{equation}
			\left[-\frac{\partial^2}{\partial \tau^2} + \omega^2(\tau) \right] \psi_1(\tau) = 0 \quad \text{and} \quad -\frac{\partial^2}{\partial \tau^2} \psi_2(\tau) = 0
		\end{equation}
		subject to the initial conditions $\psi_i(-T) = 0$ and $\psi_i'(-T) = 1$, and forming the ratio
		\begin{equation}
			\frac{\psi_1(T)}{\psi_2(T)} = \frac{\det(-\partial_\tau^2 + \omega^2(\tau))}{\det(-\partial_\tau^2)}\,.
		\end{equation}
		These initial value problems can be easily solved to yield
		\begin{equation}
			\psi_1(\tau) = \begin{dcases}
					\tfrac{1}{\omega_\subp} \cosh \omega_\subm T \sinh \omega_\subp \tau + \tfrac{1}{\omega_\subm} \sinh \omega_\subm T \cosh \omega_\subp T \,, &\tau > 0 \\
					\tfrac{1}{\omega_\subm} \sinh \omega_\subm(\tau + T) & \tau < 0 
					\end{dcases}
		\end{equation}
		and $\psi_2(\tau) = \tau + T$, so the effective action is
		\begin{equation}
			\Gamma[\omega(\tau)] = \frac{1}{2} \log \!\left[\frac{1}{2 \omega_\subp T} \cosh \omega_\subm T \, \sinh \omega_\subp T + \frac{1}{2 \omega_\subm T} \sinh \omega_\subm T \cosh \omega_\subp T\right].
		\end{equation}
		In the limit $T \to \infty$, the argument of the log is dominated by
		\begin{equation}
			\Gamma[\omega(\tau)] \sim \frac{1}{2} \log\!\left[\frac{ (\omega_\subp + \omega_\subm)\lab{e}^{(\omega_{\subp} + \omega_{\subm}) T}}{8 T \omega_{\subp} \omega_\subm}\right]\,.
		\end{equation}
		Ignoring terms that do not contribute to the metric, we may write the divergence as
		\begin{equation}
			\minus \mathcal{D}(\omega_\subp, \omega_\subm) = - \frac{1}{2} \log (\omega_\subp + \omega_\subm) + \cdots
		\end{equation}
		 We thus recover the important $\omega_\subp$- and $\omega_\subm$-dependence of (\ref{eq:shoDivergence}), and the metric $g_{\omega \omega} = 1/(8 \omega^2)$.

	\subsubsection*{Interpretation of Infinite Distance}

		The harmonic oscillator's information metric has infinite distance points at both $\omega = 0$ and as $\omega \to \infty$. The statistical length
		\begin{equation}
			d(\omega_1, \omega_0) = \frac{1}{2 \sqrt{2}} \int_{\omega_0}^{\omega_1} \!\frac{\ud \omega}{\omega} = \log \frac{\omega_1}{\omega_0}\,,
		\end{equation}
		with $\omega_1 > \omega_0$, diverges as $\omega_1 \to \infty$ or $\omega_0 \to 0$, with $\omega_0$ and $\omega_1$ held constant respectively. Fortunately, this model is simple enough to know exactly why this length diverges.

		It is easy to see that the limit $\omega \to \infty$ is hyper-distinguishable, since the particle becomes \emph{fixed} to the minimum of the potential at $q = 0$. The support of the ground state probability distribution $|\langle q | \Psi(\omega) \rangle|^2 \propto \exp(- \omega q^2)$ collapses to a single point, and the Hilbert space of the theory becomes effectively one-dimensional. There is zero chance that we detect the particle away from $x = 0$, so a single observation of the particle somewhere else $q \neq 0$ is sufficient to firmly rule out the possibility that $\omega = \infty$. The information metric quantifies this qualitative change in the distribution by placing this point at infinite distance.

		A similar qualitative change occurs in the $\omega \to 0$ limit, where the Hamiltonian approaches that of the free particle. Here, the structure of the Hilbert space also undergoes a dramatic qualitative change---the eigenfunctions of the Hamiltonian are no longer nice and normalizable, and this is reflected in the tails of the vacuum state probability distribution. In fact, we can see that this is the dual of the $\omega \to \infty$ limit by considering the vacuum state probability distribution in $p$, $|\langle p | \Psi(\omega) \rangle|^2 \propto \exp(-p^2/\omega)$. As $\omega \to 0$, this distribution becomes localized about $p = 0$ and any non-zero measurement of $p$ in the ground state would rule out $\omega = 0$. This hyper-distinguishable point is thus infinitely far away from any finite $\omega$.

	\subsection{The Free Scalar} \label{sec:scalar}

		We now turn to the free scalar field in $(d+1)$-dimensional spacetime, with Euclidean action
		\begin{equation}
			S_\slab{e}[\Phi, m^2(\varphi)] = \int\!\ud^{d+1} x \left[ \frac{1}{2} (\partial_\mu \Phi)^2 + \frac{1}{2}  m^2(\varphi) \Phi^2\right]\,.
		\end{equation}
		We take the mass of the field to depend on the fixed parameters $\varphi^a$, and our goal is to compute the information metric $g_{ab}(\varphi)$ with respect to them. 

		As described in \S\ref{sec:metricReg}, we will regulate this theory by placing it both in a box of side-length $L \equiv \Lambda_{\slab{ir}}^{\sminus 1}$ and on a lattice with spacing $\epsilon = \Lambda_{\slab{uv}}^{\sminus 1}$. As one might expect, the major qualitative difference between the scalar field and the simple harmonic oscillator is that the number of degrees of freedom $\mathcal{N}$ in the former diverges as we take $\mathcal{N} = (L \Lambda_\slab{uv})^d \to \infty$. We can anticipate the structure of these divergences by considering the massless correlator,
		\begin{equation}
			\langle \Phi(\tau_1, \mb{x}_1) \Phi(\tau_2, \mb{x}_2)\rangle = \frac{\Gamma\big(\frac{d-1}{2}\big)}{4 \pi^{(d-1)/2}} \frac{1}{\left[(\tau_1 - \tau_2)^2 + (\mb{x}_1 - \mb{x}_2)^2\right]^{\mathrlap{(d-1)/2}}}\mathrlap{\qquad\quad .}
		\end{equation}
		The scaling dimension of a mass perturbation is $\Delta =d -1$, and so from the general analysis we expect this metric to diverge super-extensively when $2\Delta < d + 2$ or $d < 4$. Specifically, at the critical point the metric should be proportional to $g_{ab} \propto (L \Lambda_\slab{uv})^{4}$, which for $d < 4$ is super-extensive. With these expectations in place, we will derive this metric in a number of ways, to help us understand different aspects of the metric and its singularities.

		\subsubsection*{From the Divergence}

			Since a scalar field can be considered as a set of non-interacting harmonic oscillators labeled by the wavevector $\mb{k} \in 2 \pi \mathbb{Z}^d/L$ with frequency $\omega_{\mb{k}}(\varphi) = \sqrt{\mb{k}^2 + m^2(\varphi)}$, we can use (\ref{eq:shoDivergence}) to quickly find the divergence
			\begin{equation}
				\minus \mathcal{D}(\varphi_\subp, \varphi_{\subm}) = \frac{1}{2} \sum_{\mb{k}} \log\left[\frac{\Lambda_{\textsc{ir}}}{\omega_{\mb{k}}(\varphi_+) + \omega_{\mb{k}}(\varphi_-)}\right]\,. \label{eq:scalarDivergence}
			\end{equation}
			For a complex scalar field, this divergence is doubled to account for the doubling of the number of degrees of freedom. The metric is then
			\begin{equation}
				g_{ab}(\varphi) = - \!\left.\frac{\partial^2 \mathcal{D}(\varphi_+, \varphi_-)}{\partial \varphi^a_+ \partial \varphi_-^b} \right|_{\varphi} = \frac{1}{32}  \sum_{\mb{k}} \frac{\partial_a m^2(\varphi)\, \partial_b m^2(\varphi)}{(\mb{k}^2 + m^2(\varphi))^2} 
			\end{equation}
			where the $|_{\varphi}$ denotes that the expression is evaluated at $\varphi_+^a = \varphi_-^a = \varphi^a$. In the continuum limit, this becomes
			\begin{equation}
				g_{ab} = \frac{L^d}{32} \int\!\!\frac{\ud^d k}{(2 \pi)^d} \frac{\partial_a m^2 \partial_b m^2}{(\mb{k}^2 + m^2)^2}\,, \label{eq:scalarQIM}
			\end{equation}
			where we have suppressed $\varphi$-dependence of $m^2$. In converting the sum over $\mb{k}$ to an integral, we have already introduced an IR cutoff $\Lambda_\slab{ir} = L^{\sminus 1}$. We will further regulate this using a hard UV cutoff, restricting the range of the integral to $\Lambda_\slab{ir} \leq |\mb{k}| \leq \Lambda_\slab{uv}$.

		\subsubsection*{From the Correlators}

			We can similarly derive the metric from its expression in terms of correlation functions (\ref{eq:qimTwoPoint}). The Euclidean action changes by
			\begin{equation}
				\frac{\partial S_{\slab{e}}^{\pm}}{\partial \varphi^a_{\pm}} = \pm \frac{1}{2} \partial_a m^2(\varphi_\pm) \int_{0}^{\pm \infty}\!\!\ud \tau\, \ud^d x\, \Phi^2(\tau, \mb{x})\,,
			\end{equation}
			so the information metric is
			\begin{equation}
				g_{ab}(\varphi) = \frac{1}{4} \partial_a m^2 \partial_b m^2\,  \mathcal{I}_\lab{s}(\varphi)
			\end{equation}
			where we have introduced the quantity
			\begin{equation}
				\mathcal{I}_\lab{s}(\varphi) = \int_\epsilon^{\infty}\!\ud \tau_1 \!\int_{\sminus \infty}^{\sminus \epsilon} \!\ud \tau_2 \int_{L^d} \ud^d x_1\, \ud^d x_2 \left[\langle \Phi^2(\tau_1, \mb{x}_1) \Phi^2(\tau_2, \mb{x}_2) \rangle - \langle \Phi^2(\tau_1, \mb{x}_1)\rangle \langle \Phi^2(\tau_2, \mb{x}_2)\rangle\right]. \label{eq:scalarIntegral}
			\end{equation}
			We have regulated this expression by both introducing a short-time cutoff $\epsilon$ and by placing the theory in a box of side-length $L$. Since we will only be interested in the asymptotic behavior of this expression as $L \to \infty$ and $\epsilon \sim \Lambda_\slab{uv}^{\sminus 1} \to 0$, we will use the standard free-field correlation functions calculated in the continuum theory. Any differences between the continuum correlation functions and the appropriately discretized ones will be sub-leading in the limit $L \to \infty$ and $\epsilon \to 0$, and so we can thankfully ignore the distinction.

			The correlation functions of these composite operators must be further regulated, and we will use point-splitting to define 
			\begin{align}
				\langle \Phi^2(\tau_1, \mb{x}_1)& \Phi^2(\tau_2, \mb{x}_2) \rangle - \langle \Phi^2(\tau_1, \mb{x}_1)\rangle \langle \Phi^2(\tau_2, \mb{x}_2)\rangle \nonumber \\
				\quad = \lim_{\tilde{\epsilon} \to 0} \Big[ \, &\big\langle \Phi(\tau_1 + \tilde{\epsilon}, \mb{x}_1) \Phi(\tau_1, \mb{x}_1)  \Phi(\tau_2, \mb{x}_2) \Phi(\tau_2 - \epsilon, \mb{x}_2) \big\rangle \nonumber \\
				&- \big\langle \Phi(\tau_1 + \tilde{\epsilon}, \mb{x}_1) \Phi(\tau_1, \mb{x}_1)\big\rangle \big\langle \Phi(\tau_2, \mb{x}_2) \Phi(\tau_2 - \tilde{\epsilon}, \mb{x}_2)\big\rangle\Big]\,.
			\end{align}
			This is again a free theory so Wick's theorem reduces this to
			\begin{equation}
				\langle \Phi^2(\tau_1, \mb{x}_1) \Phi^2(\tau_2, \mb{x}_2) \rangle - \langle \Phi^2(\tau_1, \mb{x}_1)\rangle \langle \Phi^2(\tau_2, \mb{x}_2)\rangle = 2 \langle \Phi(\tau_1, \mb{x}_1) \Phi(\tau_2, \mb{x}_2)\rangle^2\,.
			\end{equation}
			The divergence generated by the two operators coinciding cancel, and we arrive at the same result we found find using other regularization schemes, i.e. defining the perturbation through normal-ordering.

			Using the Euclidean correlator
			\begin{equation}
				\langle \Phi(\tau_1, \mb{x}_1) \Phi(\tau_2, \mb{x}_2) \rangle = \int\!\!\frac{\ud^d k}{(2 \pi)^d} \frac{1}{2 \omega_{\mb{k}}} \lab{e}^{i \mb{k} \cdot(\mb{x}_1 - \mb{x}_2) -\omega_{\mb{k}}(\tau_1 - \tau_2)},
			\end{equation}
			with $\omega_{\mb{k}} = \sqrt{\mb{k}^2 + m^2}$, our expression for $\mathcal{I}$ reduces to
			\begin{equation}
				\mathcal{I}_{\lab{s}}(\varphi) = \int_{\epsilon}^{\infty}\!\ud \tau_1 \int_{\sminus \infty}^{\sminus \epsilon}\!\ud \tau_2 \int_{L^d} \!\ud^d x_1\, \ud^d x_2 \int\!\frac{\ud^d k_1}{(2 \pi)^d} \int\!\frac{\ud^d k_2}{(2 \pi)^d}\, \frac{\lab{e}^{i(\mb{k}_1 + \mb{k}_2)\cdot(\mb{x}_1 - \mb{x}_2) - (\omega_{\mb{k}_1} + \omega_{\mb{k}_2})(\tau_1- \tau_2)}}{4 \omega_{\mb{k}_1} \es \omega_{\mb{k}_2}}\,.
			\end{equation}
			Integrating over both spatial coordinates produces a factor of $(2\pi)^d L^d \delta^{(d)}(\mb{k}_1 + \mb{k}_2)$, and using $\omega_{\sminus \mb{k}} = \omega_{\mb{k}}$, we may write
			\begin{equation}
				g_{ab}(\varphi) = \frac{L^d}{32} \int\!\!\frac{\ud^d k}{(2\pi)^d} \frac{\partial_a m^2 \partial_b m^2}{(\mb{k}^2 + m^2)^2} \lab{e}^{-4 \epsilon \sqrt{\mb{k}^2 + m^2}}\,. 
			\end{equation}
			As expected, we see that $\epsilon$ plays the role of a UV cutoff. It will be simpler to take $\epsilon \to 0$ and instead regulate this expression with a hard UV cutoff as we did in the previous section, and we again recover (\ref{eq:scalarQIM}).

		\subsubsection*{From the Spectrum}

			Before we move onto actually evaluating the metric, it will be useful to derive it directly from the spectral representation (\ref{eq:qimPert}). This will make it obvious exactly which states contribute to the metric, which will be useful in understand its behavior in various limit we consider below.

			\newpage 
			We again regulate the theory by placing it in a box and imposing periodic boundary conditions. We then expand the field into creation and annihilation operators
			\begin{equation}
				\Phi(\mb{x}) = \frac{1}{L^d} \sum_{\mb{k}} \frac{1}{2 \omega_{\mb{k}}} \left[a_{\mb{k}} \lab{e}^{i \mb{k} \cdot \mb{x}} + a^\dagger_{\mb{k}} \lab{e}^{\sminus i \mb{k} \cdot \mb{x}}\right]\,,
			\end{equation}
			where the creation and annihilation operators satisfy $[\smash{a^{\vphantom{\dagger}}_{\mb{k}_1}, a^{\dagger}_{\mb{k}_2}}] = 2 \smash{L^d} \omega_{\mb{k}_1} \delta_{\mb{k}_1, \mb{k}_2}$\,, the analog of the familiar commutation relations.
			Under the shift $\varphi^a \to \varphi^a + \ud \varphi^a$, the Hamiltonian is deformed by
			\begin{equation}
				\partial_a \mathcal{H} = \frac{1}{2} \partial_a m^2 \int_{L^d} \!\ud^d x\, \Phi^2(\mb{x})\, = \frac{\partial_a m^2}{8 L^d} \sum_{\mb{k}} \frac{1}{\omega_{\mb{k}}^2} \left[a_{\mb{k}}^{\vphantom{\dagger}} a_{\sminus \mb{k}}^{\vphantom{\dagger}} + a_{\mb{k}}^{\vphantom{\dagger}} a^\dagger_{\mb{k}} + a^\dagger_{\mb{k}} a^{\vphantom{\dagger}}_{\mb{k}} + a_{\mb{k}}^\dagger a_{\sminus \mb{k}}^\dagger\right]\,.
			\end{equation}
			The spectral formula instructs us to consider the matrix element
			\begin{equation}
				\langle \alpha |\partial_a \mathcal{H} | 0 \rangle = \frac{\partial_a m^2}{8 L^d} \sum_{\mb{k}} \frac{1}{\omega_{\mb{k}}^2} \langle \alpha | a^\dagger_{\mb{k}} a^\dagger_{\sminus \mb{k}} | 0 \rangle
			\end{equation}
			for all states $|\alpha \rangle$ that are \emph{not} the vacuum state. The only states for which this matrix element is non-zero are zero momenta two-particle states,
			\begin{equation}
				\langle \mb{k}', \minus \mb{k}' | \partial_a \mathcal{H} | 0 \rangle = \frac{\partial_a m^2}{2 \omega_{\mb{k}'}}
			\end{equation}
			Putting this all together, we find
			\begin{equation}
				g_{ab}(\varphi) = \sum_{\alpha \neq 0} \frac{\langle 0 | \partial_a \mathcal{H} | \alpha \rangle \langle \alpha | \partial_b \mathcal{H}| 0 \rangle}{(E_\alpha - E_0)^2} = \frac{1}{2} \sum_{\mb{k} \neq 0} \frac{\langle 0 | \partial_a \mathcal{H} | \mb{k}, \minus \mb{k} \rangle \langle \mb{k}, \minus \mb{k}| \partial_b \mathcal{H} | 0 \rangle}{4 \omega_{\mb{k}}^2}\,,
			\end{equation}
			where we have included an extra factor of $\frac{1}{2}$ to avoid overcounting. This expression then reduces to
			\begin{equation}
				g_{ab}(\varphi) = \frac{1}{32} \sum_{\mb{k} \neq 0} \frac{\partial_a m^2 \, \partial_b m^2}{(\mb{k}^2 + m^2)^2}\,, \label{eq:metSum}
			\end{equation} 
			where we explicitly not included the vacuum contribution $\mb{k} = 0$. 

			Intuitively, the $\mb{k} = 0$ contribution corresponds to zero-mode of the scalar field. This is taken to be fixed in the ensemble we consider. That is, implicit in the quantization of the field is that its vacuum expectation value vanishes. Allowing this mode to contribute takes us out of the family of theories we are considering, i.e. the family formed by continuously changing the mass of the scalar field while keeping its vacuum expectation value fixed, and so we are right to explicitly exclude it.

		\subsubsection*{Evaluation of the Metric}

			We have found that the information metric for single $(d+1)$-dimensional real scalar field with mass-squared $m^2(\varphi)$ under the deformation $\varphi^a \to \varphi^a + \ud \varphi^a$ is given by
			\begin{equation}
				g_{ab}(\varphi) = \frac{L^d}{32} \frac{\partial_a m^2 \partial_b m^2}{m^4} \int\!\!\frac{\ud^d k}{(2 \pi)^d} \frac{m^4}{(\mb{k}^2 + m^2)^2}\,, \label{eq:scalarMet2}
			\end{equation}
			which we will regulate with hard momenta cutoffs $\Lambda_\slab{ir} \leq |\mb{k}| \leq \Lambda_\slab{uv}$. If the Compton wavelength $m^{\sminus 1}$ of the scalar field is comparable to the size of the regulating box, we should instead take $\Lambda_\slab{ir} \sim m$ to avoid including the vacuum state. Since we will only be interested in the parametric dependence of this metric on the mass-squared $m^2(\varphi)$ and the different cutoffs, the exact regularization scheme is unimportant, as it will only affect overall constant factors. 

			The integral can be explicitly evaluated in terms of hypergeometric functions,
			\begin{align}
				g_{ab}(\varphi) = \frac{L^d (4 \pi)^{\sminus d/2}}{64\,  \Gamma\big(\frac{d+2}{2}\big)} \frac{\partial_a m^2 \,\partial_b m^2}{m^4} \Bigg[&\,\,\Lambda_\slab{uv}^d\left[\frac{d\es m^2}{m^2 + \Lambda_\slab{uv}^2} + \left(2- d\right) {}_2 F_{1}\Big(1, \tfrac{d}{2}; \tfrac{d+2}{2}; \minus \tfrac{\Lambda_\slab{uv}^2}{m^2}\Big)\right] \nonumber \\
				&\!\!-\Lambda_\slab{ir}^d\left[\frac{d\es m^2}{m^2 + \Lambda_\slab{ir}^2} + \left(2- d\right) {}_2 F_{1}\Big(1, \tfrac{d}{2}; \tfrac{d+2}{2}; \minus \tfrac{\Lambda_\slab{ir}^2}{m^2}\Big)\right]\,\Bigg]\,. \label{eq:scalarMetEval}
			\end{align}
			Because we exclude the vacuum contribution, $\Lambda_{\slab{ir}} \leq m$, the integral never diverges in the IR---we can safely set $\Lambda_{\slab{ir}} = 0$, and so the second term in the above vanishes. Said differently, this term is subdominant in the $L , \Lambda_\slab{uv} \to \infty$ limit. The only divergence arises in the UV and, in the limit $\Lambda_\slab{uv}/m \to \infty$, the metric takes the leading form 
			\begin{equation}
				g_{ab}(\varphi) \propto \frac{\partial_a m^2 \, \partial_b m^2}{m^4} \begin{dcases}
											(L m)^d & d < 4 \\
											(L m)^4 \log \tfrac{\Lambda_\slab{uv}}{m} & d = 4 \\
											(L \Lambda_\slab{uv})^d (m/\Lambda_\slab{uv})^4 & d > 4
										\end{dcases}
			\end{equation} 
			where we have dropped unimportant constant factors. In the limit $m \to 0$ with $\Lambda_\slab{ir} \sim m$, the metric takes the same form, so this scaling captures the correct asymptotic behavior for all $m$. 

			While we assumed that the parameter space spanned by $\varphi^a$ was $n$-dimensional, we see that the information metric only has one non-trivial eigenvalue. This makes sense, as the theory only depends on a single parameter $m^2$. Choosing our coordinate as the Wilson coefficient $\varphi = m^2/\Lambda_\slab{uv}^2$ and ignoring the other directions, the metric reduces to a single component,
			\begin{equation}
				g(\varphi) \propto \begin{dcases}
											(L \Lambda_\slab{uv})^d \varphi^{\frac{d-4}{2}} & d < 4 \\
											\tfrac{1}{2}(L \Lambda_\slab{uv})^4 \log \tfrac{1}{\sqrt{\varphi}} & d = 4 \\
											(L \Lambda_\slab{uv})^d  & d > 4
										\end{dcases} \label{eq:scalarMetDim}
			\end{equation}
			which matches the scaling from the general analysis (\ref{eq:metAwayFromCrit}) since $\xi \Lambda_\slab{uv} \propto \Lambda_\slab{uv}/m = \varphi^{\sminus 1/2}$. As expected, the point at $\varphi = 0$ is never at infinite distance in this metric. That is, unless $d = 0$, in which case the theory is just the harmonic oscillator.

			We should note that the metric never actually diverges as $m^2 \to 0$ at finite volume. The divergence naively comes from the vacuum contribution, but since the information metric purposefully excludes this, the divergence is instead regulated by the IR cutoff. In the basis with $\varphi = m^2/\Lambda_\slab{uv}^2$ as our coordinate, (\ref{eq:metSum}) reduces to
			\begin{equation}
					g(\varphi) = \frac{\Lambda_\slab{uv}^4}{32} \sum_{\mb{k} \neq 0} \frac{1}{\mb{k}^4} = \frac{(L \Lambda_\slab{uv})^4}{2 (4 \pi)^4} \sum_{\mb{n} \neq 0} \frac{1}{\mb{n}^4}\,.
			\end{equation}  
			Here, the sum is over all elements of the lattice $\mb{n} \in \mathbb{Z}^d$ besides the origin $\mb{n} \neq 0$. As we might expect from our continuum analysis (\ref{eq:scalarMetDim}), this sum is UV divergent for $d \geq 4$ and must be regulated. For $d < 4$, this sum over integers is just a constant\footnote{This sum can be written as $\int_0^{\infty}\!\ud s\, s\! \left[\theta_3(0, \lab{e}^{-s})^d - 1\right]$, which converges for $d < 4$. For $d = 1, 2, 3$ it evaluates to $2\es \zeta(4) \approx 2.1647$, $4 \es\zeta(2) \es \beta(2) \approx 6.0268$, and $16.5323$, respectively. Here, $\theta_3(z, q)$ is the elliptic theta function, $\zeta(s)$ is the Riemann zeta function, and $\beta(s)$ is the Dirichlet beta function \cite{NIST:DLMF}.} and so the metric scales super-extensively at the critical point.

		\subsection*{Interpreting the Metric}

			The scaling analysis of \S\ref{sec:metricScaling} excluded the presence of infinite distance singularities in field theories about  ``normal'' critical points, where the correlation length diverged and standard scaling arguments applied. The free scalar field in the massless limit provides an example of such a critical point, where we have found that $m^2 = 0$ is at  finite distance for $d \geq 1$. However, the free scalar does have an infinite distance point as $m^2 \to \infty$, independent of the spatial dimension~$d$. In this limit, the mass dominates the momentum integral in  (\ref{eq:scalarMet2}) and effectively turns the integrand into a constant, so that the metric behaves as
			\begin{equation}
				g_{ab}(\varphi) \propto (L \Lambda_\slab{uv})^d \, \frac{\partial_a m^2 \, \partial_b m^2}{m^4}\,,
			\end{equation}
			as $m^2(\varphi) \to \infty$. In Section~\ref{sec:towers}, we will find that a tower of fields relies on essentially the same mechanism to realize an infinite distance point as the tower degenerates in mass.

			Intuitively, it is clear why the point $m^2 \to \infty$ is at infinite distance. The massive scalar correlation function
			\begin{equation}
				\langle \Phi(x) \es \Phi(0) \rangle = \frac{m^{d-1}}{(2 \pi)^{\frac{d+1}{2}}} \frac{K_{(d-1)/2}(m x)}{(m x)^{(d-1)/2}} \sim \frac{m^{d-1}}{2 (2 \pi)^{d/2}} \frac{\lab{e}^{-m x}}{(m x)^{d/2}}\,, \mathrlap{\quad m \to \infty\,,}
			\end{equation}
			decays exponentially quickly over the correlation length $\xi = m^{\sminus 1}$, and as $m^2 \to \infty$ these correlations become ultra-local. Similar to how we could easily distinguish between finite and infinite frequency $\omega$ in the harmonic oscillator by asking a simple yes-or-no question (i.e. ``Do I detect the particle away from $x = 0$?''), finite and infinite $m^2$ can be distinguished by detecting whether separated points are correlated. This is a question that involves only a few local degrees of freedom.

			In contrast, the point at $m^2 = 0$ is not at infinite distance in the intensive metric. It is clarifying to consider the vacuum probability distribution associated with each individual Fourier mode, $p(\Phi_{\mb{k}}) \propto \exp(-\omega_\mb{k} |\Phi_{\mb{k}}|^2/L^d)$. The variance of this Gaussian is controlled by the frequency of each Fourier mode, $\omega_{\mb{k}} = \sqrt{\mb{k}^2 + m^2}$. For small mass, a shift $m^2 \to m^2 + \ud m^2$ only appreciably affects the variance of the low-momentum modes and so predictions for the high-momentum modes remain relatively unaffected. This is a reflection of the fact that, in order to measure the diverging correlation length of the field, we need to measure the fall-off of the correlator over very long distances. This reduces our ability to distinguish between the two theories at $m^2$ and $m^2 + \ud m^2$, especially when we consider that the intensive metric measures our ability to distinguish the two theories, per degree of freedom. The $m^2 \to \infty$ limit is not affected by this problem, since the mass almost entirely determines the variance for modes at all momenta.

	\subsection{The Free Dirac Fermion} \label{sec:dirac}

		Having studied the quantum information metric for the mass of a free scalar field, we now consider the metric for a free Dirac fermion in arbitrary dimension. Since we have already illustrated most of the interesting aspects of this metric using the scalar, we will not be nearly as thorough for the Dirac field, and instead make a beeline towards the answer.

		The Euclidean action for a $(d+1)$-dimensional Dirac fermion is\footnote{We follow the conventions of \cite{Srednicki:2007qs,Wess:1992cp,Martin:2012us,Dreiner:2008tw} which take the Lorentzian gamma matrices to obey $\{\gamma^\mu, \gamma^\nu\} = -2 \eta^{\mu \nu}$ and the Dirac Lagrangian to read $\mathcal{L} = \bar{\Psi}(i \gamma^\mu \partial_\mu - m) \Psi$. The Euclidean action then follows upon Wick rotation $t \to -i \tau$, with $\tau$ Euclidean time. Some authors, e.g. \cite{ZinnJustin:1989mi}, instead choose the Euclidean gamma matrices to satisfy $\{\tilde{\gamma}_\mu, \tilde{\gamma}_\nu\} = 2 \delta_{\mu \nu}$ with Euclidean Lagrangian $\mathcal{L}_\slab{e} = -\bar{\Psi}(\tilde{\gamma}_\mu \partial_\mu + m) \Psi$.}
		\begin{equation}
			S_\slab{e}[\Psi, m(\varphi)] = \int\!\ud^{d+1} x\, \bar{\Psi} \left[-i \tilde{\gamma}_\mu \partial_\mu + m(\varphi) \right] \Psi
		\end{equation}
		where $\tilde{\gamma}_\mu$ are the Euclidean gamma matrices, which satisfy $\{\tilde{\gamma}_\mu, \tilde{\gamma}_\nu\} = - 2\delta_{\mu \nu}$, and $\bar{\Psi} \equiv \minus \Psi^\dagger \tilde{\beta}$ is the Dirac adjoint, where numerically the matrix $\tilde{\beta} = \tilde{\gamma}_0$. These $(d+1)$-dimensional Dirac spinors $\Psi_\alpha$ have $d_\lab{f} \equiv 2^{\lfloor \frac{d+1}{2} \rfloor}$ components, which we will label using Greek indices $\alpha, \beta, \gamma, \sigma = 1, \dots, d_\lab{f}$. We take the mass of the field to depends on the parameters $\varphi^a$ and we will derive the associated information metric $g_{ab}(\varphi)$.

		It will be easiest to derive the information metric via (\ref{eq:qimTwoPoint}) using the Dirac field's correlation functions,
		\begin{equation}
			g_{ab}(\varphi) = \partial_a m\, \partial_b m \,\es \mathcal{I}_\lab{f}(\varphi, L, \epsilon)\,,
		\end{equation}
		where we have introduced the fermionic analog of (\ref{eq:scalarIntegral}),
		\begin{equation}
			\mathcal{I}_\lab{f}(\varphi)= \int_{\epsilon}^{\infty}\!\!\ud \tau_1 \int_{\sminus \infty}^{\sminus \epsilon}\!\!\ud \tau_2 \int_{L^d} \!\ud^d x_1 \,\ud^d x_2 \left[\langle \bar{\Psi} \Psi(\tau_1, \mb{x}_1) \bar{\Psi}\Psi(\tau_2, \mb{x}_2)\rangle -\langle \bar{\Psi} \Psi(\tau_1, \mb{x}_1)\rangle \langle \bar{\Psi}\Psi(\tau_2, \mb{x}_2)\rangle\right].
		\end{equation}
		We again employ point-splitting to regularize this expression. The one-point functions cancel the divergences generated as each of the operator pairs coincide, so we are left with
		\begin{equation}
			\mathcal{I}_\lab{f}(\varphi) = \int_{\epsilon}^{\infty}\!\!\ud \tau_1 \int_{\sminus \infty}^{\sminus \epsilon} \!\!\ud \tau_2 \int_{L^d}\!\ud^d x_1\, \ud^d x_2\,  \langle \bar{\Psi}_{\alpha}(\tau_1, \mb{x}_1)  \Psi_\beta(\tau_2, \mb{x}_2) \rangle \langle \Psi_{\alpha}(\tau_1, \mb{x}_1) \bar{\Psi}_{\beta}(\tau_2, \mb{x}_2)\rangle\,,
		\end{equation}
		where repeated spinor indices are implicitly summed over.

		Using the Euclidean correlator
		\begin{align}
			\langle 0 | \Psi_\alpha(\tau_1, \mb{x}_1) \bar{\Psi}_\beta(\tau_2, \mb{x}_2) | 0 \rangle  &= \int\!\!\es \frac{\ud^{d+1} k}{(2 \pi)^{d+1}}  \frac{(\tilde{\gamma}_\mu k_\mu - m)_{\alpha \beta}}{k^2 + m^2}\lab{e}^{i k (x_1 - x_2)}\,,
		\end{align}
		we can rewrite the integrand of $\mathcal{I}_\lab{f}(\varphi)$ as
		\begin{equation}
			\langle \bar{\Psi}_{\alpha}(\tau_1, \mb{x}_1) \Psi_\beta(\tau_2, \mb{x}_2) \rangle \langle \Psi_{\alpha}(\tau_1, \mb{x}_1) \bar{\Psi}_{\beta}(\tau_2, \mb{x}_2)\rangle =  \int\!\frac{\ud^{d+1} k_1}{(2 \pi)^{d+1}}\frac{\ud^{d+1} k_2}{(2 \pi)^{d+1}} \frac{d_\lab{f} (k_1 \cdot k_2-m^2)}{(k_1^2 + m^2)(k_2^2 + m^2)}\,.
		\end{equation}
		After a little $\tilde{\gamma}$-matrix algebra, the metric may written as single integral of spatial momenta,
		\begin{equation}
			g_{ab}(\varphi) = \frac{d_\lab{f} L^d}{8} \frac{\partial_a m \, \partial_b m}{m^2} \int\!\!\frac{\ud^d k}{(2 \pi)^d} \frac{m^2 \, \mb{k}^2}{(\mb{k}^2 + m^2)^2}\,, \label{eq:diracMet}
		\end{equation}
		a form analogous to (\ref{eq:scalarQIM}) for the scalar field.

		Imposing hard cutoffs $ \Lambda_\slab{ir} \leq |\mb{k}| \leq \Lambda_\slab{uv}$, the leading term in the large $L$ limit is,
		\begin{equation}
			g_{ab} = \frac{d_\lab{f} (L \Lambda_\slab{uv})^d}{8 (4 \pi)^{d/2} \Gamma\big(\frac{d}{2}\big)} \frac{\partial_a m \, \partial_b m}{m^2}\left[ \frac{\Lambda_{\slab{uv}}^2}{\Lambda_{\slab{uv}}^2 + m^2} - \frac{d}{d + 2} \frac{\Lambda_{\slab{uv}}^2}{m^2}\, {}_2 F_1 \!\left(1, \tfrac{d+2}{2}; \tfrac{d+4}{2}; \minus \tfrac{\Lambda_{\slab{uv}}^2}{m^2}\right)\right]\,,
		\end{equation}
		which further reduces to
		\begin{equation}
				g_{ab}(\varphi) \propto \frac{\partial_a m \, \partial_b m}{m^2} \begin{dcases}
											L m & d = 1 \\
											(L m)^2 \log \tfrac{\Lambda_\slab{uv}}{m} & d = 2 \\
											(L \Lambda_\slab{uv})^d (m/\Lambda_\slab{uv})^2 & d > 3
										\end{dcases}
		\end{equation}
		in the limit $\Lambda_\slab{uv} /m \to \infty$. This displays similar behavior to the scalar metric, though it is super-extensive for a smaller range of dimensions $d$. Importantly, it also does not exhibit an infinite distance singularity as $m \to 0$. 

		This concludes our study of free fields and the information metric associated to their mass parameters. We found that, for single fields with non-zero spatial dimension, infinite distance singularities do not appear in the information metric. This agrees with the scaling analysis presented in \S\ref{sec:metricScaling}. In Section~\ref{sec:towers}, we will find that infinite distance singularities can emerge when towers of fields compress in mass. Before that, though, we will study how the information metric is related to the other metrics that are naturally defined in both conformal field theories and nonlinear sigma models.

	\subsection{The Conformal Manifold} \label{sec:cft}

		Conformal field theories related to one another by exactly marginal deformations, generated as in (\ref{eq:defLag}) by the marginal local operators $\mathcal{O}_a(x)$ with scaling dimensions $\Delta_a = d+1$,  form a conformal manifold or a moduli space of conformal field theories. Each point of this space represents a different conformal field theory. The geometry of this moduli space can be characterized via the two-point functions
		\begin{equation}
			\langle \mathcal{O}_a(x) \mathcal{O}_b(y) \rangle = \frac{G_{ab}(\varphi)}{|x - y|^{2 d+ 2}}\,,
		\end{equation}
		where the normalization is given by the \emph{Zamolodchikov metric} on moduli space. How is the metric related to the quantum information metric associated with these operators?

		As with the scalar and Dirac fields above, we can compute the information metric using (\ref{eq:qimTwoPoint}),
		\begin{equation}
			g_{ab}(\varphi) =  \int_{\epsilon}^{\infty}\!\!\ud \tau_1 \int^{\sminus \epsilon}_{\sminus \infty}\!\ud \tau_2 \int_{L^d} \!\ud^d x_1 \, \ud^d x_2\,\frac{G_{ab}(\varphi)}{\left[(\tau_1 - \tau_2)^2 + (\mb{x}_1 - \mb{x}_2)^2\right]^{d+1}} \sim  \mathcal{C}_d \left(L \Lambda_{\textsc{uv}}\right)^d G_{ab}(\varphi)\,,
		\end{equation}
		where $\mathcal{C}_d = {(\pi/4)^{\frac{d+1}{2}}}/{ \left(d \, \Gamma\big(\frac{d+3}{2}\big)\right)}$ is an unimportant numerical factor. The intensive quantum information metric (\ref{eq:intMet}) for marginal deformations of a conformal field theory is equal (up to an overall constant) to the Zamolodchikov metric. This connection is perhaps not so surprising, as the Zamolodchikov metric can be derived from a K\"{a}hler potential in a way very similar to the information metric (\ref{eq:kahlerPotential}). This has been done explicitly in low-dimensional and supersymmetric theories, where the vacuum overlap it is related to a sphere partition function~\cite{Cecotti:1991me,Ranganathan:1993vj,Hori:2003ic,deBoer:2008ss,Papadodimas:2009eu,Gerchkovitz:2014gta}. The Zamolodchikov metric also has an expansion similar to (\ref{eq:qimPert}), in which energies and transition elements are replaced with scaling dimensions and OPE coefficients, respectively  \cite{Ranganathan:1993vj,deBoer:2008ss,Papadodimas:2009eu,Baggio:2017aww}.

		Infinite distances points in the Zamolodchikov metric thus derive an information theoretic interpretation\footnote{This is not the first time the Zamolodchikov metric has been used as a sort of information metric, as \cite{Moore:2015bba} used Zamolodchikov volumes to define a probability measure of sequences of conformal field theories, much in the same way the Fisher information metric is used to define the so-called Jeffrey's prior.} from the quantum information metric. They describe when the theory is undergoing a drastic, qualitatively distinct change. This matches the holographic picture of these points, wherein an infinite tower of higher spin fields become light \cite{Seiberg:1999xz,Baume:2020dqd,Perlmutter:2020buo}.

	\subsection{The Nonlinear Sigma Model} \label{sec:nlsm}

		Let us now turn to an arbitrary, $(d+1)$-dimensional bosonic nonlinear sigma model specified by the Euclidean action 
		\begin{equation}
			S_\slab{e}[\phi] = \frac{\Lambda_\slab{uv}^{d-1}}{2}\!  \int\!\ud^{d+1} x\, G_{ab}(\phi)\, \partial_\mu \phi^a \partial^\mu \phi^b\,,
		\end{equation}
		and defined with respect to the UV cutoff $\Lambda_\slab{uv}$. Here, we have scaled the fields $\phi^a$ by the cutoff so that they are dimensionless. The class of theories we are interested in can be defined by the partition function
		\begin{equation}
			\mathcal{Z}[\varphi^a] = \int\!\mathcal{D} \phi \,\sqrt{\det G_{ab}(\phi^a)}\, \lab{e}^{-S_\slab{e}[\phi^a]}
		\end{equation}
		together with the constraint on the vacuum expectation values $\langle \phi^a \rangle = \varphi^a$, where $\mathcal{D}\phi = \prod_{a=1}^n \mathcal{D}\phi^a$. The integration measure is chosen so that the path integral preserves field space covariance, where the $\phi^a$ transforms as field space coordinates and $\partial_\mu \phi^a$ transforms as both a field space vector and a spacetime 1-form. Our goal is to compute the quantum information metric associated to the parameters $\varphi^a$, which measures how distinguishable the theory at $\varphi^a$ is from its neighbors. One might expect that the intensive quantum information metric $\tilde{g}_{ab}(\varphi)$ and the metric on field space $G_{ab}(\varphi)$ are proportional to one another, and we will confirm that this is indeed the case, at least at tree level.\footnote{See \cite{Trivella:2016brw} for a holographic derivation.} Furthermore, we will argue indirectly that this relationship survives the inclusion of quantum corrections---the quantum-corrected metric on field space is proportional to the quantum information metric, and vice versa.

		We will compute the information metric by quantizing this theory about the time-dependent background
		\begin{equation}
			\varphi^a(\tau) = \begin{dcases} 
				\varphi^a_\subp & \tau \to \infty \\
				\varphi^a_\subm & \tau \to \minus \infty
				\end{dcases}\,,
		\end{equation}
		where we take $\varphi^a(\tau)$ to vary between its two limits on a time-scale of order $\epsilon \sim \Lambda_\slab{uv}^{\sminus 1}$ at $\tau = 0$. The exact functional form of this background will only affect the constant numerical factors appearing in front of the metric. Since we are only interested in its parametric scaling, we will take 
		\begin{equation}
			\varphi^a(\tau) = \varphi^a_\subm + \tfrac{1}{2}(\varphi^a_\subp - \varphi^a_\subm)\big(1 + \tanh \Lambda_\slab{uv} \tau\big)\,. \label{eq:nlsmProfile}
		\end{equation}
		Similar to how we extracted the harmonic oscillator's information metric from the path integral in \S\ref{sec:sho}, our goal is to compute the effective action about this background,
		\begin{equation}
			\lab{e}^{-\Gamma[\varphi(\tau)]} = \int_{\varphi(\tau)}\!\!\!\mathcal{D} \phi \, \sqrt{\det G_{ab}(\phi)}\, \lab{e}^{-S_\slab{e}[\phi]}\,, \label{eq:nlsmPathInt}
		\end{equation}
		in the limit $\Lambda_\slab{uv} \to \infty$ to derive the divergence $\mathcal{D}(\varphi_\subp, \varphi_\subm) = \Gamma[\varphi(\tau)]$, from which we can extract the metric.

		It is easiest to covariantly quantize\footnote{Other quantization methods exist and give the same results as this covariant quantization method, as long as one restricts to on-shell quantities. However, it is not clear how ``on-shell'' the information metric is, since it characterizes all possible correlation functions we could ask of the theory. We thus take this covariant quantization as a necessary ingredient in specifying the theory (and making predictions), just like the regularization and renormalization schemes.} this theory using the covariant background field method~\cite{Ketov:2000dy,AlvarezGaume:1981hn,Mukhi:1985vy,Howe:1986vm,Hull:1986hn,Goon:2020myi}. The goal is to covariantly expand the quantum field $\phi^a(x)$ into fluctuations about the background field value $\varphi^a(x)$. To do this, we define $\phi^a(x) = \Phi^a(x; s=1)$ as the endpoint of a geodesic~$\Phi^a(x; s)$ in the field space manifold emanating from the point $\varphi^a(x) = \Phi^a(x; s=0)$ with tangent vector $\xi^a(x) = \tfrac{\ud}{\ud s} \Phi^a(x; s=0)$, such that
		\begin{equation}
			\frac{\ud^2}{\ud s^2} \Phi^a(x; s) + \Gamma^{a}_{b c}[\Phi^d(x; s)] \frac{\ud}{\ud s}\Phi^b(x; s) \, \frac{\ud}{\ud s} \Phi^c(x; s) = 0\,,
		\end{equation} 
		where the $\Gamma^a_{bc}$ are Christoffel symbols associated to the field space metric $G_{ab}$. In practice, this consists of the nonlinear field redefinition
		\begin{equation}
			\phi^a = \varphi^a + \xi^a - \frac{1}{2} \Gamma^{a}_{bc}(\varphi^d) \xi^b \xi^c + \cdots\,.
		\end{equation}
		Introducing a pair of vector-valued Grassmann-valued auxiliary fields $(\theta^a, \bar{\theta}^a)$ and vector-valued bosonic auxiliary field $\alpha^a$ (so-called ``Lee-Yang ghosts'' \cite{Bastianelli:1992ct}), we can rewrite the path integral measure so that 
		\begin{align}
			\lab{e}^{-\Gamma[\varphi(\tau)]} = \int\!\mathcal{D} \xi \, \mathcal{D}\bar{\theta}\,\mathcal{D}\theta\,\mathcal{D}\alpha\, &\exp\!\left(-\frac{\Lambda_\slab{uv}^{d-1}}{2}\!\int\!\ud^{d+1} x\, G_{ab}(\phi) \partial_\mu \phi^a \partial_\mu \phi^b\right) \nonumber \\
			\times&\exp\left(-\frac{\Lambda_\slab{uv}^{d-1}}{2} \!\int\!\ud^{d+1} x\, G_{ab}(\phi)\!\left[ \frac{\partial \phi^a}{\partial \xi^c} \frac{\partial \phi^b}{\partial \xi^d} \left(\bar{\theta}^c \theta^d + \alpha^c \alpha^d\right)\right]\right) \label{eq:nlsmPath}
		\end{align}
		The action, to quadratic order in fluctuations $\xi^a$, can be written as \cite{Ketov:2000dy}
 			\begin{align}
 				S_\slab{e}[\varphi, \xi] &= \frac{\Lambda_\slab{uv}^{d-1}}{2}\int\!\ud^{d+1} x\left[ G_{ab}(\varphi) \partial_\mu \varphi^a \partial^\mu \varphi^b -2 \es G_{ab}(\varphi) \xi^a D_\mu \partial_\mu \phi^b \right] \\
 				&+ \frac{\Lambda_\slab{uv}^{d-1}}{2}\int\!\ud^{d+1} x\, \left[G_{ab}(\varphi)  D_\mu \xi^a D_\mu \xi^b + R_{abcd}(\varphi) \xi^b \xi^c \partial_\mu \varphi^a \partial^\mu \varphi^d + G_{ab}(\varphi) \left(\bar{\theta}^a \theta^b + \alpha^a \alpha^b\right)\right] \nonumber 
 			\end{align}
 			where we have introduced the covariant derivative $D_\mu \xi^a = \partial_\mu \xi^a + \Gamma^a_{bc} \partial_\mu \varphi^b \xi^c$. This is what we will take as our definition of the family of theories. Since this construction is manifestly covariant, higher-order corrections involve contractions of field space curvature tensors and their derivatives, the fluctuation $\xi^a$, spacetime derivatives of both the background field $\partial_\mu \varphi^a$ and fluctuation $\mathcal{D}_\mu \xi^a$, and the ghost fields $\theta^a$, $\bar{\theta}^a$, and $\alpha^a$ \cite{Ketov:2000dy}.

 			As expected, at tree-level effective action is simply the classical action of the background field
 			\begin{equation}
 				\Gamma[\varphi] = \frac{\Lambda_\slab{uv}^{d-1}}{2} \int\!\ud^{d+1}\, g_{ab}(\varphi) \, \partial_\mu \varphi^a \partial_\mu \varphi^b = \frac{\Lambda_\slab{uv}^{d-1} L^d}{2} \int\!\ud \tau\, g_{ab}(\varphi(\tau)) \dot{\varphi}^a \dot{\varphi}^b
 			\end{equation}
 			If we take $\varphi^a(\tau)$ to be a geodesic, then we can recognize that the divergence $\mathcal{D}(\varphi_\subp, \varphi_\subm) = \Gamma[\varphi(\tau)]$ is proportional to Synge's worldfunction \cite{Poisson:2011nh}, i.e. half the geodesic distance between the points $\varphi_\subp^a$ and $\varphi_\subm^a$. Infinitesimally, it should not matter whether or not $\varphi^a(\tau)$ is a geodesic, and so the information metric derived from the divergence should be proportional to the metric on field space. Indeed, using (\ref{eq:nlsmProfile}) confirms that
 			\begin{equation}
 				g_{ab}(\varphi) = - \!\left.\frac{\partial^2 \mathcal{D}(\varphi_\subp, \varphi_\subm)}{\partial \varphi^a_\subp \partial \varphi^b_\subm}\right|_{\varphi} \propto (L\Lambda_\slab{uv})^d G_{ab}(\varphi)\,, 
 			\end{equation}
 			where we have again used $|_{\varphi}$ to denote that the derivative is evaluated at $\varphi^a_\subp = \varphi^a_\subm = \varphi^a$. Intuitively, the divergence quantifies the separation between the distributions at $\varphi_\subp$ and $\varphi_\subm$ and it is not surprising that it is related to the geodesic distance between the two points, especially when the two points are infinitesimally separated. 

 			\newpage
 			Since the path integral (\ref{eq:nlsmPath}) is manifestly covariant, including quantum fluctuations and interactions modifies the tree-level result to\footnote{We should note that this is not, strictly speaking, the ``quantum effective action'' normally considered, i.e. it is not the generator of single-particle-irreducible (1PI) diagrams. This takes a form very similar to the quantum effective action evaluated using the background field method, but it is missing the term $\int\!\varphi^a \delta \Gamma/\delta \varphi^a$ in the exponent that ensures diagrams remain 1PI \cite{Abbott:1981ke}. It is true that this term vanishes when evaluated on a solution to the corrected equations of motion, i.e. a profile $\varphi^a(\tau)$ that satisfies $\delta \Gamma/\delta \varphi^a = 0$. Since we are interested in infinitesimally separated points $\varphi_\subp$ and $\varphi_\subm$, we expect that $\varphi(\tau)$ is ``good enough'' and that the relevant contributions come only from 1PI diagrams. We will leave this for future work.}
 			\begin{equation}
 				\Gamma[\varphi] = \frac{\Lambda_\slab{uv}^{d-1}}{2} \int\!\ud^{d+1} x \! \left[\tilde{G}_{ab}(\varphi) \es \partial_\mu \varphi^a \es\partial_\mu \varphi^b  + \frac{1}{\Lambda_\slab{uv}^2} \tilde{M}_{abcd}(\varphi) (\partial_\mu \varphi^a \es \partial_\mu \varphi^b) (\partial_\nu \varphi^c \es \partial_\nu \varphi^d) + \dots\right]
 			\end{equation}
 			where $\tilde{G}_{ab}(\varphi)$ and $M_{abcd}(\varphi)$ are covariant tensors constructed out of $g_{ab}(\varphi)$ and its curvature tensors, and the $\dots$ represent similar terms with more powers of $\partial_\mu \varphi^a \partial_\mu \varphi^b$ (and other contractions of spacetime derivatives). Fortunately, any term with more than two powers of $\partial_\mu \varphi^a$ will not contribute to the metric, and we similarly find that the intensive part of the metric is
 			\begin{equation}
 				\tilde{g}_{ab}(\varphi) \propto  \tilde{G}_{ab}(\varphi)\,.
 			\end{equation}
 			That is, the effect of fluctuations and interactions is to renormalize both the metric on field space and thus the information metric. Because of the general structure used here, we expect this conclusion to also hold in supersymmetric nonlinear sigma models---as well as those that include gravitational interactions---and it would be interesting to confirm this in future work.

 			As in the previous section, we find that the information metric is related in a simple way to the natural metric of the theory. There, we found that the information metric was proportional to the Zamolodchikov metric. Here, we find that it is proportional to the metric on field space. Again, this is perhaps not so surprising from the holographic picture, as the metric on the moduli space of the bulk theory is proportional to the Zamolodchikov metric $(L_\lab{AdS} M_\lab{pl})^3 G^\lab{Bulk}_{ab} \sim G^\lab{Zam}_{ab}$ of the boundary theory~\cite{Baume:2020dqd}, where $L_\lab{AdS}$ is the AdS radius and $M_\lab{pl}$ the Planck mass.\footnote{It is curious to note that the prefactor $(L_\lab{AdS} M_\lab{pl})^3$ is of the same form as the number of degrees of freedom $\mathcal{N} = (L \Lambda_\slab{uv})^d$ we so often run into in this work. We will encounter similar factors in the next section when we use a tower of fields to ``restore'' a dimension via a decompactification limit.}  

 			From our (bottom-up) perspective, both of these metrics are predetermined and infinite distance points are necessarily an input. We argued in \S\ref{sec:metricScaling} that such points necessitated departures from the naive scaling typically found around quantum critical points, i.e. when a gap closes for a finite number of fields. We found supporting evidence for this behavior in \S\ref{sec:scalar} and \S\ref{sec:dirac} where we studied the information metric associated with the mass of a scalar and fermionic field, respectively. In the next section, we will show how infinite distance singularities can emerge from infinite towers of fields.

\newpage

\section{Towers of Fields and the Emergence of Infinite Distance Singularities} \label{sec:towers}

	In the previous section, we found that neither the free scalar nor the free Dirac fermion exhibited an infinite distance point as the correlation length diverged. Interestingly, the simple harmonic oscillator did. Can infinite distance points occur in theories with non-zero spatial dimension? The analysis of \S\ref{sec:metricScaling} suggested that they could not unless there was some violation of the typical scaling features found around non-trivial critical points. In this section, we show that infinite distance points in the information metric can emerge from the collective behavior of a tower of these fields.

	We first consider a theory with a tower of $(d+1)$-dimensional scalar fields $\Phi_{\mb{q}}(\tau, \mb{x})$, each of which is labeled by an element $\mb{q}$ of a $\dc$-dimensional charge lattice $\Gamma$.\footnote{The $\dc$-dimensional lattice $\Gamma = \{\sum_i n_i \mb{b}_i \, |\, n_i \in \mathbb{Z}\}$ is spanned by the basis vectors $\mb{b}_i \in \mathbb{R}^{\dc}$, where $i = 1, \dots, \dc$.  Vectors $\bar{\mb{q}}$ in the dual lattice $\bar{\Gamma}$ satisfy $\mb{q} \cdot \bar{\mb{q}} \in \mathbb{Z},\,\forall \mb{q} \in \Gamma$. The dual lattice $\bar{\Gamma}$ is similarly spanned by the dual basis vectors $\bar{\mb{d}}_i$, which satisfy $\mb{b}_i \cdot \bar{\mb{d}}_j = \delta_{ij}$.  }  We will assume that the mass of each field is of the form
	\begin{equation}
		m_\mb{q}^2(\varphi) = m_0^2(\varphi) + \mu^2(\varphi) f(\mb{q}) \equiv m_0^2(\varphi) + \mu_{\mb{q}}^2(\varphi)\, \label{eq:massEq}
	\end{equation}
	where $m_0^2(\varphi)$ and $\mu^2(\varphi)$ are functions of the parameters of our theory, $\varphi^a$, and $f(\mb{q})$ is a positive, smoothly varying function on the charge lattice that describes how the mass-squared of the fields varies from site to site. The Euclidean action of this theory is then
	\begin{equation}
		S_\slab{e}(\varphi) = \sum_{\mb{q} \in \Gamma} \int\!\ud^{d+1} x \!\left[\frac{1}{2} \left(\partial_\mu \Phi_{\mb{q}}\right)^2 + \frac{1}{2} m_\mb{q}^2(\varphi) \,\Phi_{\mb{q}}^2\right]\,.
	\end{equation}
	and our goal is to compute the information metric $g_{ab}(\varphi)$ with respect to the parameters $\varphi^a$, specifically in the limit that the mass-spacing between the fields vanishes, $\mu^2(\varphi) \to 0$.

	Since the fields do not interact with one another, the information metric is simply the sum of the
	 metrics (\ref{eq:scalarQIM}) associated to each field, i.e.
	\begin{equation}
		g_{ab}(\varphi) = \frac{L^d}{32} \sum_{\mb{q} \in \Gamma}\es  \int\!\!\frac{\ud^d k}{(2\pi)^d} \frac{\partial_a m_{\mb{q}}^2 \, \partial_b m_{\mb{q}}^2}{(\mb{k}^2 + m^2_{\mb{q}})^2}\,. \label{eq:infMetTowerInt}
	\end{equation}
	The specific form of the mass (\ref{eq:massEq}) allows us to rewrite the metric as a sum of terms
	\begin{equation}
		g_{ab}(\varphi) = \frac{L^d}{32}\left[\frac{\partial_a m_0^2 \, \partial_b m_0^2}{m_0^4} \, \mathcal{S}_{0}(\varphi) + \frac{2 \partial_{(a} m_0^2 \, \partial_{b)} \mu^2}{m_0^2 \, \mu^2} \, \mathcal{S}_{1}(\varphi) + \frac{\partial_a \mu^2\, \partial_b \mu^2}{\mu^4} \mathcal{S}_{2}(\varphi)\right] \label{eq:genTowerMet}
	\end{equation}
	where we have defined the functions
	\begin{equation}
		\mathcal{S}_{\nu}(\varphi) = \sum_{\mb{q} \in \Gamma} \int\!\!\frac{\ud^d k}{(2 \pi)^d} \frac{m_0^{4 - 2 \nu} \mu_{\mb{q}}^{2 \nu}}{(\mb{k}^2 + m_{\mb{q}}^2)^2}\,. \label{eq:metricS}
	\end{equation}
	Our goal now is to understand their behavior as $\mu^2(\varphi) \to 0$.

	\begin{figure}
		\centering
 			\includegraphics[scale=1]{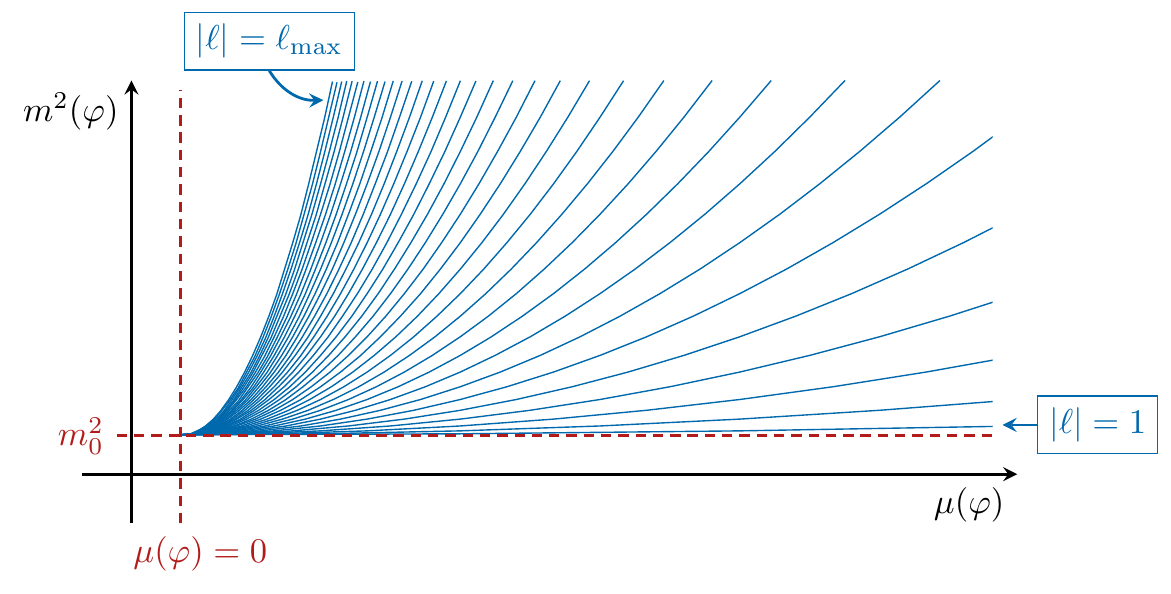}
		\caption{ A schematic representation of the tower of fields with $m^2_{\ell}(\varphi) = m_0^2 + \mu^2(\varphi) \ell^2$, as a function of the mass-spacing parameter $\mu$. As this spacing shrinks, $\mu(\varphi) \to 0$, the entire tower degenerates towards the fixed fiducial mass~$m_0$.\label{fig:tower}}
	\end{figure}

	Before we move onto the more general case, let us first get our bearings by restricting to a one-dimensional lattice where $m_{\ell}^2 = m_0^2 + \mu^2(\varphi) \ell^2$ with $\ell \in \mathbb{Z}$ and constant $m_0^2$, illustrated in Figure~\ref{fig:tower}. This is, for example, the case for the Kaluza-Klein reduction of a scalar field on $\mathbb{R}^{1, d}\times \lab{S}^1$, where $\mu^2 \propto 1/R^2$ and $R$ the circumference of the circle $\lab{S}^1$, where $\mu^2 \to 0$ corresponds to the decompactification limit.
	 The only non-vanishing term in (\ref{eq:genTowerMet}) is the third, and so we are faced with evaluating
	\begin{equation}
		\mathcal{S}_2(\varphi) = \sum_{\ell \in \mathbb{Z}} \int\!\!\frac{\ud^d k}{(2 \pi)^d} \frac{\mu^4 \ell^4}{(\mb{k}^2 + m_0^2 + \mu^2 \ell^2)^2}\,.
	\end{equation}
	This has two types of divergences. The first is by now familiar: the integral over momenta must be regulated by a UV cutoff $\Lambda_\slab{uv}$, which is a symptom of the fact that we are dealing with an infinite number of degrees of freedom in the quantum field. The second is similar: we must regulate the sum because we are now dealing with an infinite number of quantum fields. We will discuss two different ways of regulating this expression.

	In the limit $\mu^2(\varphi) \to 0$, we can convert the sum over $\ell$ into an integral over the suggestively named $k_{d+1} \equiv \mu \ell$,
	\begin{equation}
		\mathcal{S}_2 = \frac{1}{\mu} \int_{\sminus \infty}^{\infty}\!\ud k_{d+1} \int\!\!\frac{\ud^d k}{(2 \pi)^d} \frac{k_{d+1}^4}{(\mb{k}^2 + m_0^2 + k_{d+1}^2)^2}\,. \label{eq:interimS}
	\end{equation}
	As we might expect, we can reinterpret the sum over the lattice as integration over an additional spatial dimension. However, note that we do not find the same form of the information metric as we considered in (\ref{eq:scalarQIM}), albeit with $d \to d+1$, because here we are considering a different type of perturbation of the theory.

	\newpage
	Transforming into $(d+1)$-dimensional hyperspherical coordinates with $k_{d+1} = k \cos \theta$ and imposing a hard UV cutoff $\Lambda_\slab{uv}$ on the magnitude of this higher-dimensional momentum, (\ref{eq:interimS}) becomes
	\begin{equation}
		\mathcal{S}_2 = \frac{2(4 \pi)^{\sminus d/2}}{ \mu\,  \Gamma\big(\frac{d}{2}\big)} \int_{0}^{\pi} \!\ud \theta\, \sin^{d-1} \theta \cos^4 \theta \int_0^{\Lambda_\slab{uv}} \!\!\! \frac{\ud k\, k^{d+4}}{(k^2 + m_0^2)^2} \sim \frac{3 (4 \pi)^{\frac{1-d}{2}}}{4 (d+1) \Gamma\big(\frac{d+5}{2}\big)} \left(\frac{\Lambda_\slab{uv}}{\mu}\right)\Lambda_\slab{uv}^{d}\,.
	\end{equation}
	We need to be careful about how we have regularized this expression. In Sections~\ref{sec:qim} and~\ref{sec:examples}, we introduced IR and UV cutoffs so that we could study the intensive part of the metric (cf.~\S\ref{sec:metricReg}), which effectively quantified the distinguishability per degree of freedom. We also expect the metric to be extensive in the number of fields and so we must also regulate that divergence. That is, the factor $\Lambda_\slab{uv}/\mu \sim R\Lambda_\slab{uv}$ has the exact same origin as the factors of $L \Lambda_\slab{uv}$, and only exists because we are considering an infinite number of degrees of freedom. In this regularization scheme, we have roughly imposed that the $|k_{d+1}| \lesssim \Lambda_\slab{uv}$, or equivalently that the sum only runs over $|\ell| \lesssim \ell_\lab{max}$, with $\ell_\lab{max} \propto \Lambda_\slab{uv}/\mu$.  If we keep this number fixed as we vary the $\varphi$, we find that~(\ref{eq:genTowerMet}) becomes
	\begin{equation}
		g_{ab}(\varphi) \propto \ell_\lab{max} (L \Lambda_\slab{uv})^d \, \frac{\partial_a \mu^2 \, \partial_b \mu^2}{\mu^4}\,.
	\end{equation}
	The metric is again extensive in the number of degrees of freedom and, and the intensive metric
	\begin{equation}
		\tilde{g}_{ab}(\varphi) \propto \frac{\partial_a \mu^2\, \partial_b \mu^2}{\mu^4}\,, \label{eq:towerIntensive}
	\end{equation}
	exhibits an infinite distance point as $\mu^2 \to 0$. The theory evidently undergoes a drastic, ultra-distinguishable change in this decompactification limit. Note that if we do not keep the number of fields fixed, the information metric enjoys an even more severe infinite distance singularity. 

	We are now in a position to understand \emph{why} this tower generates an infinite distance point. As in \S\ref{sec:scalar}, we can consider the vacuum probability distribution associated to each Fourier mode, whose variance is now controlled by the frequency $\omega_{\ell, \mb{k}} = \sqrt{\mb{k}^2 + m_0^2 + \mu^2 \ell^2}$. Under a shift of the mass-spacing $\mu^2 \to \mu^2 + \ud \mu^2$, the change in the variance is essentially \emph{enhanced} by a factor of $\ell^2$: the masses increase quadratically with $\ell$, and so a small change of the mass-spacing can translate into a very large change of the total mass for the high-$\ell$ modes. As the mass-spacing vanishes, $\mu^2 \to 0$, these high-$\ell$ modes dominate. The intensive metric, which measures the average distinguishability per degree of freedom, is dominated by these modes that are very sensitive to changes to $\mu^2$ and thus diverges as $\mu^2 \to 0$.

	Let us now generalize to an arbitrary lattice $\Gamma$ and assume that $f(\mb{q}) = f(|\mb{q}|)$ in (\ref{eq:massEq}) is only a function of the standard Euclidean $2$-norm. We will also assume $f(|\mb{q}|)$ is a positive function whose leading behavior $|\mb{q}|^{\alpha}$ as $|\mb{q}| \to \infty$, with $\alpha > 0$. To evaluate (\ref{eq:metricS}), we can introduce a Schwinger parameter $s$ and rewrite $\mathcal{S}_\nu(\varphi)$ as
	\begin{equation}
		\mathcal{S}_{\nu}(\varphi) = m_0^{4 - 2 \nu} \int\!\!\frac{\ud^d k}{(2 \pi)^d} \int_{0}^{\infty}\!\ud s\, s \, \lab{e}^{-s(\mb{k}^2 + m_0^2)}\,  (-\partial_s)^{\nu} \!\es\Bigg[\es \sum_{\mb{q} \in \Gamma} \lab{e}^{-s \mu^2 f(|\mb{q}|)}\es\Bigg]\,. \label{eq:sInterim}
	\end{equation}
	In the limit $\mu^2(\varphi) \to 0$, the sum over $\Gamma$ becomes very poorly convergent. However, we can use Poisson resummation to recast it into a much more convergent sum over the dual lattice $\bar{\Gamma}$,
	\begin{equation}
		\sum_{\mb{q} \in \Gamma}  \lab{e}^{-s \mu^2 f(\mb{q})}  = \frac{1}{|\Gamma|} \sum_{\bar{\mb{q}} \in \bar{\Gamma}} \int\!\ud^{\dc} \mb{q}  \, \lab{e}^{-s \mu^2 f(\mb{q}) + 2 \pi i \mb{q} \cdot \bar{\mb{q}}}\,.
	\end{equation}
	As $\mu^2(\varphi) \to 0$, we can truncate this sum to the origin of the dual lattice, $\bar{\mb{q}} = 0$. This is not a bad approximation as---for instance, if $f(\mb{q}) = \mb{q}^2$, then this consists of dropping terms of order $\exp\left(\minus 2\pi |\bar{\mb{q}}| \Lambda_\slab{uv}/\mu\right)$. This is equivalent to approximating the sum over the lattice as an integral, as we did in (\ref{eq:interimS}).

	We can also approximate the continuum contribution in the limit $\mu^2(\varphi) \to 0$ as
	\begin{equation}
		\frac{1}{|\Gamma|} \int\!\ud^\dc \mb{q}  \, \lab{e}^{-s \mu^2 |\mb{q}|^\alpha} = \frac{2 \pi^{\dc/2} \Gamma\big(\frac{\dc}{\alpha}\big)}{\alpha\,  \Gamma\big(\frac{\dc}{2}\big)} \frac{1}{(s \mu^2)^{\dc/\alpha}}\,, 
	\end{equation}
	where we have dropped terms that are subleading in $\mu$. Then, 
	\begin{equation}
		S_\nu(\varphi) \sim  m_0^{4 - 2 \nu} \left[\frac{2 \pi^{\dc/2}  \Gamma\big(\frac{\dc}{\alpha} + \nu\big)}{\alpha \, \mu^{ 2 \dc /\alpha}\, \Gamma\big(\frac{\dc}{2}\big)}\right] \int\!\!\frac{\ud^d k}{(2 \pi)^d} \int_{0}^{\infty}\!\ud s\, s^{1- \nu - \dc/\alpha} \, \lab{e}^{-s(\mb{k}^2 + m_0^2)} \,,\mathrlap{\qquad \mu^2 \to 0}\,.
	\end{equation}
	The integral over $s$ diverges and we must regulate it. This is the same divergence we ran into in our previous example---there are an infinite number of fields and the metric is extensive in the total number of degrees of freedom. Since we are only interested in scaling with the UV cutoff $\Lambda_\slab{uv}$, the mass-spacing $\mu^2$, and the fiducial mass $m_0$, we will regulate the integral by keeping $\nu$ arbitrary, evaluating the integral, and throwing away the (generally infinite) prefactor $\Gamma(2 - \nu - \frac{\dc}{\alpha})$, leaving us with
	\begin{equation}
		S_\nu (\varphi) \propto \frac{m_0^{4 - 2 \nu}}{\mu^{2 \dc/\alpha}}  \int\!\!\frac{\ud^d k}{(2 \pi)^d} \frac{1}{(\mb{k}^2 + m_0^2)^{2 - \nu - \frac{\dc}{\alpha}}} \propto \Lambda_\slab{uv}^{d}  \left(\frac{\Lambda_\slab{uv}}{\mu}\right)^{\!\frac{2 \dc}{\alpha}}\! \left(\frac{m_0}{\Lambda_\slab{uv}}\right)^{4-2 \nu}\,.
	\end{equation}
	As $\mu^2(\varphi) \to 0$, the $\nu = 2$ contribution dominates and the metric simplifies to
	\begin{equation}
		g_{ab}(\varphi) \propto (L \Lambda_\slab{uv})^d \left({\Lambda_\slab{uv}}/{\mu}\right)^{\frac{2\dc}{\alpha}} \frac{\partial_a \mu^2 \, \partial_b \mu^2}{\mu^4}\,. \label{eq:scalarTowerMet}
	\end{equation}
	The intensive metric again takes the form (\ref{eq:towerIntensive}) and has an infinite distance point, independent of the spatial dimension and the dimension of the lattice.\footnote{It is fun to note that, if we replace $(\Lambda_\slab{uv}/\mu)^{2 \dc/\alpha} \to (R \Lambda_\slab{uv})^{2 \dc/\alpha}$, we can ``see'' the appearance of $2 \dc/\alpha$ spatial dimensions. For instance, in the case of $f(|\mb{q}|) \propto \mb{q}^2$ with $\alpha = 2$, then the dimension of this new space is evidently the dimension $\dc$ of the lattice. This matches what actually happens in, for instance, toroidal compactifications.} 

	It is simple to extend this result to a lattice of non-interacting Dirac fermions in $(d+1)$-dimensions, whose theory is specified by the Euclidean action (cf. \S\ref{sec:dirac})
	\begin{equation}
		S_\slab{e}[\varphi] = \sum_{\mb{q} \in \Gamma} \int\!\ud^{d+1} x\, \bar{\Psi}_{\mb{q}}\big[\minus i \tilde{\gamma}_\mu \partial_\mu + m_{\mb{q}}(\varphi)\big]\Psi_{\mb{q}}\,.
	\end{equation}
	As in the scalar case, each fermionic field $\Psi_\mb{q}(\tau, \mb{x})$ is labeled by an element $\mb{q}$ of the charge lattice~${\Gamma}$, with mass
	\begin{equation}
		m_{\mb{q}}(\varphi) = m_0 + \mu(\varphi) f(\mb{q})\,.
	\end{equation} 
	We again assume that $f(\mb{q}) = f(|\mb{q}|)$ is a positive, smooth function on the lattice which asymptotes to $f(|\mb{q}|) \sim |\mb{q}|^\alpha$ as $|\mb{q}| \to \infty$. Since $m_0$'s $\varphi$-dependence played no role in our discussion of the scalar tower's metric, we will take the reference mass to be independent of $\varphi$.

	The metric associated with $\varphi^a \to \varphi^a + \ud \varphi^a$ is again the sum the individual metrics~(\ref{eq:diracMet}),
	\begin{equation}
		g_{ab}(\varphi) = \frac{d_\lab{f} L^d}{8} \sum_{\mb{q} \in \Gamma}  \frac{\partial_a \mu \, \partial_b \mu}{\mu^2} \int\!\!\frac{\ud^d k}{(2 \pi)^d} \frac{\mu_\mb{q}^2 \, \mb{k}^2}{(\mb{k}^2 + m_0^2 + \mu_{\mb{q}}^2 )^2}\,,
	\end{equation}
	which, with the techniques we used previously, we can rewrite as
	\begin{equation}
		g_{ab}(\varphi) = \frac{d_\lab{f} L^d}{8} \frac{\partial_a \mu \, \partial_b \mu}{\mu^2}    \int\!\!\frac{\ud^d k}{(2 \pi)^d} \int_{0}^{\infty}\!\ud s\, s\,  \mb{k}^2 \lab{e}^{-s(\mb{k}^2 + m_0^2)} (\minus \partial_s)\!\es \Bigg[\es\sum_{\mb{q} \in \Gamma} \lab{e}^{-s \mu_{\mb{q}}^2}\es\Bigg]\,.
	\end{equation}
	This is very similar to the scalar analog (\ref{eq:sInterim}), aside from an additional factor of $\mb{k}^2$. We again assume that the continuum approximation is valid so that we can restrict the lattice sum to the origin of the dual lattice,
	\begin{equation}
		g_{ab}(\varphi) \propto \frac{L^d}{\mu^{2\dc/\alpha}} \frac{\partial_a \mu \, \partial_b \mu}{\mu^2}    \int\!\!\frac{\ud^d k}{(2 \pi)^d} \int_{0}^{\infty}\!\ud s\, s^{-\dc/\alpha} \,  \mb{k}^2 \lab{e}^{-s(\mb{k}^2 + m_0^2)}\,,
	\end{equation}
	in which case the metric takes the same form as (\ref{eq:scalarTowerMet}),
	\begin{equation}
		 g_{ab}(\varphi) \propto (L \Lambda_\slab{uv})^d \left({\Lambda_\slab{uv}}/{\mu}\right)^{\frac{2 \dc}{\alpha}} \frac{\partial_a \mu \, \partial_b \mu}{\mu^2}\,.
	\end{equation}
	Evidently, the fermionic nature of the tower has little effect on the behavior of the information metric as we approach the infinite distance point. That is, the parametric behavior of the intensive metric is independent of the spatial dimension, the lattice dimension, and the spin/statistics of the fields that comprise the lattice, at least for free scalar and fermionic fields. Such details affect the constants of proportionality, but we are not interested in those regularization scheme-dependent quantities here.

	Before we move on, it is interesting to understand how the tower sidesteps the conclusions of the scaling analysis performed in \S\ref{sec:metricScaling}. If we set the fiducial mass to zero, $m_0^2 = 0$, then the correlation length of the theory seemingly diverges as the mass-spacing vanishes $\mu^2(\varphi) \to 0$, and so we might expect that the scaling analysis applies. Let us again restrict to a simple one-dimensional lattice with $m_\ell^2 = \mu^2(\varphi) \ell^2$. Upon $\varphi^a \to \varphi^a + \ud \varphi^a$, the action is deformed (\ref{eq:defLag}) by the collective operator
	\begin{equation}
		\mathcal{O}_a(\tau, \mb{x}) = \frac{1}{2}\es\partial_a \mu^2 \sum_{\ell \in \mathbb{Z}} \ell^2 \,\nord{\Phi^2_{\ell}}(\tau, \mb{x})
	\end{equation}
	where we have used normal-ordering to define the composite operator and ensure that the one-point function vanishes identically, $\langle \mathcal{O}_a(\tau, \mb{x}) \rangle = 0$. Each summand has scaling dimension $\Delta = d-1$, as does the entire sum. The relevant correlation function, cf. (\ref{eq:eucScaleInvarCorr}), is
	\begin{equation}
		\langle \mathcal{O}_a(\tau, \mb{x}) \mathcal{O}_b(0) \rangle = \frac{1}{4}\es \partial_a \mu^2 \,\partial_b \mu^2 \sum_{\ell, \ell' \in \mathbb{Z}} \ell^2 \ell'^2 \langle \nord{\Phi_{\ell}^2}(\tau, \mb{x}) \nord{\Phi_{\ell'}^2}(0) \rangle\,,
	\end{equation}
	which, using Wick's theorem and the free-field Euclidean correlator, can be written as
	\begin{equation}
		\langle \mathcal{O}_a(\tau, \mb{x}) \mathcal{O}_b(0) \rangle
		= \frac{\partial_a \mu^2 \, \partial_b \mu^2}{2 \es \mu^4} \sum_{\ell \in \mathbb{Z}} \frac{(\mu |\ell|)^{d+3}}{(2 \pi)^{d+1} x^{d-1}} K_{\frac{d-1}{2}}( \mu |\ell| x)^2
	\end{equation}
	where $K_\nu(z)$ is the second modified Bessel function and $x\equiv \sqrt{\tau^2 + \mb{x}^2}$ is the Euclidean norm. In the $\mu^2(\varphi) \to 0$ limit, we can again approximate the sum as an integral and find
	\begin{equation}
		\langle \mathcal{O}_a(x) \mathcal{O}_b(0) \rangle \sim \frac{3\es\Gamma\big(\frac{d+4}{2}\big) \Gamma\big(\frac{2d+3}{2}\big)}{2^5 (2 \pi)^d \Gamma\big(\frac{d+5}{2}\big)} \frac{\partial_a \mu^2 \es \partial_b \mu^2}{\es\mu^4} \frac{1}{\mu x} \frac{1}{x^{2 d +2}}.
	\end{equation}
	In the limit of vanishing mass-spacing, the tower effectively acts like a perturbation with scaling dimension $\Delta = d + \frac{3}{2}$, where $\mu^2$ is used as the appropriate dimensionful scale. Comparing with~(\ref{eq:eucCorrAsymp}), we see that the asymptotic form in the $\mu^2(\varphi) \to 0$ limit does not approach what we would naively expect, especially since we are naively perturbing the action by a mass perturbation with  $\Delta = d-1$. This change in the effective scaling dimension of the perturbation and the fact that the relevant scale is $\mu$ and not $\Lambda_\slab{uv}$, both of which cannot be inferred from a simple scaling analysis, is ultimately what enables the infinite distance singularity to emerge.

\section{Implications for the Swampland} \label{sec:swamp}
	
	While we were generally interested in understanding the geometry of arbitrary continuous families of quantum theories, which were not necessarily coupled to gravity, one of the primary motivations for this work was to gain a better intuitive understanding of the Swampland Distance Conjecture~\cite{Ooguri:2006in} and its variants.\footnote{See \cite{Brennan:2017rbf,Palti:2019pca,vanBeest:2021lhn} for helpful reviews on the Swampland program and its interrelated conjectures.} In the previous section, we saw that both the metric on field space and the Zamolodchikov metric were proportional to the (intensive) quantum information metric. The distance conjectures are statements about the behaviors of these metrics, and so it is interesting to understand what their interpretation as information metrics implies about these conjectures. We will find that it suggests a simple bottom-up interpretation of the distance conjecture. In particular, it will suggest an intuitive explanation of why points at which global symmetries are restored are necessarily placed at infinite distance in consistent quantum gravitational theories.

	Let us consider a $(d+1)$-dimensional effective field theory of $n$ massless scalar fields $\phi^a$ coupled to gravity, with field space metric $G_{ab}(\phi)$ and action~\cite{vanBeest:2021lhn}
	\begin{equation}
		S = \frac{ M_\lab{pl}^{d-1}}{2} \int\!\ud^{d+1} x\, \sqrt{\minus h} \left(R - G_{ab}(\phi)\es  \nabla_\mu \phi^a \, \nabla^\mu \phi^b\right)\,. \label{eq:gravNLSM}
	\end{equation}
	Here, $h_{\mu \nu}$ is the metric on spacetime and $R$ is its associated Ricci scalar.
	The most basic form of the Swampland Distance Conjecture \cite{Ooguri:2006in} states that an infinite tower of fields become exponentially light as we approach any infinite distance limit in the metric on field space, such that
	\begin{equation}
		M(\varphi_1) \sim M(\varphi_0)\, \lab{e}^{-\lambda\es d(\varphi_1, \varphi_0)}\,,\mathrlap{\qquad d(\varphi_1, \varphi_0) \to \infty\,,}
	\end{equation}
	where $d(\varphi_1, \varphi_0)$ is the length
	\begin{equation}
		d(\varphi_1, \varphi_0) = \int_{0}^1 \!\ud t\, \sqrt{G_{ab}(\varphi) \es \dot{\varphi}^a(t) \es \dot{\varphi}^b(t)}
	\end{equation}
	of the geodesic curve $\varphi^a(t)$  connecting the two points $\varphi^a(1) = \varphi^a_1$ and $\varphi^a(0) = \varphi^a_0$ in field space. There is a similar conjecture on the behavior of the Zamolodchikov metric near infinite distance points, where a tower of higher spin fields appears \cite{Seiberg:1999xz,Baume:2020dqd,Perlmutter:2020buo}.

	The effective theory (\ref{eq:gravNLSM}) is chosen to capture the low-energy dynamics of a UV complete quantum gravitational theory. Given that quantum gravitational theory and an impressive amount of computational power, we could compute the quantum information metric associated to changing the moduli $\langle \phi^a \rangle = \varphi^a$ in any of the ways discussed in Section~\ref{sec:qim}. We could also compute the information metric using (\ref{eq:gravNLSM}), at least at low energies, in which case we would deduce that
	\begin{equation}
		g_{ab}(\varphi) \propto (L M_\lab{pl})^d G_{ab}(\varphi)\,.
	\end{equation}
	We might be worried that the UV cutoff used in (\ref{eq:nlsmProfile}) should be taken to be smaller than the Planck scale, like at the string scale $\Lambda_\slab{uv} = M_s$ or the Kaluza-Klein scale $\Lambda_\slab{uv} = M_\slab{kk}$, in which case $g_{ab}(\varphi) \propto (L \Lambda_\slab{uv})^{d} (M_\lab{pl}/\Lambda_\slab{uv})^{d-1} G_{ab}(\varphi)$, but this will not change the fact that the intensive information metric is proportional to the metric on field space even though it depends precisely on how (\ref{eq:gravNLSM}) is defined. A similar conclusion holds for the Zamolodchikov metric, though it is much easier to think about the relevant scales in that context.

	Since the intensive information metric measures, on average, how distinguishable two nearby theories are if each are in their vacuum state, it may provide a simple bottom-up interpretation of why infinite distance points arise in quantum gravity---these points are infinite distance because they are hyper-distinguishable. This would be satisfying, as it agrees with the intuition the theories at infinite distance are qualitatively \emph{different}---they correspond to an emerging dimension, or the appearance of a tensionless string \cite{Seiberg:1999xz,Grimm:2018ohb,Lee:2018urn}. Indeed, we might expect that the reason theories with an exact, continuous global symmetry are at infinite distance is because they are so readily distinguishable: a single measurement of charge non-conservation is enough information to completely rule out the theory. The quantum gravity novelty is that these points are truly at infinite distance, and that it only takes a small number of local measurements---in the sense discussed in \S\ref{sec:metricReg}---to distinguish them. It would be worthwhile to make this picture more precise in the future by better clarifying the different degrees of distinguishability and identifying the relevant observables that may most easily discern between the different theories.

	Much like in the emergence proposal \cite{Heidenreich:2018kpg,Grimm:2018ohb}, the tower of exponentially light fields ensures that these points are actually at infinite distance in the intensive metric.\footnote{See \cite{Heidenreich:2017sim,Harlow:2015lma} for similar ideas applied to the Weak Gravity Conjecture. We should note that our story differs slightly from these works on emergence, in that they take the Planck scale $M_\lab{pl}$ to be fixed for all $\varphi^a$. This forces the UV cutoff $\Lambda_\slab{uv}(\varphi)$ of the theory to vary, possibly dramatically, as one moves around parameter space. In contrast, we consider the emergence of an infinite distance point at fixed $\Lambda_\slab{uv}$. However, we stress that there is no obstruction to considering the former using these information theoretic techniques---they are just different family of theories. We thank Matt Reece for bringing this point to our attention.} Such a bottom-up explanation would thus require arguing that the only way an infinite distance point can appear is via such a tower of fields. It is unclear whether or not a theory can become strongly-coupled enough to generate these infinite distance points, and this a question that deserves future work. We expect, however, that this is not possible. Usually, if the coupling is strong enough the Hilbert space rearranges itself into weakly-interacting degrees of freedom where the ``strong force'' is screened, and there is a weakly-coupled dual description that becomes more useful.

	This perspective could help resolve a few conceptual puzzles surrounding the Swampland Distance Conjecture and its variants, and help extend it beyond the different highly symmetric lampposts under which most of our understanding has taken place. The definition of this information metric does not rely upon any sort of (super)symmetry and can be defined with respect to arbitrary deformations of the theory, as long they do not introduce an inconsistency. This means that there is no trouble in defining this metric in a class of theories with a potential---it still yields the same interpretation in terms of distinguishability. The information metric simply provides a method of studying families of physical theories, which encodes the relations between them geometrically. In particular, the expectation that various other distinct limits of quantum gravity should correspond to infinite distance limits in \emph{some} metric, like the AdS distance conjecture \cite{Lust:2019zwm}, should be easily translatable into this language.
	
	Furthermore, while the information metric is only defined for continuous families of theories, other statistical divergences could be used to study families that depend on discrete parameters. Without resorting to contrived thought experiments involving balls and urns, it is difficult to come up a family of classical distributions that provides an example of this.  However, such families are ubiquitous in string theory in the context of flux vacua. As long as members of a family make predictions for the same observables, so that it makes sense to compare them, we can define a statistical distance between them. In light of our computation of the log fidelity using the Euclidean path integral, presented in \S\ref{sec:sho} and \S\ref{sec:nlsm}, we might expect that this distance is related to the action of an instanton that interpolates between these different discrete vacua. It would be interesting to explore this further in the future.

\section{Conclusions} \label{sec:conclusions}

	In retrospect, it is not so surprising that information-theoretic tools show up naturally when we study effective theories of quantum gravity. At their heart, effective field theories are reduced descriptions of the UV theory. How do we find a probability distribution (specified by an effective action) which yields predictions close to that of the true distribution (the UV theory), albeit simpler? This is not unlike the general task of statistical inference or machine learning, in which one tries to approximate the true probability distribution from which a data set is drawn using an approximate distribution that is both close by and, crucially, structured so that it can be easily computed. It is natural to expect that tools developed to study one would be helpful to study the other.

	The goal of this paper was to study the information metric associated to the vacua of general, continuous families of quantum field theories, with a specific interest in decoding the meaning of infinite distance limits in this metric. The information metric bestows a notion of distance on a space of theories that reduces to the familiar Zamolodchikov and moduli space metrics when restricted appropriately. However, it can be defined more generally and allows us to understand infinite distance in terms of hyper-distinguishability. The further apart two theories are, the easier it is to distinguish them. 

	We ran into an interesting puzzle: why is the zero frequency limit of the simple harmonic oscillator at infinite distance, but not the massless limit of a free quantum field? We found that typical quantum critical points, i.e. those that obey the scaling hypothesis, are never at infinite distance and we saw how, at least for the free scalar field, the spatial dependence of the field could wash out this distinguishability. Finally, we studied how a tower of free fields could collectively restore the hyper-distinguishability, and thus the infinite distance, of this limit.

	The point of view presented here raises many interesting questions and future directions: 
	\begin{itemize}
		\item In order to make the connection between the Swampland Distance Conjecture and hyper-distinguishability more rigorous, it would be helpful to identify an optimal set of measurements that allow one to readily distinguish between an infinite distance point and others. That is, it is not clear \emph{which} measurements are useful and this paper does not identify them. When the infinite distance point is associated to the restoration of a global symmetry, it seems intuitively obvious that such measurements should be associated to charge conservation. However, there must be a qualitative differences between finite and infinite distance points and it is not clear which measurements optimally reflect this. For instance, which observables most readily distinguish the decompactification limit?

		\item Our focus on infinite distance points allowed us to ignore the constant factors in front of the metric. That is, we only studied the meaning of the infinite distance points themselves from the bottom-up, and not the rate at which the distance diverged. Are there other information theoretic aspects of quantum gravitational theories that force physical distances to be measured with respect to the Planck scale $M_\lab{pl}$? We know that quantum gravity places limits on \emph{what} can be measured---for instance, we cannot probe extremely small length scales due to black hole production---and it is reasonable to expect that it also limits how \emph{well} we can measure something. Similarly, is there a preferred regularization/renormalization scheme for computing the information metric? Can the constant factors we ignored be given a precise bottom-up interpretation?  

		\item The quantum information metric, as well as general distances between states, show up naturally in the study of complexity and have well-defined holographic analogs \cite{MIyaji:2015mia,Alishahiha:2015rta,Chapman:2017rqy,Belin:2018bpg}. Could complexity be used to justify swampland constraints? Do swampland constraints provide complexity bounds?

		\item In Section~\ref{sec:cim}, we saw that there was a duality between the two parameterizations of the exponential family. Specifically, when the information metric with respect to one parameterization vanished or degenerated so that model at different values of the parameter became indistinguishable,\footnote{Such statistical models are called \emph{singular statistical models}~\cite{watanabe2009algebraic}. Almost all statistical models or learning machines are singular. The structure of these singularities govern, for example, how quickly these statistical models can learn a given data set.} the metric with respect the other diverged and different models became hyper-distinguishable. Can this duality, or other information-theoretic techniques, be used to identify useful weakly-coupled descriptions of these infinite distance points? Can we use information-theoretic techniques to determine when a theory has a weakly-coupled description? Is this related to the fact that (\ref{eq:scalarTowerMet}) can be interpreted as the sum of $\mathcal{N}$ Gaussian metrics? How is this related to the holographic picture of~\cite{Grimm:2020cda}? 

		\item Are there other ways of realizing infinite distance points in families of quantum field theories, not necessarily coupled to quantum gravity? Such points require a violation of standard scaling behavior---what violations are consistent in relativistic theories? Non-relativistic field theories? Non-field theories? As described in Section~\ref{sec:qim}, a K\"{a}hler potential can always be defined for the vacuum submanifold associated to a set of quantum theories. The classification of infinite distance points leveraged in~\cite{Grimm:2018ohb} only relies this complex geometry, and so we expect that this classification extends beyond supersymmetric string compactifications and towards families of general quantum theories. There is a well-defined classification and understanding of finite-distance quantum critical points in terms of their critical exponents and the renormalization group. Infinite-distance critical points are obviously not as well-understood---do the periods of~\cite{Grimm:2018ohb} provide a complete classification or, like their finite-distance counterparts, is there a more refined description? What organizing principle should we use to analyze them?

	\end{itemize}

	We hope to address some of these fascinating questions in the future.

\subsection*{Acknowledgments}
	It is a pleasure to thank Alex Alemi, Tarek Anous, Alex Belin, Alex Cole, Markus Dierigl, Austin Joyce, Cody Long, Liam McAllister, Miguel Montero, Matt Reece, Irene Valenzuela for many helpful conversations. We are particularly indebted to Alex Alemi, Cody Long, and Matt Reece for comments on a draft. This work is supported by NASA grant \texttt{80NSSC20K0506}.

\newpage

\phantomsection
\addcontentsline{toc}{section}{References}
\bibliographystyle{utphys}
\bibliography{sdc}

\end{document}